\documentclass[aps, pra, twocolumn, tightenlines, letterpaper, amsmath, amssymb, preprintnumbers, superscriptaddress, floatfix, longbibliography, nofootinbib]{revtex4-2}
\usepackage[T1]{fontenc}
\usepackage{xspace}
\usepackage[dvipsnames]{xcolor}
\usepackage{graphicx}
\usepackage{tikz}
\usepackage[caption=false]{subfig}
\usepackage{multirow}
\usepackage{bbold}
\usepackage[mathscr]{eucal}
\usepackage{multirow}
\usepackage{pdfpages}

\makeatletter
\AtBeginDocument{\let\LS@rot\@undefined}
\makeatother

\def\cbar{\mathscr{C}}

\usepackage[hypertexnames=false]{hyperref}
\hypersetup{
    colorlinks=true,       
    linkcolor=blue,          
    citecolor=blue,        
    filecolor=blue,      
    urlcolor=blue           
}
\usepackage{mathtools}
\usepackage{orcidlink}
\usepackage{mmacells}
\usepackage[breakable]{tcolorbox}

\definecolor{incolor}{HTML}{303F9F}
\definecolor{outcolor}{HTML}{D84315}
\definecolor{cellborder}{HTML}{CFCFCF}
\definecolor{cellbackground}{HTML}{F7F7F7}
\definecolor{cellbackgroundmma}{HTML}{FFEEEA}
\makeatletter
\def\PY@reset{\let\PY@it=\relax \let\PY@bf=\relax%
    \let\PY@ul=\relax \let\PY@tc=\relax%
    \let\PY@bc=\relax \let\PY@ff=\relax}
\def\PY@tok#1{\csname PY@tok@#1\endcsname}
\def\PY@toks#1+{\ifx\relax#1\empty\else%
    \PY@tok{#1}\expandafter\PY@toks\fi}
\def\PY@do#1{\PY@bc{\PY@tc{\PY@ul{%
    \PY@it{\PY@bf{\PY@ff{#1}}}}}}}
\def\PY#1#2{\PY@reset\PY@toks#1+\relax+\PY@do{#2}}

\@namedef{PY@tok@w}{\def\PY@tc##1{\textcolor[rgb]{0.73,0.73,0.73}{##1}}}
\@namedef{PY@tok@c}{\let\PY@it=\textit\def\PY@tc##1{\textcolor[rgb]{0.25,0.50,0.50}{##1}}}
\@namedef{PY@tok@cp}{\def\PY@tc##1{\textcolor[rgb]{0.74,0.48,0.00}{##1}}}
\@namedef{PY@tok@k}{\let\PY@bf=\textbf\def\PY@tc##1{\textcolor[rgb]{0.00,0.50,0.00}{##1}}}
\@namedef{PY@tok@kp}{\def\PY@tc##1{\textcolor[rgb]{0.00,0.50,0.00}{##1}}}
\@namedef{PY@tok@kt}{\def\PY@tc##1{\textcolor[rgb]{0.69,0.00,0.25}{##1}}}
\@namedef{PY@tok@o}{\def\PY@tc##1{\textcolor[rgb]{0.40,0.40,0.40}{##1}}}
\@namedef{PY@tok@ow}{\let\PY@bf=\textbf\def\PY@tc##1{\textcolor[rgb]{0.67,0.13,1.00}{##1}}}
\@namedef{PY@tok@nb}{\def\PY@tc##1{\textcolor[rgb]{0.00,0.50,0.00}{##1}}}
\@namedef{PY@tok@nf}{\def\PY@tc##1{\textcolor[rgb]{0.00,0.00,1.00}{##1}}}
\@namedef{PY@tok@nc}{\let\PY@bf=\textbf\def\PY@tc##1{\textcolor[rgb]{0.00,0.00,1.00}{##1}}}
\@namedef{PY@tok@nn}{\let\PY@bf=\textbf\def\PY@tc##1{\textcolor[rgb]{0.00,0.00,1.00}{##1}}}
\@namedef{PY@tok@ne}{\let\PY@bf=\textbf\def\PY@tc##1{\textcolor[rgb]{0.82,0.25,0.23}{##1}}}
\@namedef{PY@tok@nv}{\def\PY@tc##1{\textcolor[rgb]{0.10,0.09,0.49}{##1}}}
\@namedef{PY@tok@no}{\def\PY@tc##1{\textcolor[rgb]{0.53,0.00,0.00}{##1}}}
\@namedef{PY@tok@nl}{\def\PY@tc##1{\textcolor[rgb]{0.63,0.63,0.00}{##1}}}
\@namedef{PY@tok@ni}{\let\PY@bf=\textbf\def\PY@tc##1{\textcolor[rgb]{0.60,0.60,0.60}{##1}}}
\@namedef{PY@tok@na}{\def\PY@tc##1{\textcolor[rgb]{0.49,0.56,0.16}{##1}}}
\@namedef{PY@tok@nt}{\let\PY@bf=\textbf\def\PY@tc##1{\textcolor[rgb]{0.00,0.50,0.00}{##1}}}
\@namedef{PY@tok@nd}{\def\PY@tc##1{\textcolor[rgb]{0.67,0.13,1.00}{##1}}}
\@namedef{PY@tok@s}{\def\PY@tc##1{\textcolor[rgb]{0.73,0.13,0.13}{##1}}}
\@namedef{PY@tok@sd}{\let\PY@it=\textit\def\PY@tc##1{\textcolor[rgb]{0.73,0.13,0.13}{##1}}}
\@namedef{PY@tok@si}{\let\PY@bf=\textbf\def\PY@tc##1{\textcolor[rgb]{0.73,0.40,0.53}{##1}}}
\@namedef{PY@tok@se}{\let\PY@bf=\textbf\def\PY@tc##1{\textcolor[rgb]{0.73,0.40,0.13}{##1}}}
\@namedef{PY@tok@sr}{\def\PY@tc##1{\textcolor[rgb]{0.73,0.40,0.53}{##1}}}
\@namedef{PY@tok@ss}{\def\PY@tc##1{\textcolor[rgb]{0.10,0.09,0.49}{##1}}}
\@namedef{PY@tok@sx}{\def\PY@tc##1{\textcolor[rgb]{0.00,0.50,0.00}{##1}}}
\@namedef{PY@tok@m}{\def\PY@tc##1{\textcolor[rgb]{0.40,0.40,0.40}{##1}}}
\@namedef{PY@tok@gh}{\let\PY@bf=\textbf\def\PY@tc##1{\textcolor[rgb]{0.00,0.00,0.50}{##1}}}
\@namedef{PY@tok@gu}{\let\PY@bf=\textbf\def\PY@tc##1{\textcolor[rgb]{0.50,0.00,0.50}{##1}}}
\@namedef{PY@tok@gd}{\def\PY@tc##1{\textcolor[rgb]{0.63,0.00,0.00}{##1}}}
\@namedef{PY@tok@gi}{\def\PY@tc##1{\textcolor[rgb]{0.00,0.63,0.00}{##1}}}
\@namedef{PY@tok@gr}{\def\PY@tc##1{\textcolor[rgb]{1.00,0.00,0.00}{##1}}}
\@namedef{PY@tok@ge}{\let\PY@it=\textit}
\@namedef{PY@tok@gs}{\let\PY@bf=\textbf}
\@namedef{PY@tok@gp}{\let\PY@bf=\textbf\def\PY@tc##1{\textcolor[rgb]{0.00,0.00,0.50}{##1}}}
\@namedef{PY@tok@go}{\def\PY@tc##1{\textcolor[rgb]{0.53,0.53,0.53}{##1}}}
\@namedef{PY@tok@gt}{\def\PY@tc##1{\textcolor[rgb]{0.00,0.27,0.87}{##1}}}
\@namedef{PY@tok@err}{\def\PY@bc##1{{\setlength{\fboxsep}{\string -\fboxrule}\fcolorbox[rgb]{1.00,0.00,0.00}{1,1,1}{\strut ##1}}}}
\@namedef{PY@tok@kc}{\let\PY@bf=\textbf\def\PY@tc##1{\textcolor[rgb]{0.00,0.50,0.00}{##1}}}
\@namedef{PY@tok@kd}{\let\PY@bf=\textbf\def\PY@tc##1{\textcolor[rgb]{0.00,0.50,0.00}{##1}}}
\@namedef{PY@tok@kn}{\let\PY@bf=\textbf\def\PY@tc##1{\textcolor[rgb]{0.00,0.50,0.00}{##1}}}
\@namedef{PY@tok@kr}{\let\PY@bf=\textbf\def\PY@tc##1{\textcolor[rgb]{0.00,0.50,0.00}{##1}}}
\@namedef{PY@tok@bp}{\def\PY@tc##1{\textcolor[rgb]{0.00,0.50,0.00}{##1}}}
\@namedef{PY@tok@fm}{\def\PY@tc##1{\textcolor[rgb]{0.00,0.00,1.00}{##1}}}
\@namedef{PY@tok@vc}{\def\PY@tc##1{\textcolor[rgb]{0.10,0.09,0.49}{##1}}}
\@namedef{PY@tok@vg}{\def\PY@tc##1{\textcolor[rgb]{0.10,0.09,0.49}{##1}}}
\@namedef{PY@tok@vi}{\def\PY@tc##1{\textcolor[rgb]{0.10,0.09,0.49}{##1}}}
\@namedef{PY@tok@vm}{\def\PY@tc##1{\textcolor[rgb]{0.10,0.09,0.49}{##1}}}
\@namedef{PY@tok@sa}{\def\PY@tc##1{\textcolor[rgb]{0.73,0.13,0.13}{##1}}}
\@namedef{PY@tok@sb}{\def\PY@tc##1{\textcolor[rgb]{0.73,0.13,0.13}{##1}}}
\@namedef{PY@tok@sc}{\def\PY@tc##1{\textcolor[rgb]{0.73,0.13,0.13}{##1}}}
\@namedef{PY@tok@dl}{\def\PY@tc##1{\textcolor[rgb]{0.73,0.13,0.13}{##1}}}
\@namedef{PY@tok@s2}{\def\PY@tc##1{\textcolor[rgb]{0.73,0.13,0.13}{##1}}}
\@namedef{PY@tok@sh}{\def\PY@tc##1{\textcolor[rgb]{0.73,0.13,0.13}{##1}}}
\@namedef{PY@tok@s1}{\def\PY@tc##1{\textcolor[rgb]{0.73,0.13,0.13}{##1}}}
\@namedef{PY@tok@mb}{\def\PY@tc##1{\textcolor[rgb]{0.40,0.40,0.40}{##1}}}
\@namedef{PY@tok@mf}{\def\PY@tc##1{\textcolor[rgb]{0.40,0.40,0.40}{##1}}}
\@namedef{PY@tok@mh}{\def\PY@tc##1{\textcolor[rgb]{0.40,0.40,0.40}{##1}}}
\@namedef{PY@tok@mi}{\def\PY@tc##1{\textcolor[rgb]{0.40,0.40,0.40}{##1}}}
\@namedef{PY@tok@il}{\def\PY@tc##1{\textcolor[rgb]{0.40,0.40,0.40}{##1}}}
\@namedef{PY@tok@mo}{\def\PY@tc##1{\textcolor[rgb]{0.40,0.40,0.40}{##1}}}
\@namedef{PY@tok@ch}{\let\PY@it=\textit\def\PY@tc##1{\textcolor[rgb]{0.25,0.50,0.50}{##1}}}
\@namedef{PY@tok@cm}{\let\PY@it=\textit\def\PY@tc##1{\textcolor[rgb]{0.25,0.50,0.50}{##1}}}
\@namedef{PY@tok@cpf}{\let\PY@it=\textit\def\PY@tc##1{\textcolor[rgb]{0.25,0.50,0.50}{##1}}}
\@namedef{PY@tok@c1}{\let\PY@it=\textit\def\PY@tc##1{\textcolor[rgb]{0.25,0.50,0.50}{##1}}}
\@namedef{PY@tok@cs}{\let\PY@it=\textit\def\PY@tc##1{\textcolor[rgb]{0.25,0.50,0.50}{##1}}}

\def\PYZus{\char`\_}
\def\PYZob{\char`\{}
\def\PYZcb{\char`\}}

\def\PYZlt{\char`\<}
\def\PYZgt{\char`\>}

\def\PYZhy{\char`\-}
\def\PYZsq{\char`\'}
\def\PYZdq{\char`\"}

\makeatother

\makeatletter
\newcommand{\boxspacing}{\kern\kvtcb@left@rule\kern\kvtcb@boxsep}
\makeatother
\newcommand{\prompt}[4]{
    {\ttfamily\llap{{\color{#2}[#3]:\hspace{3pt}#4}}\vspace{-\baselineskip}}
}

\makeatletter
\renewcommand\onecolumngrid{
\do@columngrid{one}{\@ne}
\def\set@footnotewidth{\onecolumngrid}
\def\footnoterule{\kern-6pt\hrule width 1.5in\kern6pt}%
}
\makeatother

\newcommand{\beginsupplement}{%
        \renewcommand{\thefigure}{S\arabic{figure}}
        \renewcommand{\theHfigure}{S\arabic{figure}}
        \setcounter{figure}{0}
        \renewcommand{\thetable}{S\arabic{table}}
        \renewcommand{\theHtable}{S\arabic{table}}
        \setcounter{table}{0}
        \renewcommand{\theequation}{S\arabic{equation}}
        \renewcommand{\theHequation}{S\arabic{equation}}
        \setcounter{equation}{0}
        \renewcommand{\thesection}{S\arabic{section}}
        \renewcommand{\theHsection}{S\arabic{section}}
        \setcounter{section}{0}
     }
     
\makeatletter
\def\maketitle{
\@author@finish
\title@column\titleblock@produce
\suppressfloats[t]}
\makeatother

\begin{document}

\title{Basic Elements for Simulations of Standard Model Physics with Quantum Annealers: Multigrid and Clock States
}

\author{Marc~Illa\,\orcidlink{0000-0003-3570-2849}}
\email{marcilla@uw.edu}
\affiliation{InQubator for Quantum Simulation (IQuS), Department of Physics, University of Washington, Seattle, WA 98195}
\author{Martin J.~Savage\,\orcidlink{0000-0001-6502-7106}}
\email{mjs5@uw.edu}
\affiliation{InQubator for Quantum Simulation (IQuS), Department of Physics, University of Washington, Seattle, WA 98195}

\preprint{IQuS@UW-21-020}
\date{\today}

\begin{abstract} 
We explore the potential of D-Wave's quantum annealers for computing some of the basic components required for quantum simulations of Standard Model physics.
By implementing a basic multigrid (including ``zooming'') and specializing Feynman-clock algorithms, D-Wave's {\tt Advantage} is used to study harmonic and anharmonic oscillators relevant for lattice scalar field theories and effective field theories, the time evolution of a single plaquette of SU(3) Yang-Mills lattice gauge field theory, and the dynamics of flavor entanglement 
in four-neutrino systems.
\end{abstract}

\maketitle

\let\oldaddcontentsline\addcontentsline
\renewcommand{\addcontentsline}[3]{}
\section{Introduction}
\noindent
Simulations of the dynamics of quantum matter, from neutron stars to materials, which are beyond the reach of classical computation, are expected to become possible through continued advances in quantum computation.
While universal quantum computation~\cite{5392446,5391327,Benioff1980,Manin1980,Feynman1982,Fredkin1982,Feynman1986,doi:10.1063/1.881299} is essential in this quest
to simulate Standard Model physics,
the near-term devices that define the noisy intermediate-scale quantum (NISQ) era~\cite{Preskill2018quantumcomputingin}, without high-fidelity qubits and error correction, will be challenged to provide  results that can be quantitatively compared with experiment (see, for example, Refs.~\cite{Banuls:2019bmf,PRXQuantum.2.017001,Klco:2021lap}).
Much of the current research in this area is performed on gate-based quantum computers, and the alternative, adiabatic quantum computing~\cite{Finnila:1994,PhysRevE.58.5355,farhi2000quantum,Farhi:2001}, has not been explored with as much detail.
Applications for such devices, such as D-Wave's quantum annealers (QAs)~\cite{johnson2011quantum}, are optimization~\cite{Boixo:2013,smolin2014a,venturelli2016quantum,neukart2017traffic,Nguyen:2020,Irie:2021}, high energy physics~\cite{mott2017solving,das2020track}, machine learning~\cite{neven2008training,Pudenz:2012,denchev2012robust,Gorman:2015,adachi2015application,benedetti2016estimation,Amin:2018,crawford2019reinforcement,Perdomo:2018,caldeira2019restricted,vinci2019path,rocutto2020quantum,Dixit:2021}, spin systems~\cite{gardas2018defects,doi:10.1126/science.aat2025,King:2018,PhysRevLett.124.090502,King:2021,PhysRevResearch.2.033369,PRXQuantum.1.020320,lanting2020probing,PhysRevA.102.042403,doi:10.1126/science.abe2824,king2022coherent}, quantum chemistry~\cite{hernandez2016novel,hernandez2017enhancing,xia2017electronic,doi:10.1021/acs.jctc.9b00402,Teplukhin:2020,teplukhin2021computing}, biology~\cite{PhysRevA.78.012320,perdomoortiz2012finding,Li:2018,babej2018coarsegrained}, finance~\cite{Rosenberg:2016,Grant:2021a,bouland2020prospects}, graph equations~\cite{Zick:2015,10.1145/3149526.3149531,Vert:2021}, multivariate equations~\cite{PhysRevResearch.4.013096}, integer equations~\cite{,chang2020integer}, linear equations~\cite{Chang:2019}, and factorization problems~\cite{schaller2009role,Peng:2008,PhysRevLett.108.130501,Dridi:2017,8314195,jiang2018quantum}, and more, but a quantum advantage for scientific applications remains to be demonstrated.

The first steps toward simulating quantum field theories using QAs have been taken by finding the ground state of modest SU(2) plaquette systems and the time evolution of these systems~\cite{ARahman:2021ktn} using the Feynman-clock algorithm~\cite{McCleanE3901}.
Given that, modulo emergent fine-tunings, many Standard Model systems of interest are gapped, with finite correlation lengths, quantum circuits for universal quantum computers are expected to be able to have localized control structures for which domain decomposition will be effective. 
This suggests that QAs may provide efficient pre-conditioners for preparing parametrizations of ground states and excited states for universal quantum computers (e.g., fast-forwarding time evolution~\cite{Cristina:2020}) in the future.
This, of course, remains to be demonstrated.

Using D-Wave's QAs (and simulators), we explore building blocks that are required for quantum simulations of Standard Model physics and its descendant low-energy effective field theories.
Building upon the works of Refs.~\cite{Chang:2019,ARahman:2021ktn},
a ``zooming'' algorithm is used to converge  coefficients of the basis states defining 
annealing problem instances for ground states.
Second, we use a simple multigrid
procedure that iteratively employs course  grids to provide starting conditions for  finer grids to converge wavefunctions.
Third, we generalize a previously implemented
Feynman clock algorithm~\cite{ARahman:2021ktn} to arbitrary Hermitian matrices.
These algorithms are used to simulate the harmonic oscillators (HOs) and anharmonic oscillators (AHOs), the time evolution of the SU(3) Yang-Mills plaquette~\cite{Ciavarella:2021nmj}, and neutrino flavor evolution~\cite{Hall:2021rbv}, which have been previously simulated using IBM's superconducting  quantum computers.

\section{Mapping a Hamiltonian onto a QUBO Problem}
\label{sec:qubo}
\noindent
In order to find the ground-state energy and wavefunction of a given Hamiltonian 
using D-Wave's QAs, 
a minimization problem is mapped onto a quadratic unconstrained binary optimization (QUBO) problem 
$f_Q(q)=\sum\nolimits_{ij}Q_{ij}q_iq_j$, 
where $q_i$ are binary variables. 
Following techniques and protocols for using D-Wave's systems~\cite{DwaveSystemDocs} and specific methods that users have developed~\cite{doi:10.1021/acs.jctc.9b00402, ARahman:2021ktn}, 
an objective function of the form
\begin{equation}
    F=\langle \Psi \rvert \mathcal{\hat H} \lvert \Psi \rangle -\eta \langle \Psi| \Psi \rangle 
    \ \ ,
    \label{eq:Ffun}
\end{equation}
is minimized,
where $\eta$ is a parameter that is tuned to avoid the null solution ($\langle \Psi | \Psi \rangle=0$). 
Expanding or approximating the wavefunction $\lvert \Psi \rangle$ 
in a finite-dimensional orthonormal basis 
$\lvert \psi_{\alpha}\rangle$, 
$\lvert \Psi \rangle = \sum_\alpha^{n_s} a_\alpha \lvert \psi_{\alpha} \rangle$, 
with $a_\alpha$ real numbers, 
$F$ can be written as
\begin{align}
    F=&\sum_{\alpha\beta}^{n_s}
    a_\alpha a_\beta(\langle \psi_\alpha \rvert \mathcal{\hat H} \lvert \psi_\beta \rangle -\eta \langle \psi_\alpha| \psi_\beta \rangle) \nonumber\\
    =&\sum_{\alpha\beta}^{n_s}
    a_\alpha a_\beta (\mathcal{\hat H}_{\alpha\beta} -\eta \delta_{\alpha\beta})=\sum_{\alpha\beta}^{n_s}
    a_\alpha a_\beta h_{\alpha\beta} \ \ .
    \label{eq:QAE}
\end{align}
Mapping 
the minimization of $F$ 
onto a QUBO problem appropriate for solution using an annealer requires expressing
$a_\alpha$ in terms of binary variables. 
Following previous works, the fixed-point representation~\cite{doi:10.1021/acs.jctc.9b00402} of each $a_\alpha$ in terms of $K$ bits $q^\alpha_i$ is used,
\begin{equation}
    a_\alpha=-q^{\alpha}_K + \sum_{i=1}^{K-1} \frac{q^{\alpha}_i}{2^{K-i}}\ \ ,
    \label{eq:defa}
\end{equation}
where $a_\alpha\in [-1,1)$.
Finer digitizations of $a_\alpha$, accomplished by the use of larger values of $K$, provide better resolution of the $a_\alpha$ and consequently higher precision and accuracy in solution, but are limited by device performance with increasing size of the QUBO matrix. 
This digitization of $a_\alpha$ puts the expression in Eq.~\eqref{eq:QAE} into QUBO form,
\begin{equation}
\begin{aligned}
    F=&\sum_{\alpha\beta,ij}
    Q_{\alpha,i;\beta,j}
    q^{\alpha}_i q^{\beta}_j
    \ \ , \\
    &\text{with} \quad Q_{\alpha,i;\beta,j}=2^{i+j-2K}(-1)^{\delta_{iK}+\delta_{jK}}h_{\alpha\beta}\ \ .
\label{eq:QUBO_simple}
\end{aligned}
\end{equation}
The QUBO matrix $Q_{\alpha,i;\beta,j}$, with dimensions $K n_s\times K n_s$, is subsequently passed from the D-Wave API to a simulator or D-Wave's QAs.

\begin{figure}[!t]
\centering
\includegraphics[width=0.9\columnwidth]{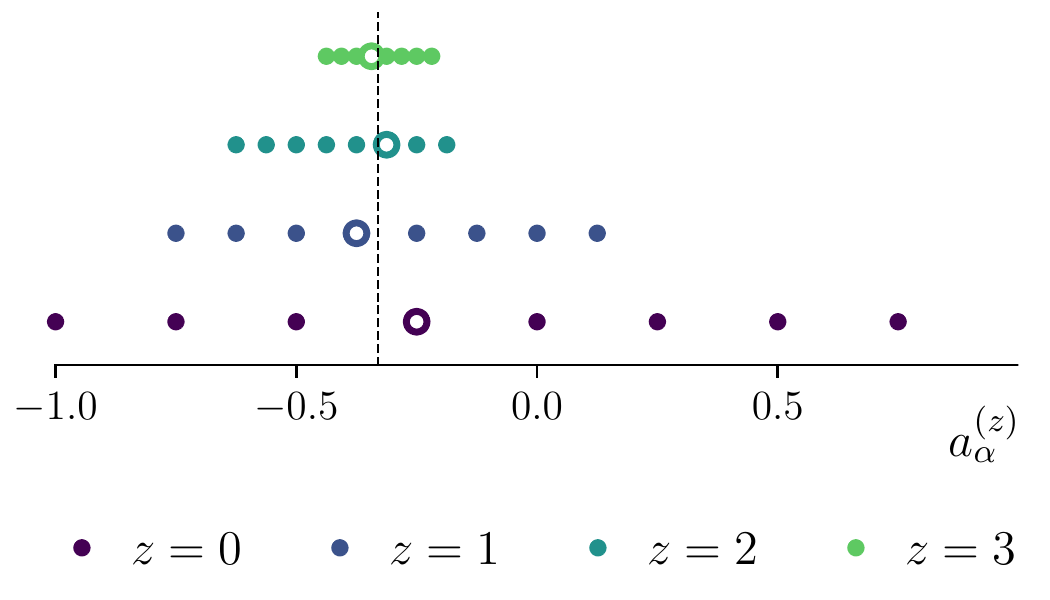}
\caption{
Example of the convergence of $a^{(z)}_\alpha$ 
from the iterative zooming method discussed in the text using $K=3$, for a true value of $a_\alpha=-0.33$ (vertical dashed line).
The closed circles denote the sampled values 
of $a^{(z)}_\alpha$ at each zoom step.
The open circles correspond to the points closest to the true value of $a_\alpha$ and are used as input for the subsequent zoom step.}
\label{fig:diffz}
\end{figure}

An adaptive QA eigenvalue (AQAE) solver,
implemented in Refs.~\cite{Chang:2019,ARahman:2021ktn},
incorporates an algorithmic improvement 
that reduces the required value of $K$ to reach a given solve precision, and hence increases the size of problem instances that can be addressed using any given QA (a similar idea was applied in Ref.~\cite{,Zlokapa:2019lvv} for machine learning).
After the initial solve for coefficients 
$a^{(z=0)}_\alpha$, 
the range of search values for 
$a^{(z+1)}_\alpha$ are systematically reduced, guided by the previously obtained 
$a^{(z)}_\alpha$.
This permits not only a reduced value for $K$ but also a reduced number of anneals at each zoom level.
We implement a relation between successive zoom steps similar to Ref.~\cite{Chang:2019}, of the form
\begin{equation}
    a^{(z+1)}_\alpha=a^{(z)}_\alpha-2^{-z}q^{\alpha}_K+\sum_{i=1}^{K-1}\frac{q^{\alpha}_i}{2^{K-i+z}}\ \ .
    \label{eq:defa_AQAE_improved_2}
\end{equation}
An example of the progressive decimation of a coefficient $a^{(z+1)}_\alpha$ with increasing zoom step is shown in Fig~\ref{fig:diffz}.
In forming the QUBO matrix at each zoom step, 
the  contributions that are naively linear in $q_i^\alpha$ 
in the product $a^{(z+1)}_\alpha a^{(z+1)}_\beta$
are changed to quadratic via 
$q_i^\alpha = (q_i^\alpha)^2$.
The derivation of the QUBO matrix can be found in Appendix~\ref{app:derivation}, yielding
\begin{align}
    Q_{\alpha,i;\beta,j} =&\, 2^{i+j-2K-2z} (-1)^{\delta_{iK}+\delta_{jK}} h_{\alpha\beta} \nonumber\\
    & + 2 \delta_{\alpha\beta} \delta_{ij} 2^{i-K-z} (-1)^{\delta_{iK}} \sum_\gamma a^{(z)}_\gamma h_{\gamma\beta}
    \ \ .
\label{eq:QUBOmatrix}
\end{align}

As considered in Ref.~\cite{Teplukhin:2020},
excited states can also be addressed with this same construction by including chemical potentials for the states that are lower in the spectrum.
The $N^{\rm th}$ state in the spectrum can be obtained by including $N-1$ chemical potentials $\mu_n$ that give an energy shift to each of the $n^{\rm th}$ states that place them higher in the spectrum than the $N^{\rm th}$ state.
To accomplish this,
an effective Hamiltonian of the form
\begin{equation}
\mathcal{\hat H}^{(N)}\ =\ 
\mathcal{\hat H}\ +\ \sum_n^{N-1}\ \mu_n\ 
|\Psi_n\rangle\langle \Psi_n |
\ \ ,
\label{eq:chems} 
\end{equation}
is used.
The $\mu_n$ depend upon the energy eigenvalues of 
$\lvert \Psi_{n<N} \rangle$, both of which are determined in earlier problem solutions 
using $\mathcal{H}^{(n<N)}$
in the workflow, and an approximate knowledge of the energy
of $\lvert \Psi_{N} \rangle$, determined possibly in tuning or from other approximate solutions.
Explicitly, to implement Eq.~(\ref{eq:chems}), 
and find the $n+1^{\rm th}$ excited state,
the wavefunctions of $n^{\rm th}$ lowest-lying states are used to generate a matrix contribution from the outer products $|\Psi_n\rangle\langle \Psi_n |$, which are multiplied by $\mu_n$ and added to the existing Hamiltonian (see Supplemental Material for the practical implementation of these and subsequent algorithms).

\subsection{Analyzing Results from Annealing Simulator and Quantum Devices}
\label{subsec:analysis}
\noindent
For any given QUBO matrix, annealing simulators or QAs perform $N_A$ anneals to locate the lowest-energy configuration(s) and associated wavefunction(s). 
Outputs of the annealing workflow include an ensemble of $N_A$ results, and 
generation of such ensembles can be repeated $N_{\rm run}$ times to provide estimates of associated uncertainties.
As the lowest energy configuration in an ensemble provides the lowest upper bound to the true energy of the target state,
$N_{\rm run}$ sets of such measurements can yield a global minimum energy and wavefunction, and also
a mean and standard deviation, or a median and 68\% confidence interval
(for robustness).
These estimators provide a measure of some uncertainties, including those associated with zooming and fluctuations in the annealing process.

Systematic studies of uncertainties associated with D-Wave's QAs have been previously performed, e.g., Refs.~\cite{Pearson2019,Zaborniak:A2021,Oshiyama:2021lmr}.
These contain detailed sets of measurements and discussions of device configuration and noise.
The D-Wave online documentation~\cite{DwaveNeal}, in particular that related to D-Wave's quantum simulator {\tt neal}, provides algorithms to simulate the annealers, discussions of the noise model, and provides codes.

To differentiate between results obtained with a noisy simulator and a QA that are shown in figures in the text,
we will assign one of the icons introduced in Ref.~\cite{Klco:2019xro}: the yellow square icon for results obtained using {\tt neal} and the blue diamond icon for those obtained using D-Wave's QA {\tt Advantage system 4.1} (which we will refer to simply as {\tt Advantage}).

\section{Harmonic and Anharmonic Oscillators: Eigenstates and Energies}
\label{sec:scalarfield}
\noindent
Perhaps the simplest quantum field theory to consider is $\lambda \phi^4$ scalar field theory.
Jordan, Lee and Preskill have shown that state preparation and simulating $S$-matrix elements resides in the BQP-complete complexity class~\cite{Jordan:2017lea}.
As a starting point for exploring lattice scalar field theory using QAs, we examine a single site harmonic oscillator with and without the non-linear $\lambda\phi^4$ interaction.
The Hamiltonian for a single site has the form
\begin{equation}
\mathcal{\hat H}=
\frac{1}{2}\hat{\Pi}^2
+\frac{1}{2} m_0^2\hat{\phi}^2+\frac{\lambda}{4!}\hat{\phi}^4
\ \ ,
\label{eq:hamphi4}
\end{equation}
with the bare mass $m_0$ and bare coupling $\lambda$ (all quantities are in lattice units, l.u.). In the Jordan-Lee-Preskill (JLP) basis~\cite{Jordan:2011ci,Jordan:2012xnu,Jordan:2014tma,Jordan:2017lea,Klco:2018zqz}, the field 
$\phi$ is digitized at each spatial site in a space spanned by 
$n_s$ uniformly distributed states\footnote{For a register of a universal quantum computer of $n_Q$ qubits,
$n_s=2^{n_Q}$.} with mapped values
\begin{equation}
    {\phi}=-{\phi}_{\text{max}}+\delta_{{\phi}}\beta_{{\phi}}\ \ , \quad 
    \delta_{{\phi}}=\frac{2{\phi}_{\text{max}}}{n_s-1}\ \ ,
\end{equation}
where ${\phi}_{\text{max}}$ is the maximum value of ${\phi}(\mathbf{x})$, 
and $\beta_{{\phi}}=0,1,\ldots,n_s-1$. 
In ${\phi}$-space, while two of the terms in Eq.~\eqref{eq:hamphi4} are diagonal, the conjugate momentum operator can be computed with a finite difference operator. However, this introduces polynomial $\delta_{{\phi}}$-discretization errors.
A better way to compute it is to use Quantum Fourier Transforms into and out of 
conjugate momentum space~\cite{Jordan:2011ci,Jordan:2012xnu}, since $\langle k_{{\phi}}|\hat{\Pi}^2|k'_{{\phi}} \rangle=k^2_{{\phi}}\delta_{k_{{\phi}},k'_{{\phi}}}$, with
\begin{align}
    & k_{{\phi}}=-k^{\text{max}}_{{\phi}}+\left(\beta_{{\phi}}-\frac{1}{2}\right)\delta k_{{\phi}}\ \ , \nonumber\\ 
    & k^{\text{max}}_{{\phi}}=\frac{\pi}{\delta_{{\phi}}}\ \ , \quad \delta k_{{\phi}}=\frac{2\pi}{\delta_{{\phi}} n_s}\ \ .    
\end{align}
This has been shown to eliminate power-law corrections to $\hat{\Pi}^2$, giving exponentially convergent digitization via the Nyquist-Shannon theorem~\cite{Jordan:2011ci,Jordan:2012xnu,Somma:2016:QSO:3179430.3179434,Jordan:2014tma,Jordan:2017lea,Somma:2016:QSO:3179430.3179434,Macridin:2018gdw,Macridin:2018oli,Klco:2018zqz,Macridin:2021uwn}.
A detailed comparison between this operator and finite-difference versions can be found in an appendix of Ref.~\cite{Klco:2018zqz}.
For a given number of states $n_s$, the Hamiltonian is an $n_s\times n_s$ real matrix, from which the eigenstates and energies can be found  via mappings to a QUBO-problem and annealing, as discussed in Sec.~\ref{sec:qubo}.

\subsection{Results from the Annealer Simulator {\tt neal}}
\label{subsec:neal_scalar}
\noindent
Available D-Wave annealer simulators were used to prepare for working with D-Wave's cloud-accessible QAs.
In particular, for the 1-site system, the simulator was used to perform parameter tunings and calibrations, including the maximum and minimum values of the field $\phi_{\rm max}$,
the number of states over which the field is digitized $n_s$,
$\eta$ in the objective function 
(in Eq.~\eqref{eq:Ffun}),
the chemical potentials $\mu_n$ 
(in Eq.~\eqref{eq:chems}),
the number of qubits per coefficient $K$,
the number of anneals per zoom step $N_A$, 
and the total number of zoom steps $z^{\rm max}$.
These identified values, or initial tunings, for these parameters, and measures of uncertainties, both systematic and statistical, are a subset of those that will be present for computations using QAs.
Our workflow for the simulator and quantum hardware was implemented with {\tt python}~\cite{python3} 
using {\tt jupyter} notebooks~\cite{PER-GRA:2007} after formulating the matrix problem with {\tt Mathematica}~\cite{Mathematica}.

\subsubsection{Tunings}
\begin{figure*}[!htb]
	\centering
	\begin{tikzpicture}
    \node(a){\includegraphics[width=0.96\textwidth]{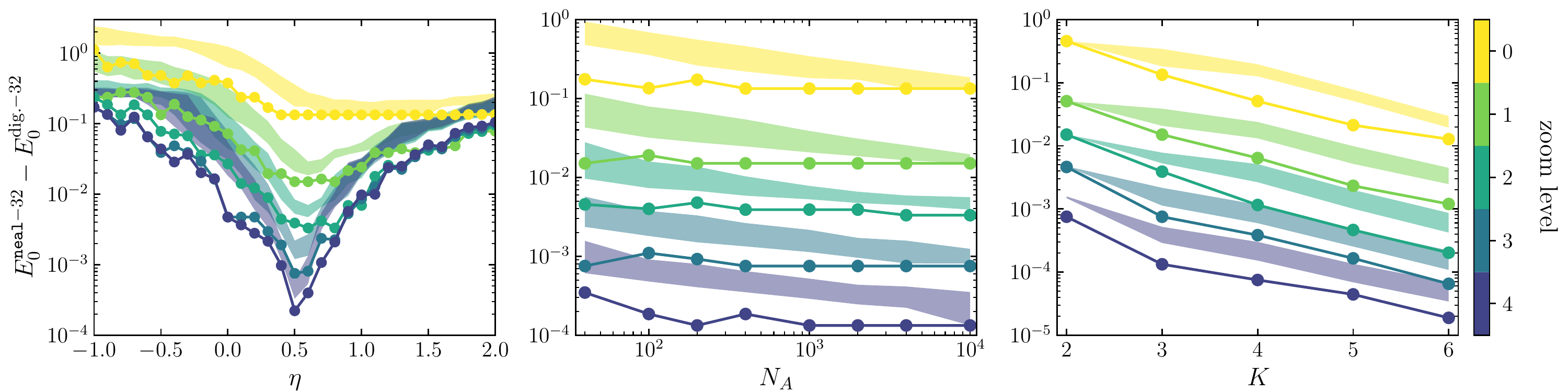}};
    \node at (a.north west) [anchor=north,xshift=2mm,yshift=0mm]
    {\includegraphics[width=0.05\textwidth]{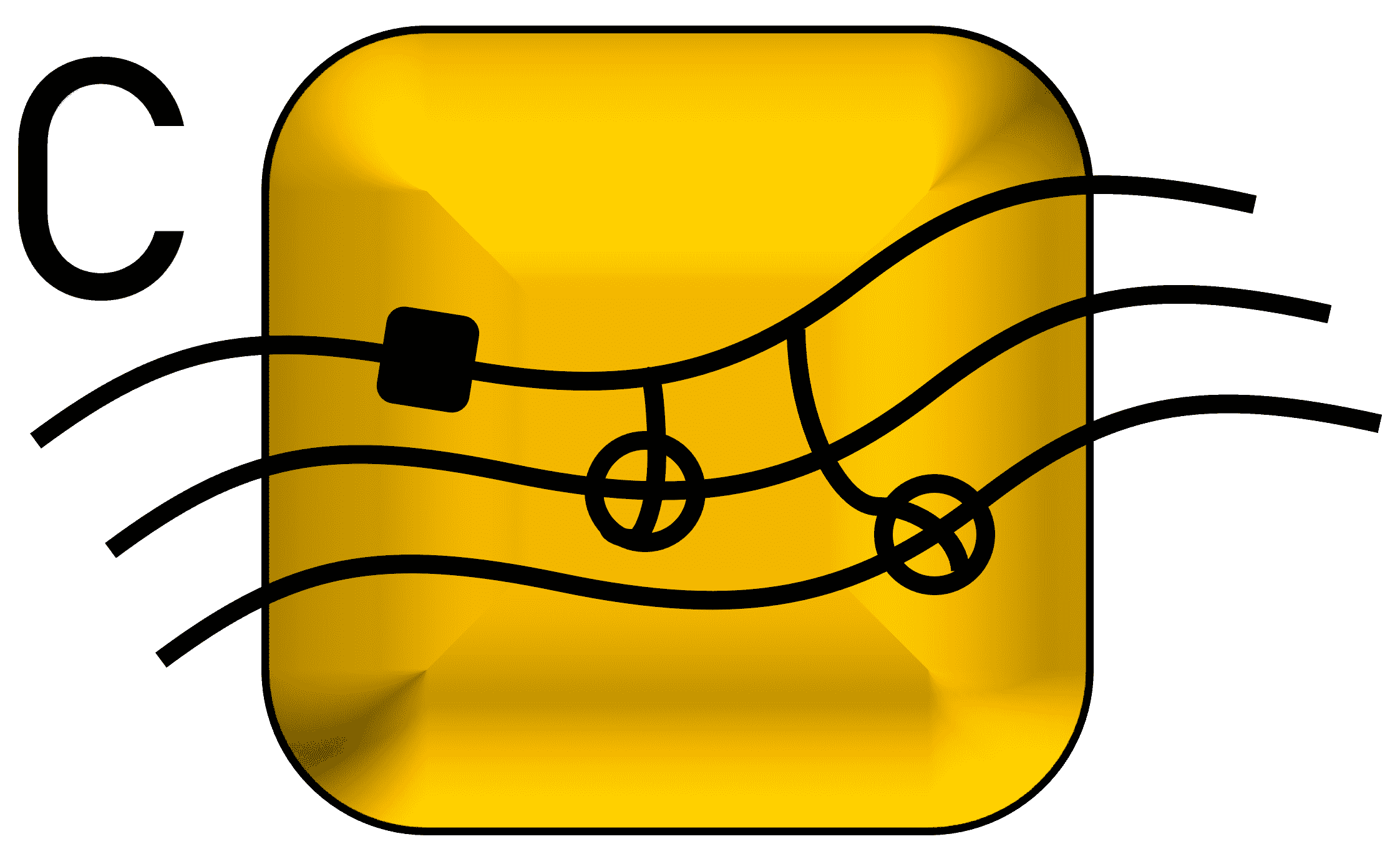}};
    \end{tikzpicture}
	\caption{
	Ground state energy (in l.u.) of the HO obtained with D-Wave's {\tt neal} as a function of the parameter $\eta$ (left panel), number of anneals $N_A$ (center panel) and $K$ parameter (right panel), for $m_0=1$, $\lambda=0$, and $\phi_{\rm max}=5$ discretized across $n_s=32$ states, setting $\eta=0.51$, $N_A=10^3$, and $K=3$ when the corresponding parameter is not varied. 
	The solid lines with points correspond to minimum energy solutions, while the solid bands correspond to the 68\% confidence intervals determined from $N_{\rm run}=200$ independent runs of the annealing workflow.
    }
	\label{fig:HOE0_scan}
\end{figure*}

\noindent
We present only highlights of parameter tunings as they generally behave as naively anticipated, or as determined previously.
The left panel of Fig.~\ref{fig:HOE0_scan} shows the systematic deviation from the true digitized ground state energy 
of the HO
determined using the annealing simulator for $m_0=1$, $\lambda=0$, and $\phi_{\rm max}=5$ digitized across $n_s=32$ states as a function of $\eta$
for $N_A=10^3$ and $K=3$.
The solid lines with points correspond to minimum energy solutions, while the solid bands correspond to the 68\% confidence intervals determined from $N_{\rm run}=200$ samples.
The accuracy in the energy is found to be optimized for $\eta\sim E_0$.
The middle panel of Fig.~\ref{fig:HOE0_scan} shows the deviations in energy for different levels of zooming as a function of $N_A$ for $K=3$ and $\eta=0.51$.
While the minimum-energy estimate is improved with an increasing number of anneals, increasing the zoom level leads to a more rapid convergence.
Similarly, the right panel of Fig.~\ref{fig:HOE0_scan} shows the energy deviation as a function of $K$ for different levels of zoom for $N_A=10^3$ and $\eta=0.51$. Although larger values of $K$ can increase the precision of the ground-state energy, as is the case for $N_A$, using more zoom steps can also reach similar levels (without increasing the number of qubits).
Overall conclusions from these explorations of parameter space are that $\eta$ should be close to the energy of the ground state and that the use of ``sloppy solutions'', where a relatively small number of anneals $N_A$ are used to iteratively estimate subsequent zoom intervals for QUBO parameters, can be used to make efficient use of computational resources.\footnote{The latter technique is in the spirit of All-Mode Averaging techniques employed in some lattice QCD generations of light quark propagators, e.g., Refs.~\cite{PhysRevD.88.094503,Shintani:2014vja}.}
We consider it to be somewhat unfortunate that the dependence on $\eta$ is that shown in the left panel of  Fig.~\ref{fig:HOE0_scan}, as ideally, quantities would be independent of $\eta$.

The systematic improvements in results 
with increasing $N_A$, $K$, and number of zoom levels 
using the annealer simulator do not persist indefinitely,
which is attributed to the white noise intrinsic to ${\tt neal}$.

\subsubsection{Harmonic Oscillator\texorpdfstring{: $V(\phi) = \frac{1}{2}\phi^2$}{}}
\begin{figure*}[!htb]
    \begin{tikzpicture}
    \node(a){\includegraphics[width=0.95\textwidth]{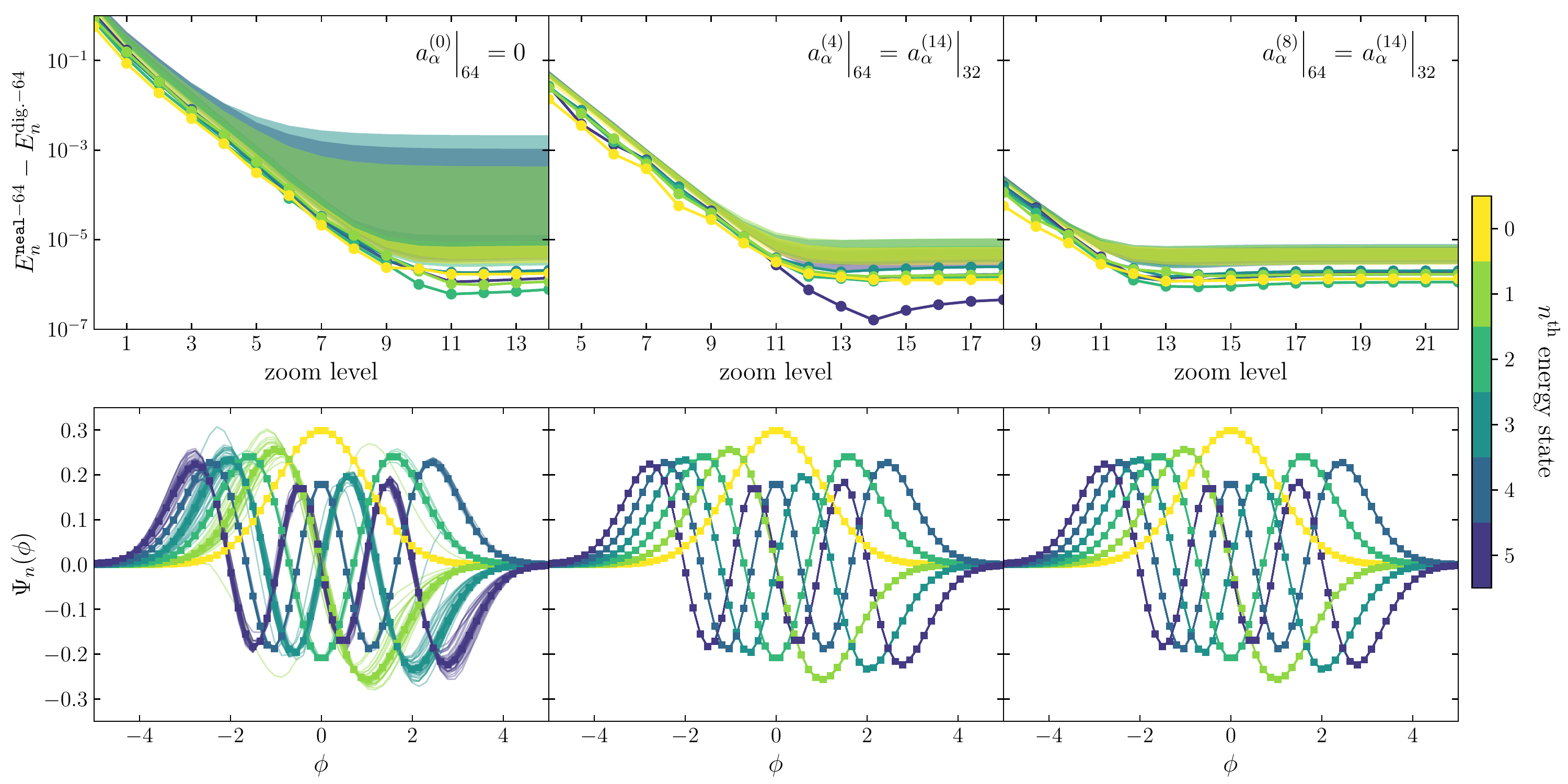}};
    \node at (a.north east) [anchor=north,xshift=-7mm,yshift=0mm]
    {\includegraphics[width=0.05\textwidth]{iconC2N_bf.png}};
    \end{tikzpicture}
	\caption{The upper panels show the convergence of the energy (in l.u.) of the first six HO states as a function of the number of zoom steps, where the solid lines with points correspond to minimum energy solutions, while the solid bands correspond to the 68\% confidence intervals determined from $N_{\rm run}=200$ independent runs of the annealing workflow.
	The lower panels show the lowest six HO wavefunctions (multiplied by $(-1)^n$).
	The squares denote the exact values from the digitized Hamiltonian, while the lines show the $N_{\rm run}=200$ independent runs obtained using D-Wave's annealer simulator {\tt neal} with the maximal number of zoom steps. The results from {\tt neal} are also sets of discrete points, and for display purposes we have shown them as joined line segments.
	The initial coefficients for the left panel are null coefficients,
	for the middle panels correspond to starting the $z^{\rm init}=4$ zooming for $n_s=64$ states from the $z^{\rm max}=14$ values and their interpolations from $n_s=32$,
	and the right panels correspond to staring with $z^{\rm init}=8$.
	The maximum value of the field is $\phi_{\rm max}=5$ digitized on $n_s=64$ states, with $K=3$ and $N_A=10^3$.}
	\label{fig:HO1site_project_mg}
\end{figure*}

\noindent
For demonstrative purposes, through the use of appropriately tuned parameters, we present the results for the lowest six eigenstates of the HO with $V(\phi) = \frac{1}{2}\phi^2$ (i.e., $m_0=1$).
The exact energies in the field-space continuum limit are
$E^{\rm exact}_n = \{ \frac{1}{2}, \frac{3}{2}, \frac{5}{2}, \frac{7}{2}, \frac{9}{2}, \frac{11}{2}, ...\}$, with eigenfunctions given by
\begin{equation}
\Psi_n(\phi) = \frac{1}{\sqrt{2^n n!}}\,\left(\frac{1}{\pi}\right)^{1/4} e^{-{\frac{1}{2}\, \phi^2}}\, {\rm H}_n( \phi)\ \ ,
\label{eq:HOexact}
\end{equation}
where 
${\rm H}_n (x)$ are the Hermite polynomials.
A systematic study of the impact of digitization on the low-lying wavefunctions and energies has been performed previously~\cite{Klco:2018zqz}, and we use that as a guide in selecting digitization parameters.
We work with $\phi_{\rm max}=5$ and $n_s=64$, resulting in $\delta\phi = 0.1587$ and $k_{\rm max} = 19.7920$, and energies $E^{\rm dig.-64}_{n\le 6}$ that are the same as $E^{\rm exact}_{n\le 6}$ to better than $\sim 10^{-5}$.
The value of $\eta_n$ was set equal to the corresponding $E^{\rm exact}_n + 0.01$, and the chemical potentials were set to $\mu_n=10$ when appropriate, to move each of the previously determined states to an energy higher than the ``next'' ground state.
\begin{table}[!tb]
\begin{center}
\begin{tabular}{|| c | c | c | c | c ||} 
 \hline
 $n$ 
 & $E^{\rm exact}_n$ 
 & $|\delta E^{\rm dig.-64}_n|$ 
 & $|\delta E^{{\tt neal}-64}_n|_{z=0}$ 
 & $|\delta E^{{\tt neal}-64}_n|_{z=8}$ \\ [0.5ex] 
 \hline\hline
 $0$ 
 & $1/2$ 
 & $3.5\times 10^{-11}$ 
 & $(4.9^{\,+2.5}_{\,-1.8})\times 10^{-6}$ 
 & $(4.0^{\,+2.6}_{\,-1.2})\times 10^{-6}$\\ 
 $1$ 
 & $3/2$ 
 & $1.8\times 10^{-9}$ 
 & $(9.1^{\,+409}_{\,-5.1})\times 10^{-6}$
 & $(5.2^{\,+2.5}_{\,-1.7})\times 10^{-6}$\\ 
 $2$ 
 & $5/2$ 
 & $4.1\times 10^{-8}$ 
 & $(4.7^{\,+4.7}_{\,-2.0})\times 10^{-6}$ 
 & $(4.4^{\,+2.2}_{\,-1.6})\times 10^{-6}$\\ 
 $3$ 
 & $7/2$ & $6.6\times 10^{-7}$ 
 & $(0.7^{\,+200}_{\,-0.3})\times 10^{-5}$ 
 & $(5.3^{\,+2.7}_{\,-1.6})\times 10^{-6}$\\ 
 $4$ 
 & $9/2$ 
 & $6.6\times 10^{-6}$ 
 & $(5.5^{\,+7.2}_{\,-2.0})\times 10^{-6}$ 
 & $(4.4^{\,+2.1}_{\,-1.6})\times 10^{-6}$\\ 
 $5$ 
 & $11/2$ 
 & $6.0\times 10^{-5}$ 
 & $(0.7^{\,+100}_{\,-0.3})\times 10^{-5}$ 
 & $(4.7^{\,+2.3}_{\,-1.6})\times 10^{-6}$\\ 
[0.5ex] 
 \hline
\end{tabular}
\caption{
Energies associated with the HO with $m_0=1$ (in l.u.).
The exact energies are shown in the second column, the difference between the diagonalization of the digitized Hamiltonian and exact energies are shown in the third column, with $\phi_{\rm max}=5$ and $n_s=64$, and the differences between the digitized energies and the corresponding results obtained using D-Wave's annealer simulator, {\tt neal}, are shown in the fourth (no MG) and fifth (with MG) columns, with the uncertainties showing the 68\% confidence intervals, with $K=3$ and $N_A=10^3$.
}
\label{tab:HOsimenergies}
\end{center}
\end{table}
The energies of the lowest states found using {\tt neal} are given  in Table~\ref{tab:HOsimenergies} and displayed in the left panel of Fig.~\ref{fig:HO1site_project_mg} (with $N_A=10^3$), and recover the digitized values with accuracy  $\sim 10^{-3}-10^{-6}$. 

In order to reduce the uncertainty bands, 
instead of increasing $N_A$ or $K$, we work with a (somewhat basic) multigrid (MG) AQAE solver,
inspired by the wide success of multigrid algorithms~\cite{10.2307/2006422}, 
where a coarser (smaller) system is solved 
first, and that solution is used as an input parameter for the finer (larger) system. 
Specific to our case, a coarser system can be defined by using a smaller value of $n_s$ for the digitization of the field. 
With the coefficients $a^{(z)}_\alpha$ extracted
from the $n_s=32$ system, 
we interpolate  (using cubic splines) to find  starting points for the $n_s=64$ system. To help guide the annealer, it has been found that starting with $z^{\rm init}\gtrsim 4$ reduces the uncertainty on the energies, as it limits the range of the values that $a^{(z)}_\alpha$ can assume. As shown in the center and right panels of Fig.~\ref{fig:HO1site_project_mg}, starting with $z^{\rm init}=4$ or 8 significantly reduces the error compared to $z^{\rm init}=0$.

To display the convergence obtained with {\tt neal} for an increasing number of zoom steps, the top panels of Fig.~\ref{fig:HO1site_project_mg} show the deviation in the energy of the lowest six states as a function of zoom steps.
The convergence is consistent with exponential in the number of zoom steps, as found in Ref.~\cite{Chang:2019}.
This result is encouraging as the HO is one of the simplest systems to consider.

\begin{figure}[!tb]
\begin{tikzpicture}
    \node(a){\includegraphics[width=0.95\columnwidth]{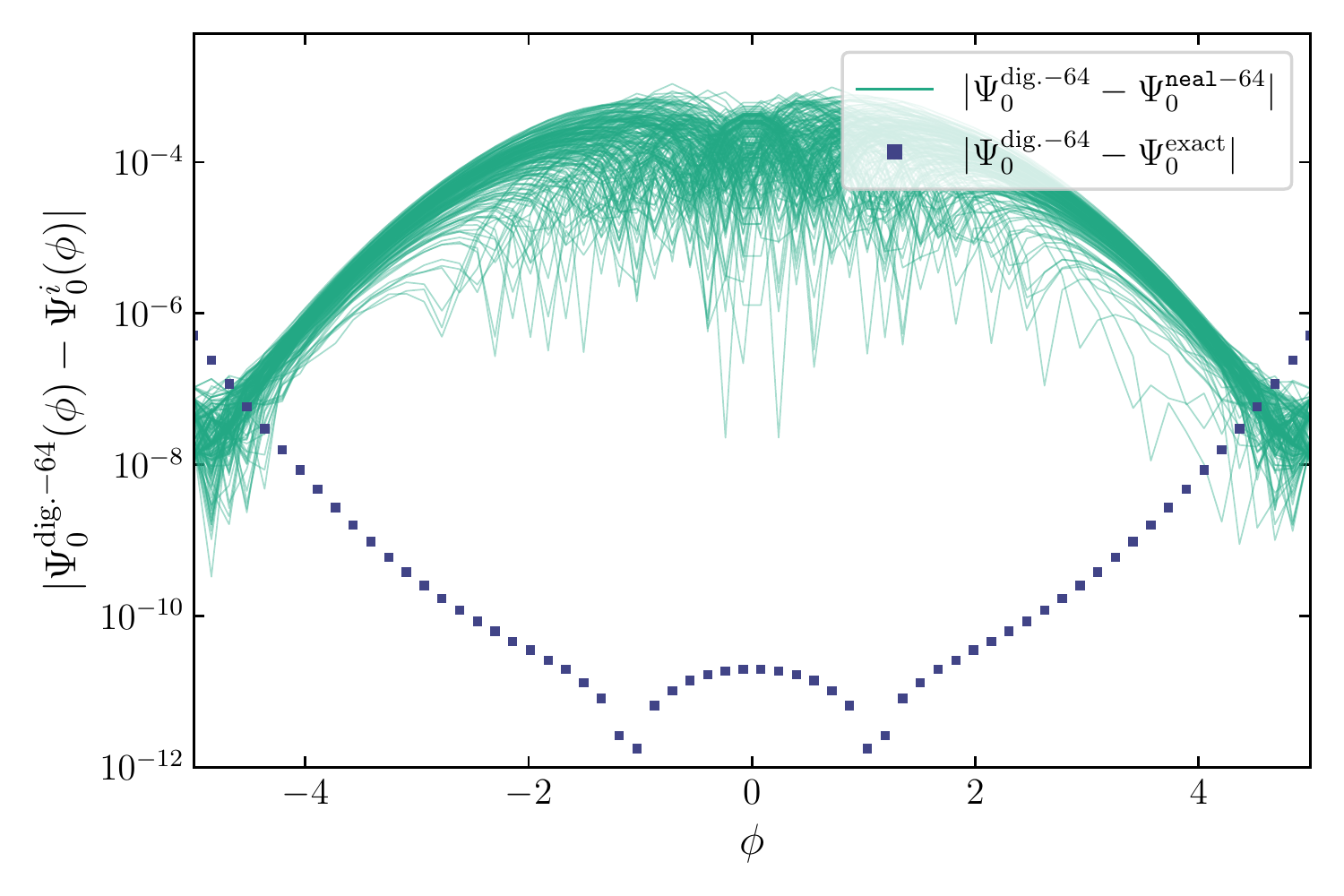}};
    \node at (a.north west) [anchor=north,xshift=5mm,yshift=0mm]
    {\includegraphics[width=0.05\textwidth]{iconC2N_bf.png}};
    \end{tikzpicture}
	\caption{
	Deviations between the digitized ground-state HO wavefunction and the i) analytic continuum-field expression (squares) and ii) the wavefunctions determined using {\tt neal} (lines, showing the $N_{\rm run}=200$ independent runs). The digitized simulation employed $\phi_{\rm max}=5$, $n_s=64$, $m_0=1$, $K=3$, $\eta=0.51$, $N_A=10^3$, and $\{z^{\rm init},z^{\rm max}\}=\{8,22\}$.
    }
	\label{fig:HOwavediffs}
\end{figure}

The wavefunctions associated with the energies in Table~\ref{tab:HOsimenergies} are shown in the lower panels of Fig.~\ref{fig:HO1site_project_mg}.
The diagonalization of the HO employed $\phi_{\rm max}=5$, $n_s=64$, and $m_0=1$ and reproduced the values of the continuum-field wavefunctions with high precision -- a well-known result.
For the ground state, the digitized wavefunction (squares) reproduce the continuum wavefunction to better than $\sim 10^{-6}$, as shown in Fig.~\ref{fig:HOwavediffs}.
\begin{figure*}[htb]
    \begin{tikzpicture}
    \node(a){\includegraphics[width=0.95\textwidth]{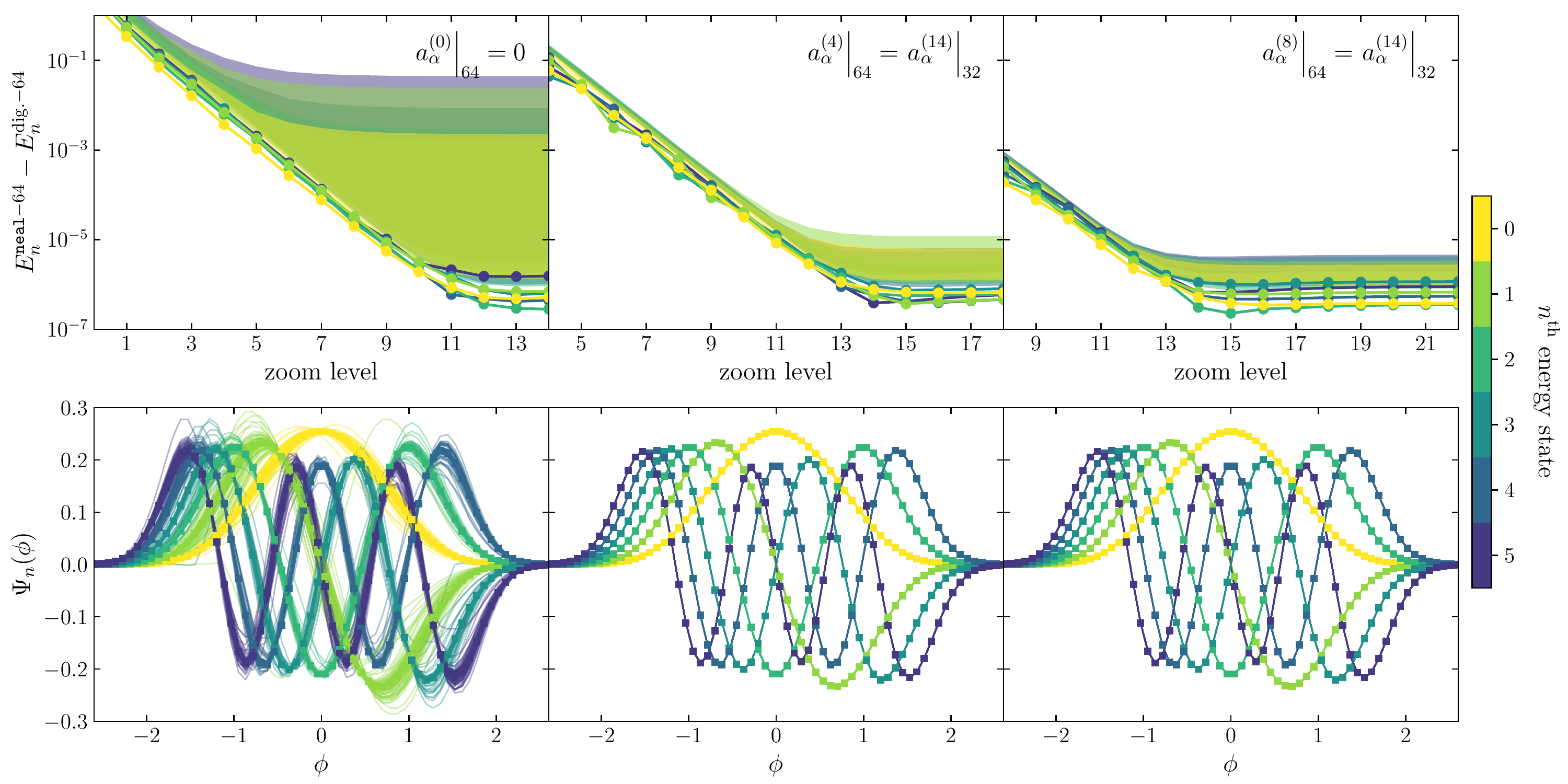}};
    \node at (a.north east) [anchor=north,xshift=-7mm,yshift=0mm]
    {\includegraphics[width=0.05\textwidth]{iconC2N_bf.png}};
    \end{tikzpicture}
	\caption{The upper panels show the convergence of the energy (in l.u.) of the first six AHO states as a function of the number of zoom steps, where the solid lines with points correspond to minimum energy solutions, while the solid bands correspond to the 68\% confidence intervals determined from $N_{\rm run}=200$ independent runs of the annealing workflow.
	The lower panels show the lowest six AHO wavefunctions (multiplied by $(-1)^n$).
	The squares denote the exact values from the digitized Hamiltonian, while the lines show the $N_{\rm run}=200$ independent runs obtained using D-Wave's annealer simulator {\tt neal} with the maximal number of zoom steps.
	The initial coefficients for the left panel are null coefficients,
	for the middle panels correspond to starting the $z^{\rm init}=4$ zooming for $n_s=64$ states from the $z^{\rm max}=14$ values and their interpolations from $n_s=32$,
	and the right panels correspond to staring with $z^{\rm init}=8$.
	The maximum value of the field is $\phi_{\rm max}=2.6$ digitized on $n_s=64$ states, with $K=3$ and $N_A=10^3$.
	}
	\label{fig:lamphi41site_project}
\end{figure*}
Using {\tt neal}, $K=3$ and $\eta=0.51$ were used with 14 levels of zoom.
The wavefunctions, shown by the lines in Fig.~\ref{fig:HO1site_project_mg}, reproduce the digitized wavefunctions to better than $\sim 10^{-3}$, as shown for the ground state in Fig.~\ref{fig:HOwavediffs}. Similar fidelity is obtained for the other wavefunctions.

Overall, the HO is amenable to simulation with {\tt neal}. With appropriate (and easy to identify) parameter tunings, the energies and wavefunctions of the lowest-lying states can be determined with precision.
Results for coarser and finer digitizations behave in ways that are consistent with expectations.
While we have not performed a systematic exploration,
we expect that the primary limitation on the number of states of the HO that can be isolated with precision is the
accumulation of errors through the iterative process of Eq.~\eqref{eq:chems}.

\subsubsection{Anharmonic Oscillator\texorpdfstring{: $V(\phi) = \frac{1}{2}\phi^2 + \frac{4}{3} \phi^4$}{}}
\noindent

Depending upon the size of the non-linear interaction, the low-lying spectrum of the AHO can differ significantly from those of the HO.
For the coupling of $\lambda=32$ that we have chosen for demonstrative purposes, important differences are present in both the energy and wavefunctions of the 1-site system.
The value of $\eta_n$ used to solve for $n^{\rm th}$ eigenstate is set, as in the HO case, to the corresponding $E^{\rm exact}_n + 0.01$, the chemical potentials are set to $\mu_n=20$, $K=3$, and the number of anneals is $N_A=10^3$.
Table~\ref{tab:lamphi4simenergies} displays the exact energies, 
the difference between the exact and digitized energies
for the system with $\phi_{\rm max}=2.6$ and $n_s=64$, 
and the difference between the results obtained using {\tt neal} and the exact digitized energies.
\begin{table}[!tb]
\begin{center}
\begin{tabular}{|| c | l | c | c | c ||} 
 \hline
 $n$ 
 & $E_n^{\rm exact}$ 
 & $|\delta E^{{\rm dig.}-64}_n|$ 
 & $|\delta E^{{\tt neal}-64}_n|_{z=0}$
 & $|\delta E^{{\tt neal}-64}_n|_{z=8}$\\ [0.5ex] 
 \hline\hline
 $0$ 
 & 0.8597427
 & $1.7\times 10^{-9}$ 
 & $(0.9^{\,+223}_{\,-0.8})\times 10^{-5}$ 
 & $(1.7^{\,+1.1}_{\,-0.7})\times 10^{-6}$ \\ 
 $1$ 
 & 2.9493637 
 & $2.4\times 10^{-8}$ 
 & $(0.5^{\,+239}_{\,-0.5})\times 10^{-4}$ 
 & $(1.8^{\,+1.6}_{\,-0.6})\times 10^{-6}$ \\ 
 $2$ 
 & 5.6096611 
 & $2.0\times 10^{-7}$ 
 & $(0.2^{\,+161}_{\,-0.1})\times 10^{-5}$ 
 & $(1.2^{\,+0.7}_{\,-0.6})\times 10^{-6}$ \\ 
 $3$ 
 & 8.6270258
 & $1.5\times 10^{-6}$ 
 & $(0.5^{\,+279}_{\,-0.3})\times 10^{-5}$ 
 & $(2.7^{\,+1.6}_{\,-0.7})\times 10^{-6}$ \\ 
 $4$ 
 & 11.930637
 & $8.0\times 10^{-6}$ 
 & $(0.2^{\,+782}_{\,-0.1})\times 10^{-5}$ 
 & $(1.4^{\,+0.7}_{\,-0.4})\times 10^{-6}$ \\ 
 $5$ 
 & 15.476155
 & $4.4\times 10^{-5}$ 
 & $(4.1^{\,+439}_{\,-4.1})\times 10^{-4}$ 
 & $(3.0^{\,+1.6}_{\,-1.0})\times 10^{-6}$ \\ 
[0.5ex] 
 \hline
\end{tabular}
\caption{
Energies associated with the AHO with 
$m_0=1$ and $\lambda=32$ (in l.u.).
The exact energies are shown in the second column, the difference between the diagonalization of the digitized Hamiltonian and exact energies are shown in the third column, with $\phi_{\rm max}=2.6$ and $n_s=64$,
and the differences between the digitized energies and the corresponding results obtained using D-Wave's annealer simulator, {\tt neal}, 
are shown in the fourth (no MG) and fifth (with MG) columns, with the uncertainties showing the 68\% confidence intervals, with $K=3$ and $N_A=10^3$.
}
\label{tab:lamphi4simenergies}
\end{center}
\end{table}

The convergence of the energy of the lowest-lying states with increasing numbers of zoom steps 
determined using {\tt neal}
are shown in the top panels of Fig.~\ref{fig:lamphi41site_project}, with
exponential convergence seen up to $\sim 12$ zoom steps, 
beyond which there are diminishing returns.
The converged wavefunctions for each level are shown in the lower panels of Fig.~\ref{fig:lamphi41site_project}.
As in the case of the HO, the MG-AQAE method reduces the uncertainties in the extracted energies.
Further, the wavefunctions converge well to the exact digitized wavefunctions (shown as the squares in Fig.~\ref{fig:lamphi41site_project}).
The lower panels of the figure show the wavefunctions resulting from $N_A=10^3$ anneals compared to the exact result.

\subsection{Implementations and Results from D-Wave Annealers}
\label{subsec:dwave_scalar}
\noindent

We have run the codes used in Sec.~\ref{subsec:neal_scalar} with {\tt neal} on D-Wave's QA {\tt Advantage}, which has 5627 physical qubits, and each qubit is connected with 15 other qubits (in a so-called Pegasus topology). 
The system is accessible through the cloud via D-Wave's website~\cite{DwaveLeap}.

The mapping between the QUBO problem and the processor topology is performed automatically, via heuristics algorithms~\cite{cai2014practical}, and is the most time-consuming part of the simulation, as discussed in Appendix~\ref{app:timings} (the embedding can be computed at the beginning, and it also can be reused for all the zoom steps).
During this process, as there is no all-to-all connectivity, several physical qubits are chained together to form a logical qubit with the required connectivity.
To enforce that the qubits in a certain chain have all the same value, an extra parameter, the chain-strength value $c_s$, is fixed.
As the elements of the QUBO matrix are re-scaled to lie in the range $[-1,1]$ when they are passed to {\tt Advantage}, if 
$c_s\gg{\rm max}(|Q|)$, the QUBO elements will be re-scaled closer to zero.
We set $c_s=\omega \,{\rm max}(|Q|)$ and scan over $\omega\in[0,1]$, finding that $\omega=0.2$ gives the lowest energies for these systems.
Another parameter that can be tuned is the annealing schedule and annealing time $t_A$,
and we have used its default value of $t_A=20\,\mu {\rm s}$ in our calculations. It has been previously observed, e.g., Ref.~\cite{Grant:2021a}, that using different annealing schedules, like reverse annealing, can increase the success rate (finding the solution with minimum energy) of the QA. The exploration of such improvements is left for future work.

\begin{figure*}[!htb]
    \begin{tikzpicture}
    \node(a){\includegraphics[width=0.95\textwidth]{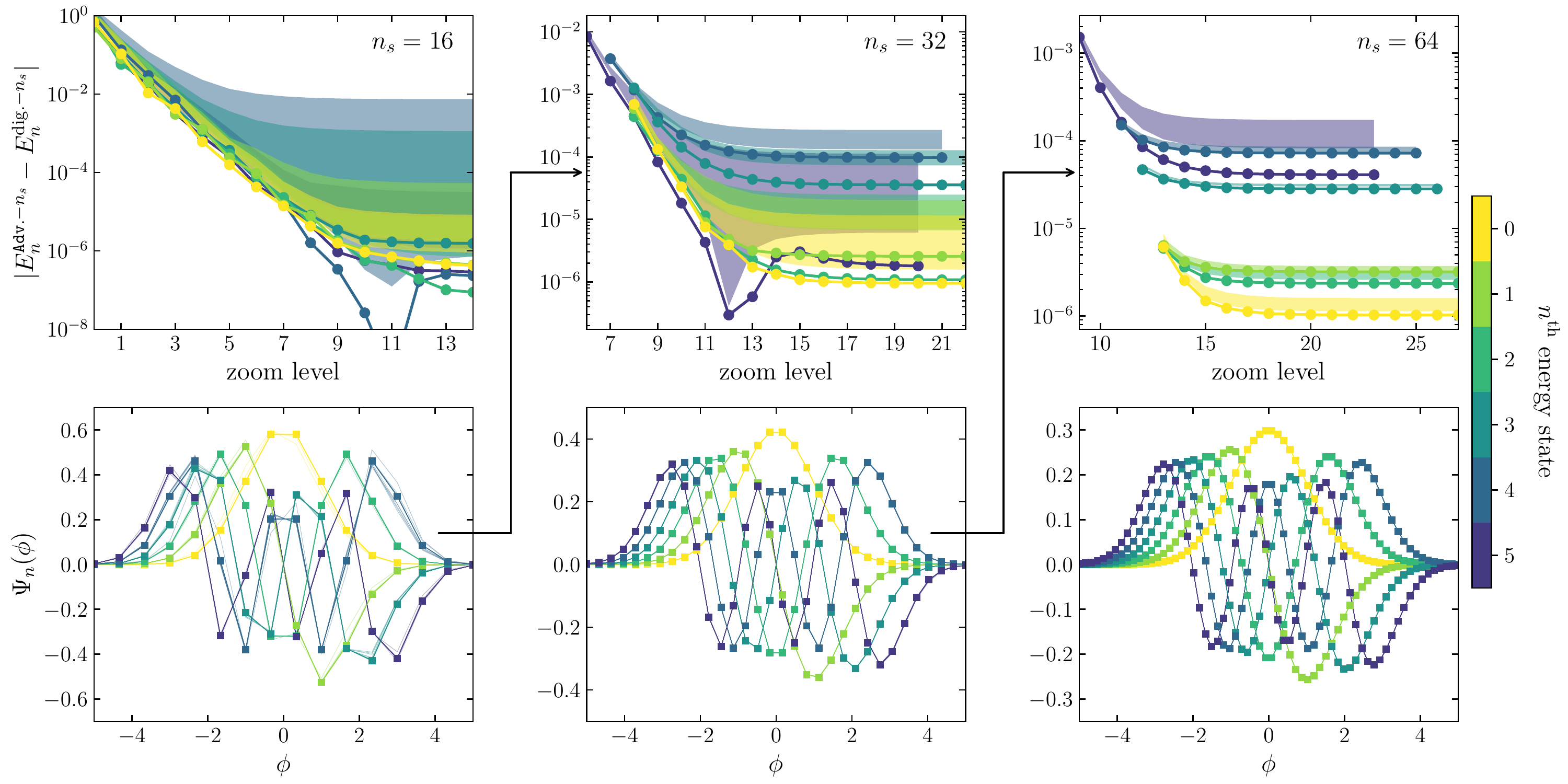}};
    \node at (a.north east) [anchor=north,xshift=-7mm,yshift=0mm]
    {\includegraphics[width=0.04\textwidth]{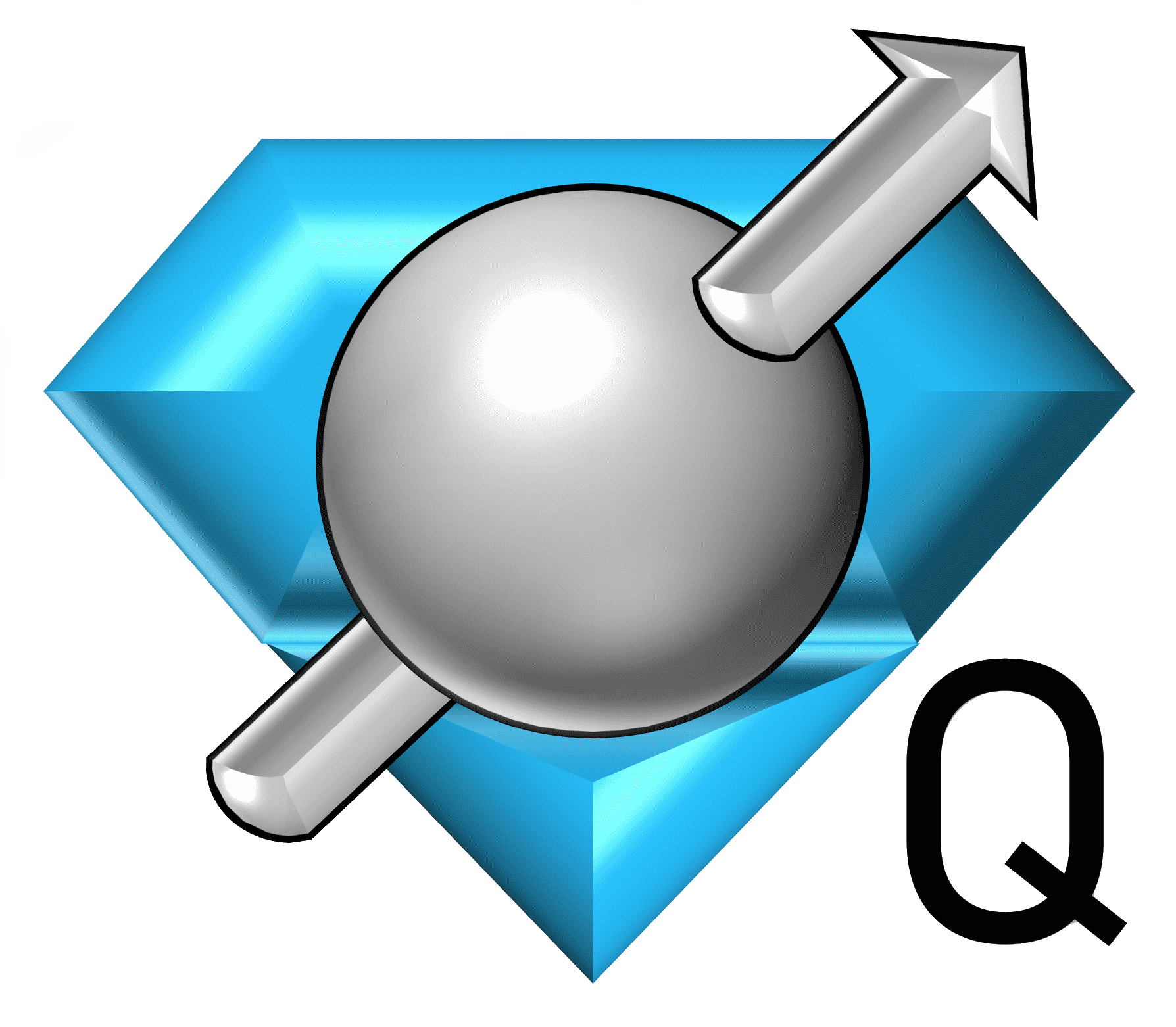}};
    \end{tikzpicture}
	\caption{The upper panels show the convergence of the energy (in l.u.) of the first six HO states as a function of the number of zoom steps, where the solid lines with points correspond to minimum energy solutions, while the solid bands correspond to the 68\% confidence intervals determined from $N_{\rm run}=20$ independent runs of the annealing workflow.
	The lower panels show the lowest six HO wavefunctions (multiplied by $(-1)^n$).
	The squares denote the exact values from the digitized Hamiltonian, while the lines show the $N_{\rm run}=20$ independent runs obtained using D-Wave's {\tt Advantage} with the maximal number of zoom steps.
	The maximum value of the field is $\phi_{\rm max}=5$ digitized on $n_s=\{16,32,64\}$ states, with $K=\{3,3,2\}$ and $N_A=10^3$.
	}
	\label{fig:HO1site_project_dwave}
\end{figure*}
\begin{figure*}[!htb]
    \begin{tikzpicture}
    \node(a){\includegraphics[width=0.95\textwidth]{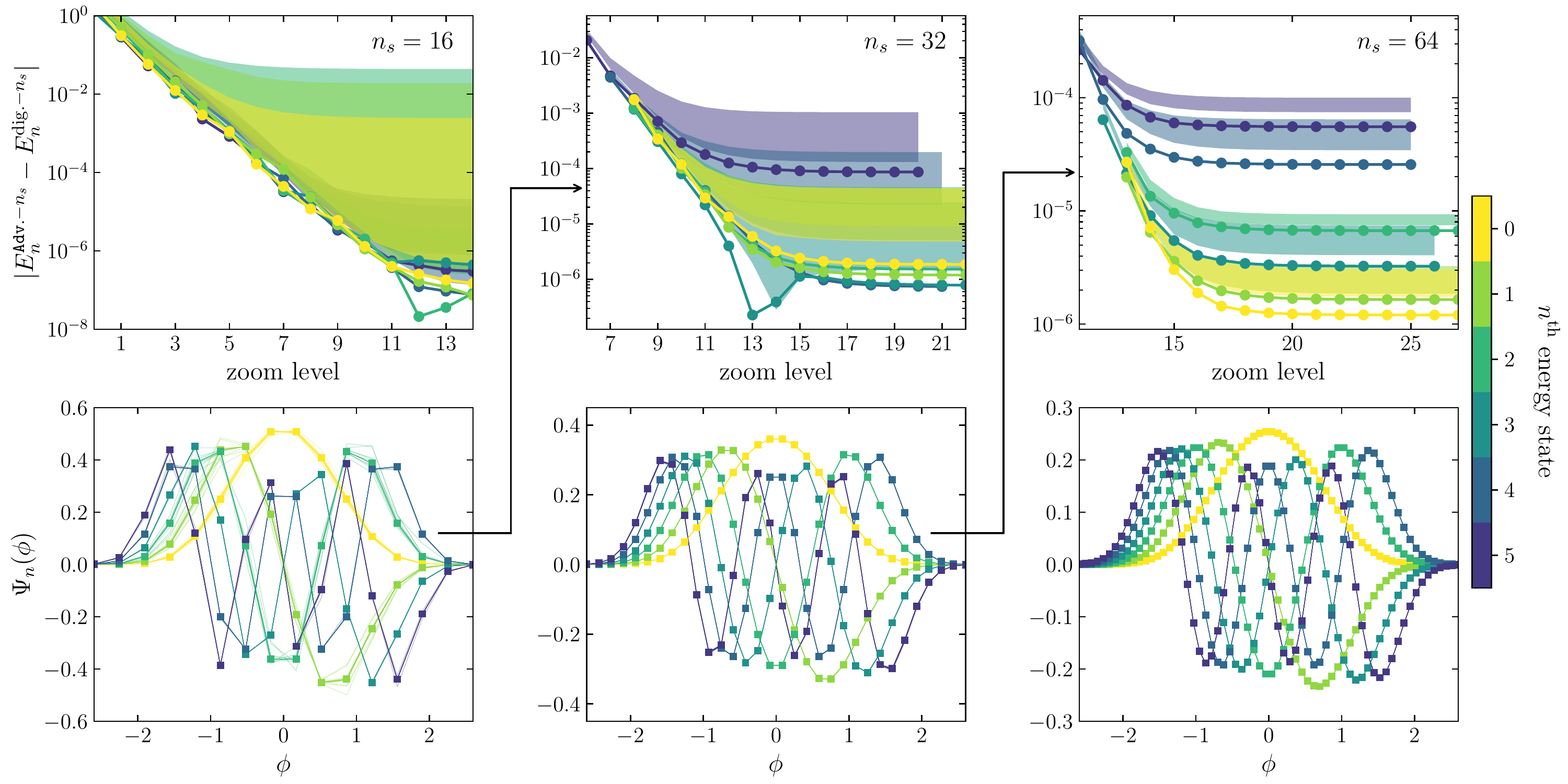}};
    \node at (a.north east) [anchor=north,xshift=-7mm,yshift=0mm]
    {\includegraphics[width=0.04\textwidth]{iconQ1_bf.png}};
    \end{tikzpicture}
	\caption{The upper panels show the convergence of the energy (in l.u.) of the first six AHO states as a function of the number of zoom steps, where the solid lines with points correspond to minimum energy solutions, while the solid bands correspond to the 68\% confidence intervals determined from $N_{\rm run}=20$ independent runs of the annealing workflow.
	The lower panels show the lowest six AHO wavefunctions (multiplied by $(-1)^n$).
	The squares denote the exact values from the digitized Hamiltonian, while the lines show the $N_{\rm run}=20$ independent runs obtained using D-Wave's {\tt Advantage} with the maximal number of zoom steps.
	The maximum value of the field is $\phi_{\rm max}=2.6$ digitized on $n_s=\{16,32,64\}$ states, with $K=\{3,3,2\}$ and $N_A=10^3$.
	}
	\label{fig:lambphi41site_project_dwave}
\end{figure*}

Due to {\tt Advantage}'s intrinsic noise, extracting 
energies and wavefunctions with adequate precision for $n_s\ge 16$
requires using the MG-AQAE solver.
Compared to the {\tt neal} simulator, 
a smaller initial value of $n_s=16$ is required for {\tt Advantage} to provide meaningful results.  These results are then used as starting values for the $n_s=32$ and $n_s=64$ anneals, with results shown in Fig.~\ref{fig:HO1site_project_dwave} for the HO and Fig.~\ref{fig:lambphi41site_project_dwave} for the 
AHO.\footnote{It is interesting to note that the ground-state wavefunction obtained with $n_s=16$ when interpolated to $n_s=64$  achieves $10^{-4}$ precision in the ground-state energy, without using the MG-AQAE solver (for higher-energy states, the precision is reduced).} 
Interestingly, while a value of $K=3$ is sufficient for the $n_s=16$ and $32$ systems, $K=2$ is required for the $n_s=64$ system to permit an embedding of the QUBO matrix into {\tt Advantage}.
%

\subsection{Delocalized Fields\texorpdfstring{: $\lambda\phi^4$ with $m_0^2<0$ }{} and Reflection Symmetry in Field Space}
\label{subsec:delocalized}
\begin{figure}[!htb]
	\centering
	\begin{tikzpicture}
    \node(a){\includegraphics[width=0.95\columnwidth]{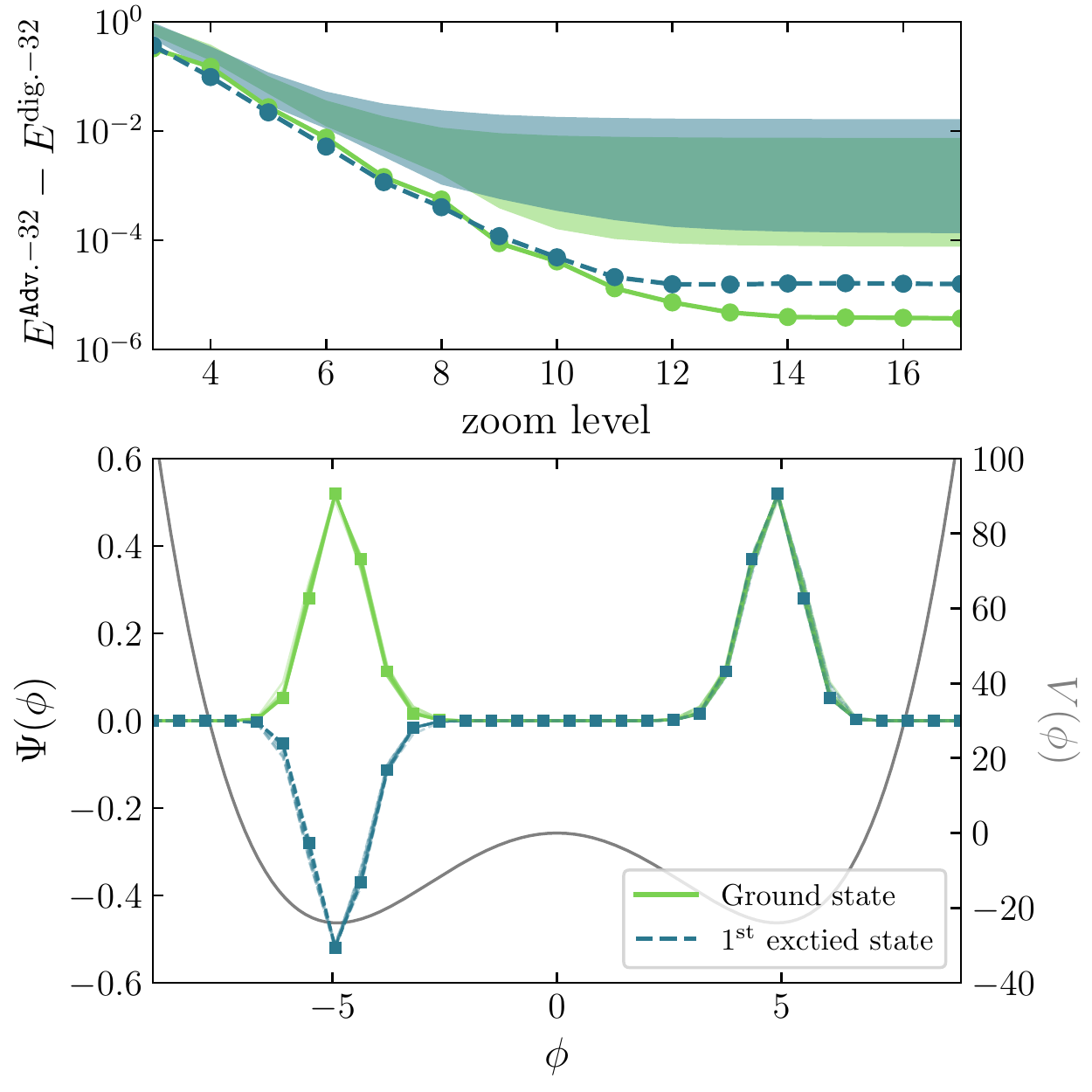}};
    \node at (a.north east) [anchor=north,xshift=-5mm,yshift=0mm] {\includegraphics[width=0.04\textwidth]{iconQ1_bf.png}};
    \end{tikzpicture}
		\caption{
	Convergence of the 
	ground state and first-excited state energies and wavefunctions of an AHO.
	The upper panel shows the convergence 
	of the energies (in l.u.) and the lower panel shows the symmetric and anti-symmetric wavefunctions (with the maximal number of zoom steps, 17) obtained using {\tt Advantage} ($N_{\rm run}=20$)
	for the system with $m_0^2=-4$, $\lambda=1$, $\phi_{\rm max}=9$ discretized across $n_s=32$ states, with $K=3$ and $N_A=10^3$.
	The half field-space Hamiltonian employed appropriate boundary conditions at the origin to independently solve for wavefunctions with definite reflection symmetry.
		}
		\label{fig:muLE0}
\end{figure}

\noindent
In the situation where $m_0^2<0$, corresponding to a double-well potential, the ground state with a symmetric wavefunction and the first-excited state with an antisymmetric wavefunction are nearly degenerate for a large region in mass-coupling space. For such parameters, the wavefunctions have support mainly in regions localized around the two minima of the potential, with exponential suppression of the energy difference as the minima become increasingly separated.
Consequently, the results obtained with {\tt neal} and  {\tt Advantage} are generally unable to uniquely identify the ground states of such systems.
As the Hamiltonian has a reflection symmetry in field space, the near degeneracy of the lowest two states can be mitigated by solving the half-space using boundary conditions at the origin consistent with a symmetric or an antisymmetric wavefunction.
Using such implementations, the ground state and first excited state 
of the systems with $m_0^2<0$ can be uniquely determined, as shown in Fig.~\ref{fig:muLE0}.
Results are obtained with {\tt Advantage} using the MG-AQAE solver (with $z^{\rm init}=3$ and using the $n_s=16$ system as the preconditioner).
It is interesting to point out that the best solution is found to have a smaller energy than one found with {\tt neal}.
The use of boundary conditions at the origin reduces the dimensionality of the 
Hamiltonian that is sent to the annealer simulator or quantum hardware, and hence the problem itself, 
to one similar to that of a HO with $m_0^2>0$.
Therefore, the implementation is essentially the same as described previously.
However, the delocalization of the wavefunction, means that the digitization of the field requires an increased number of states to recover the same level of precision in, say, the ground state energy (by maintaining a fixed $\delta_\phi$).

The results obtained for these systems
provide practical insights into the generic performance of a QA for simulating systems with (near-)degenerate ground states.
Without the half-field truncation,
each converged result of the system provided, in general, a different linear combination of the (two) degenerate states, 
as expected. 
While straightforward to perform, 
we did not undertake a study of the $m_0^2-\lambda$
parameter space to identify regions where the energy gap was sufficient for {\tt neal} and {\tt Advantage} to uniquely converge to the ground state.

While there is significant importance in simulating scalar fields exhibiting spontaneous symmetry breaking in 3+1 dimensions in high-energy (Higgs field) and 
nuclear physics 
($\sigma$ model and chiral perturbation theory), a detailed exploration is beyond the scope of the present work.

\subsection{Scaling study}
\begin{figure}[!tb]
	\centering
	\includegraphics[width=0.95\columnwidth]{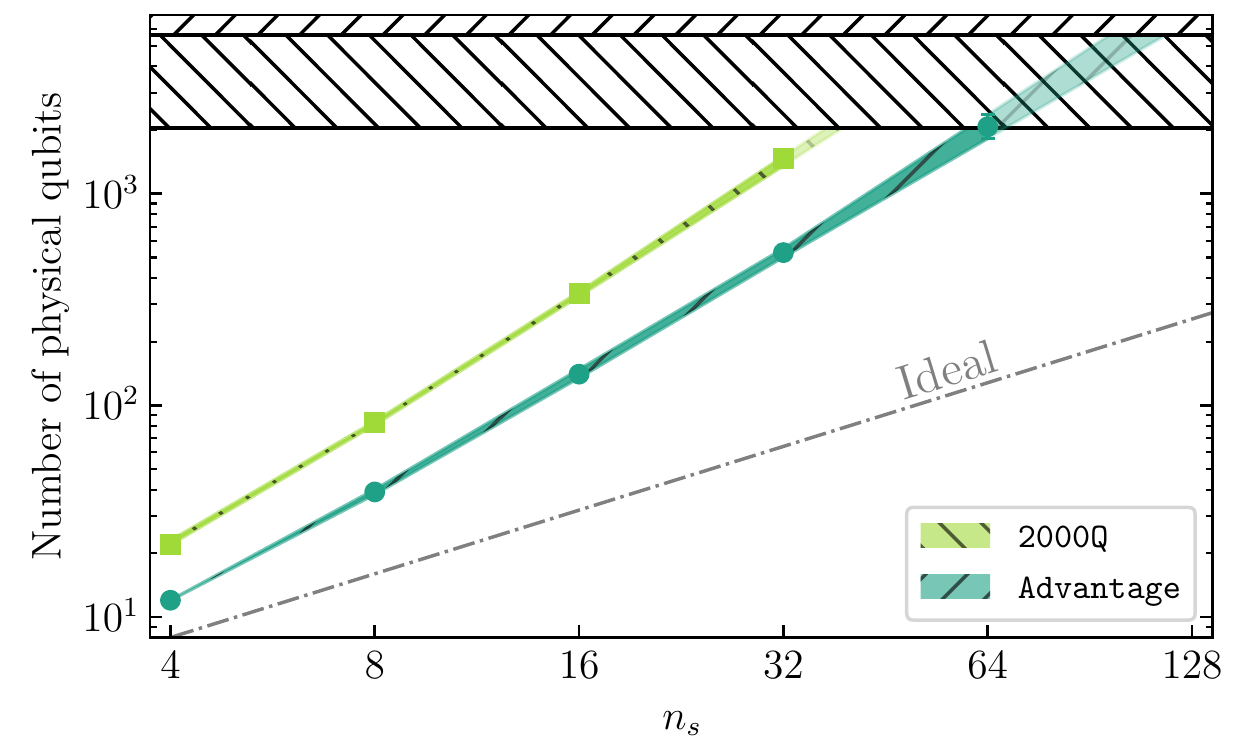}
	\caption{
	Numbers of physical qubits required to map the QUBO matrix from the Hamiltonian in Eq.~\eqref{eq:hamphi4} as a function of $n_s$ for a fixed $K=2$. The bands show results for two available D-Wave QAs, {\tt 2000Q} and {\tt Advantage}, with the lighter bands being extrapolations using the last two points. The gray dot-dashed line represents the ideal scaling, with $Kn_s$ qubits.
		}
	\label{fig:scaling}
\end{figure}

\noindent
During this NISQ era, quantum processors are being characterized to determine their strengths and weaknesses.
For this purpose, Fig.~\ref{fig:scaling} shows the number of physical qubits required for the Hamiltonian in Eq.~\eqref{eq:hamphi4} with different numbers of basis states $n_s$.
Specifically, we compute the QUBO matrix with 
$K=2$ for different values of $n_s$, find the embedding using the {\tt find\_embedding} command from {\tt minorminer}~\cite{DwaveNeal}, as discussed in Sec.~\ref{subsec:dwave_scalar}, and count the number of required physical qubits,
which are shown in Fig.~\ref{fig:scaling}.
As it uses a heuristic algorithm to find this mapping~\cite{cai2014practical}, the number of qubits is generally different each time this command is called, and the width of the bands represents a 68\% confidence interval determined from 20 different embeddings of the same problem.
Figure~\ref{fig:scaling} also shows the number of qubits in the ideal case with the gray line, assuming an all-to-all connectivity, therefore requiring only $Kn_s$ qubits.

It is interesting to note that the improvement over the previous D-Wave's QA, {\tt 2000Q}, with a total of 2048 qubits (and a connectivity of six qubits), to the current one, {\tt Advantage}, allows for a  reduction in the number of qubits by a factor of $2-3$ (directly related to the increase in connectivity between qubits, which increases from 6 to 15).
However, the points for both QAs follow a line with a similar slope, which is larger than the ideal case.

Although the current limit to $n_s$ is $64$, without the zooming algorithm described in Sec.~\ref{sec:qubo}, it would be  smaller (assuming the same precision on the eigenenergies), as $K$ would have to be larger.
Further,  using the symmetry properties of the wavefunction, as in Sec.~\ref{subsec:delocalized}, 
the value of $n_s$ could be doubled.

\section{Real-time evolution of Plaquettes and Neutrinos Using Feynman Clocks}
\label{sec:timeevol}
\noindent
The real-time dynamics of physically interesting complex quantum systems is an expected  capability of future quantum computers, which will advance the domain of sciences beyond what is possible with classical computing.
While for universal gate-based quantum computers, the challenge to implement time evolution is determining efficient quantum circuits and mappings that can be executed on available devices,
the challenge for D-Wave's QAs is in finding a viable
QUBO matrix that can be implemented.
First formalized for scientific applications in the context of quantum chemistry~\cite{McCleanE3901,Tempel_2014}, 
Feynman clock states~\cite{Feynman:85,kitaev2002classical}
provide a way to time-evolve  quantum systems in a single run of a QA (its first implementation on quantum hardware can be found in Ref.~\cite{Jalowiecki_2022}).
The constraint of real entries in the QUBO matrix
can be circumvented by an appropriate change of basis that transforms the Hamiltonian into a purely imaginary form, rendering ${\hat U}_t=e^{-it\mathcal{\hat H}}$ real, for example, as used in Ref.~\cite{ARahman:2021ktn}.
Following the formulation of McClean {\it et al.}~\cite{McCleanE3901}, called the Time-Embedded Discrete Variational Principle (TEDVP), the objective function to be minimized has the form
\begin{equation}
    F=\sum_{t,t'}\langle t' \rvert \langle \Psi_{t'} \rvert \, \hat {\cal C} \, \lvert \Psi_{t} \rangle \lvert t \rangle -\eta \sum_{t,t'}\langle t' \rvert \langle \Psi_{t'} | \Psi_{t} \rangle \lvert t \rangle \ \ ,
\end{equation}
where the $\eta$ parameter has the same purpose as in Eq.~\eqref{eq:Ffun} and $\lvert \Psi_{t}\rangle \lvert t \rangle$ are compound states formed of the physical wavefunction $\lvert \Psi_{t} \rangle$ and the time register $\lvert t \rangle$.
The clock Hamiltonian\footnote{
This is constructed in the context of superpositions of time-slices via projectors formed from
\begin{equation}
|t\rangle_- = \frac{1}{\sqrt{2}} \left(\ |t\rangle - |t+\delta t\rangle\ \right) \ \ .
\end{equation}
} $\hat {\cal C}$ is defined as
\begin{align}
    \hat {\cal C}= {\hat C_0} &\,+\,\frac{1}{2}\sum_{t}( {\hat I}\otimes | t \rangle\langle t | - {\hat U}_{\delta t}\otimes | t +\delta t \rangle\langle t | \nonumber \\
    & - {\hat U}^\dag_{\delta t} \otimes| t \rangle\langle t + \delta t |  +  {\hat I}\otimes | t + \delta t \rangle\langle t + \delta t | )\ \ ,
\end{align}
with $\hat C_0$ a penalty term to select a particular (input) state at a time $t$.
For our purposes, like previous works of others,
we will use it to select the initial state 
$\lvert \Psi_{\rm in}\rangle$ at $t=0$ with 
$\hat C_0=(I-\lvert \Psi_{\rm in}\rangle\langle \Psi_{\rm in} \rvert)\otimes \lvert 0\rangle\langle 0 \rvert$.
One issue that arises in using this method with a QA is that the matrix elements of ${\hat U}_{\delta t}$ are required to be computed explicitly.

We extend the  formulation of the Feynman clock to allow for complex values in the QUBO elements, following Ref.~\cite{D0CP04272B}.
In such a situation, Eq.~\eqref{eq:QAE} becomes
\begin{equation}
    F^{(\cbar)} = \sum^{n_T\times n_s}_{\alpha,\beta} \overline{a}_\alpha a_\beta\  \cbar_{\alpha\beta}\ \ ,
    \label{eq:QAE_cmplx}
\end{equation}
where $\overline{a}_\alpha$ is the complex conjugate of $a_\alpha$. Here
$\cbar_{\alpha\beta}$ are matrix elements of $\hat{\mathcal{C}}-\eta \hat I$ between 
basis states spanning $\lvert \Psi_t \rangle \lvert t \rangle$,
of which there are $n_T\times n_s$, where $n_T$ is the number of time slices and $n_s$ is the number of basis states of the Hamiltonian.
Since  $\hat {\cal C}$ is Hermitian, terms in Eq.~\eqref{eq:QAE_cmplx} can be written as, after the summations,
%
\begin{align}
    \overline{a}_\alpha a_\beta\ \cbar_{\alpha\beta} = & \, a^{\rm Re}_\alpha a^{\rm Re}_\beta \cbar^{\rm Re}_{\alpha\beta} - a^{\rm Re}_\alpha a^{\rm Im}_\beta \cbar^{\rm Im}_{\alpha\beta} \nonumber \\
    & + a^{\rm Im}_\alpha a^{\rm Re}_\beta \cbar^{\rm Im}_{\alpha\beta} +  a^{\rm Im}_\alpha a^{\rm Im}_\beta \cbar^{\rm Re}_{\alpha\beta}  \ \ ,
    \label{eq:QAE_cmplx_2}
\end{align}
%
which involves only real numbers.
With this form, together with Eq.~\eqref{eq:defa_AQAE_improved_2} for the fixed-point representation of the real and imaginary parts of $a_\alpha$, 
the elements of the QUBO matrix, $Q_{\alpha,i;\beta,j}$, become
\begin{widetext}
\begin{equation}
    Q_{\alpha,i;\beta,j} = \begin{cases}
    2^{i+j-2K-2z} (-1)^{\delta_{iK}+\delta_{jK}} \cbar^{\rm Re}_{\alpha\beta}+ 2\delta_{\alpha\beta} \delta_{ij} 2^{i-K-z} (-1)^{\delta_{iK}} \sum_\gamma \left(a^{{\rm Re},(z)}_\gamma \cbar^{\rm Re}_{\gamma\beta} + a^{{\rm Im},(z)}_\gamma \cbar^{\rm Im}_{\gamma\beta}\right) & 1\leq i,j \leq K \, , \\
    - 2^{i+j'-2K-2z}(-1)^{\delta_{iK} + \delta_{j'K}} \cbar^{\rm Im}_{\alpha\beta} & 1\leq i, j' \leq K \, , \\
    2^{i'+j-2K-2z}(-1)^{\delta_{i'K} + \delta_{jK}} \cbar^{\rm Im}_{\alpha\beta} & 1\leq i',j \leq K \, , \\
    2^{i'+j'-2K-2z}(-1)^{\delta_{i'K} + \delta_{j'K}} \cbar^{\rm Re}_{\alpha\beta} + 2\delta_{\alpha\beta} \delta_{i'j'} 2^{i'-K-z} (-1)^{\delta_{i'K}} \sum_{\gamma} \left(a^{{\rm Im},(z)}_\gamma \cbar^{\rm Re}_{\gamma\beta}- a^{{\rm Re},(z)}_\gamma \cbar^{\rm Im}_{\gamma\beta}\right) & 1\leq i',j' \leq K \, .
\end{cases}
\end{equation}
\end{widetext}
where, to accommodate both real and imaginary parts of $a_\alpha$,
the index $i$ resides in the range 
$1\leq i \leq 2K$, 
with $1\leq i \leq K$ for $a^{\rm Re}_\alpha$ 
and $(K+1)\leq i \leq 2K$  for $a^{\rm Im}_\alpha$,
with $i'\equiv i-K$ (the derivation of this expression can be found in Appendix~\ref{app:derivation}).
The dimension of the QUBO matrix using $n_T$ time slices with the TEDVP formalism
is $2 K n_T n_s \times 2 K n_T n_s$ 
(a factor of $(2 n_T)^2$ times larger than that used to determine wavefunctions of the Hamiltonian, as described in Sec.~\ref{sec:qubo}).

In the following subsections we examine two systems of physical interest to Standard Model research, the time evolution of a single plaquette of SU(3) Yang-Mills lattice gauge theory and of a four neutrino system, using {\tt Advantage}.
Both of these systems have been simulated previously using IBM's superconducting quantum computers, with two and four qubits.
In studying these systems with {\tt Advantage}, the following values of parameters were found to be effective: $N_A=10^3$, $N_{\rm run}=20$, $K=2$, $\eta=0$, and $\{z^{\rm init},z^{\rm max}\}=\{0,14\}$. 
The chain strength coefficient is fixed to $\omega=0.2$, and the default annealing-schedule parameters ($t_A=20\,\mu {\rm s}$) are used.

\subsection{One-Plaquette in SU(3) Yang-Mills Lattice Gauge Theory}
\label{subsec:su3}
\noindent
Quantum simulations of non-Abelian lattice gauge theories are 
anticipated to become increasingly important in Standard Model research.
Progress toward this objective is at its earliest stages, with simulations of small systems in low-dimensions underway using the available NISQ-era devices, e.g., Refs.~\cite{PhysRevA.73.022328,Zohar:2011cw,Zohar:2012ay,Tagliacozzo:2012vg,Zohar:2012ts,Zohar:2012xf,Hauke:2013jga,Wiese:2013uua,Marcos:2014lda,Kuno:2014npa,Bazavov:2015kka,Kasper:2015cca,Brennen:2015pgn,Martinez:2016yna,Kuno:2016xbf,Zohar:2016iic,Kasper:2016mzj,Muschik:2016tws,Gonzalez-Cuadra:2017lvz,Gonzalez-Cuadra:2017lvz,Banuls:2017ena,PhysRevA.98.032331,Kaplan:2018vnj,Lu:2018pjk,Stryker:2018efp,Banuls:2019bmf,Davoudi:2019bhy,PhysRevD.101.074512,Magnifico:2019kyj,Luo:2019vmi,PhysRevA.98.032331,PhysRevD.101.074512,Shaw2020quantumalgorithms,Halimeh:2020ecg,paulson2020simulating,Halimeh:2020djb,VanDamme:2020rur,Ott:2020ycj,Ciavarella:2021nmj,Atas:2021ext,Davoudi:2021ney,ARahman:2021ktn,Kan:2021nyu,Stryker:2021asy,aidelsburger2021cold,Ciavarella:2021lel}.
The Kogut-Susskind Hamiltonian~\cite{PhysRevD.11.395,RevModPhys.51.659}  developed in the 1970s, provides one concrete framework for quantum simulations of lattice gauge theories, and is being actively pursued with superconducting devices, trapped ion systems, optical systems, superconducting radio frequency cavities, and QAs.  
Simulations of small systems have been performed in one and two spatial dimensions, with simulations of the smallest three-dimensional systems barely within reach of today's devices.
Extensive efforts are underway to develop techniques to make simulations with this framework more practical, for instance, integrating over the gauge spaces at each lattice site~\cite{Banuls:2017ena,PhysRevD.101.074512}.
Other mappings of the gauge fields, for instance, quantum link models (e.g., Refs.~\cite{Brower:1997ha,Banerjee:2012xg,Tagliacozzo:2012df,Wiese:2021djl}), spin systems, and the discrete sampling of gauge fields (e.g., Refs.~\cite{Alexandru:2019nsa,Ji:2020kjk}) 
are under active exploration.
While the formal construction for quantum simulations of non-Abelian gauge theories has been established for more than a decade, and concrete protocols for implementation on quantum devices known for a comparable period of time, first implementations appeared in 2016 using trapped-ion systems~\cite{Martinez:2016yna} and soon after using superconducting~\cite{PhysRevA.98.032331} and optical systems~\cite{Lu:2018pjk}.
Last year, the first simulations of SU(3) Yang-Mills theories were performed~\cite{Ciavarella:2021nmj} of one and two plaquettes, building upon previous simulations of SU(2) plaquette systems~\cite{PhysRevD.101.074512} and one-dimensional SU(2) chains~\cite{Atas:2021ext}.
These small systems can be simulated using D-Wave's annealers, as was first demonstrated for the SU(2) plaquette systems in the work of A~Rahman and collaborators~\cite{ARahman:2021ktn}.

The time evolution of one and two plaquettes in SU(3) Yang-Mills gauge theory has been simulated using IBM’s \texttt{Athens} quantum computer~\cite{Ciavarella:2021nmj}.
Both a local basis and global bases were simulated, with the single plaquette a particularly simple system with a minimal qubit footprint in the global basis due to Gauss's Law restrictions.
In this work, we focus on a one-plaquette system in the color parity basis, including the states $\{|\mathbf{1}\rangle ,|\mathbf{3}^+\rangle , |\mathbf{6}^+\rangle,|\mathbf{8}\rangle\}$, which has the following Hamiltonian when mapped to two qubits,
\begin{align}
    \mathcal{\hat H} =&\, g^2 \left(\frac{23}{6} \hat{I} \otimes \hat{I} - \frac{5}{2} \hat{Z} \otimes \hat{I} - \frac{1}{2}
	\hat{I} \otimes \hat{Z} - \frac{5}{6} \hat{Z} \otimes \hat{Z} \right) \nonumber\\
	&- \frac{1}{2g^2} \left[ \sqrt{2}\ \hat{I}\otimes \hat{X}+\sqrt{2} \ \hat{X}\otimes \left(\frac{\hat{I}-\hat{Z}}{2}\right) \right. \nonumber\\
	& + \frac{1}{2} \hat{X}\otimes \hat{X} + \frac{1}{2} \hat{Y}\otimes \hat{Y} \nonumber\\
	& \left. + \frac{1}{4}\left(\hat{I}+\hat{Z}\right)\otimes\left(\hat{I}-\hat{Z}\right)-6 \ \hat{I} \otimes \hat{I} \right]
	\ \ ,
	\label{eq:SU3_ham}
\end{align}
where $\hat{X}$ , $\hat{Y}$, and $\hat{Z}$ are the Pauli matrices. The utility of color parity arises from the Hamiltonian containing only the symmetric combination of the plaquette operator $\Box+\Box^\dagger$ and the trivial vacuum being even under color parity transformation.
Simulations performed with IBM's {\tt Athens} 
used a strong coupling constant of $g=1$.
The system was time evolved using a Trotterized decomposition of the evolution operator to enable an efficient mapping onto 
quantum circuits~\cite{Ciavarella:2021nmj}.
Both first- and second-order Trotterizations were employed, using a single step ($\delta t=t$) and multiple steps ($\delta t=t/2$) for both. 
Applying standard error mitigation techniques (for CNOT errors) and fitting systematic error estimation, 
the vacuum-to-vacuum probability $|\langle \mathbf{1} | \hat U_t| \mathbf{1} \rangle |^2$ (the vacuum is the plaquette in the $| \mathbf{1} \rangle \equiv |0\rangle\otimes |0\rangle$ state) and the expectation value of the electric energy 
(the $g^2$ terms in Eq.~\eqref{eq:SU3_ham}) were computed as a function of time, as shown in Fig.~8 of Ref.~\cite{Ciavarella:2021nmj}.
\begin{figure}[!t]
    \begin{tikzpicture}
    \node(a){\includegraphics[width=0.9\columnwidth]{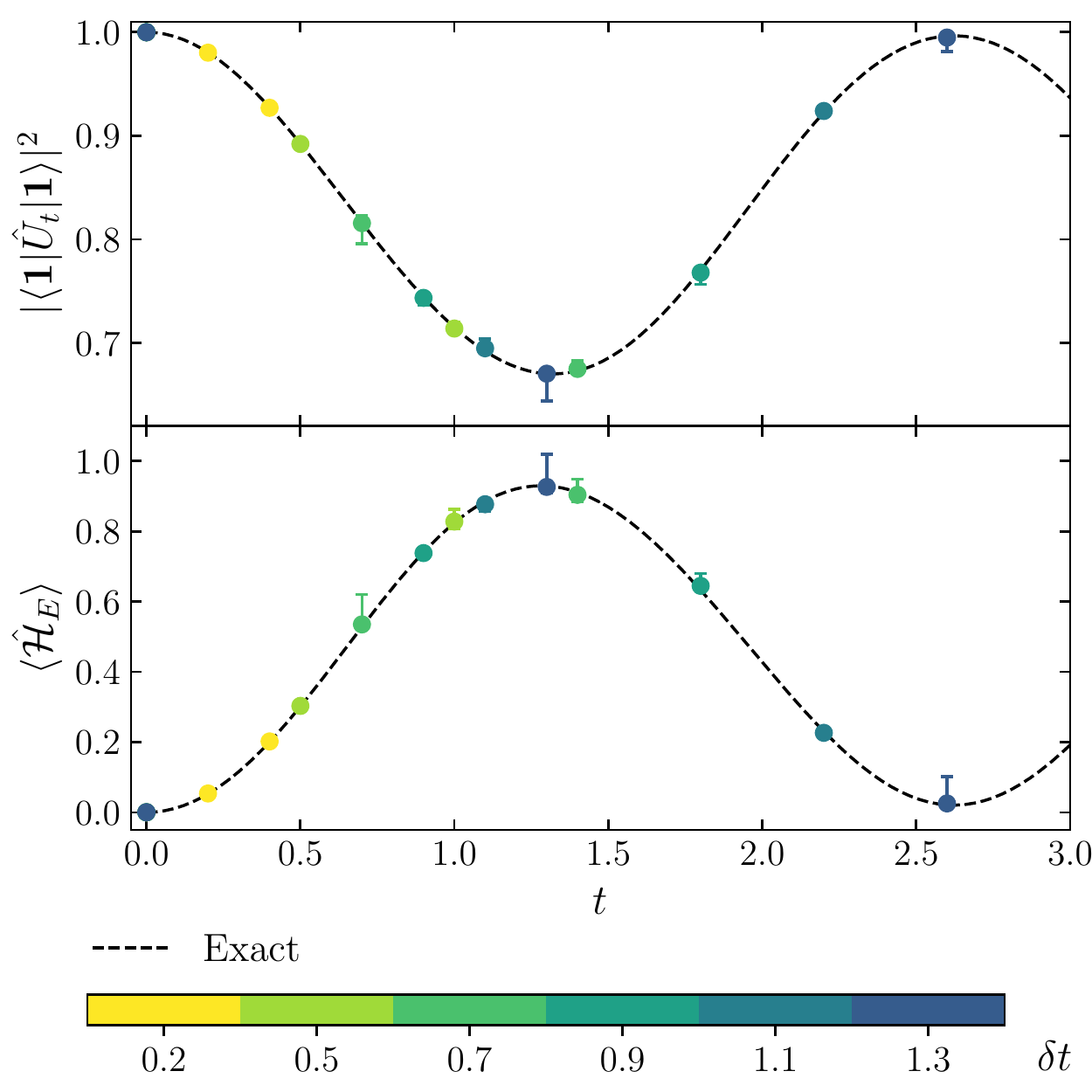}};
    \node at (a.north west) [anchor=north,xshift=1mm,yshift=0mm]
    {\includegraphics[width=0.04\textwidth]{iconQ1_bf.png}};
    \end{tikzpicture}
	\caption{Vacuum-to-vacuum probability $|\langle \mathbf{1} |\hat U_t| \mathbf{1} \rangle |^2$ (upper panel) and energy (in units of $g^2$) in the electric field (lower panel) of the one-plaquette system as a function of time (in units of $1/g^2$). The circles correspond to results obtained using D-Wave's QA {\tt Advantage} with the maximal number of zoom steps ($z^{\rm max}=14$), $K=2$, $N_A=10^3$ number of anneals, different time shifts $\delta t$, and $n_T=3$ time steps $\{0,\delta t, 2\delta t\}$. The uncertainties correspond to the 68\% confidence intervals determined from $N_{\rm run}=20$ independent runs.
	The dashed curves correspond to the exact theoretical curves.
	}
	\label{fig:SU3_dwave}
\end{figure}

In the present work, 
using exact matrix exponentiation of the Hamiltonian to determine the evolution operator over the time interval $\delta t$, 
a QUBO matrix describing the Feynman clock evolution of this system was formed.
Using the techniques described in previous sections to find eigenstates and energies, {\tt Advantage} was used to evolve the SU(3) plaquette system forward in time from an initial state of the trivial vacuum.
Results obtained for 
the vacuum-to-vacuum persistent probability and for the energy in the electric field
using {\tt Advantage} are shown in Fig.~\ref{fig:SU3_dwave}. 
These results are found to agree with exact theoretical curves within uncertainties.
Further, the precision of the results is significantly better than that obtained previously
using IBM's {\tt Athens}~\cite{Ciavarella:2021nmj}.

\subsection{Neutrino Flavor Dynamics in Beam-Beam Collisions}
\label{subsec:neutrinos}
\noindent
Neutrino flavor dynamics is a major focus of research in Standard Model physics.
While neutrinos are rendered massless by dimension-4 operators in the Standard Model, a result of the local gauge symmetries and particle content, in particular the absence of a right-handed neutrino field, the unambiguous observations of neutrino flavor dynamics, and non-zero mass differences, provides unique insight into aspects of physics beyond the Standard Model and the structure of higher-dimension operators.
The connection between lepton-number violating Majorana neutrino masses and interactions that induce neutrinoless $\beta\beta$-decay of nuclei is a strong theoretical motivation driving the current experimental program(s) searching for such processes (for a recent review, see, e.g., Ref.~\cite{Agostini:2022zub}). 
Non-zero neutrino masses, when combined with Standard Model electroweak interactions, have  implications for matter under the extreme conditions of density and temperature that are found in the early universe (see, e.g., Refs.~\cite{Savage:1990by,PANTALEONE1992128,Bruce_PhysRevD.49.2710}) and core-collapse supernova (see, e.g., Refs.~\cite{PANTALEONE1992128,Qian:1994wh})
(for recent works, see, e.g., Ref.~\cite{Capozzi:2022slf}).
Decades of work on this subject continue to uncover new  phenomena in neutrino dynamics in these environments, including the recent identification of dynamical phase transitions in collective dynamics and correlations with quantum entanglement~\cite{Bell:2003mg,Friedland:2003eh,Sawyer:2004ai,Pehlivan:2011hp,Duan:2006an,Rrapaj:2019pxz,Cervia:2019res,Roggero:2021asb,Roggero:2021fyo}.
With their importance in transport from within the core, high-precision
simulations of the evolution of supernova require the inclusion of three-dimensional neutrino distributions with detailed quantum kinetics,
a problem that has been estimated to lie beyond classical exascale computing.
This has prompted the increasing number of explorations of neutrino dynamics using quantum simulations~\cite{Arguelles:2019phs,Hall:2021rbv,Yeter-Aydeniz:2021olz} and modern theoretical tools using entanglement as an essential ingredient, e.g., tensor networks~\cite{Roggero:2021asb,Roggero:2021fyo,Cervia:2022pro},
and with, for example, classical simulations 
utilizing symmetries and matrix sparsity~\cite{Martin:2021bri} to study systems currently beyond reach of matrix product states.
These works constitute important explorations of the roles of quantum information, entanglement and real-time dynamics in dense neutrino systems, and make inroads into unifying previously identified collective phenomena while searching for new behaviors.

One such recent detailed study, which we parallel,
by Hall {\it et al.}~\cite{Hall:2021rbv}, 
performed quantum simulations of
systems of $N=4$ neutrinos, restricted to two active flavors and without the inclusion of electroweak interactions with matter (such as $e^\pm$, $\mu^\pm$ and $q, \overline{q}$'s) but with self-interactions. 
Using the known mapping of the two-flavor neutrino system to quantum spin models,
it was simulated using IBM's {\tt Vigo} superconducting quantum computer for a selection of parameters, including mass differences and neutrino densities, using the effective Hamiltonian, 
\begin{align}
     \mathcal{\hat H} =& \frac{1}{2} \sum_i^N ( -\Delta_i \cos 2 \theta_v \ \sigma_i^z + \Delta_i \sin 2 \theta_v \ \sigma_i^x) \nonumber\\
     & +\ \kappa\ \sum_{i<j}^N  ( 1 - \cos\theta_{ij} )\ {\boldsymbol\sigma}_i \cdot {\boldsymbol\sigma}_j\ \ ,
\label{eq:nuHami}
\end{align}
where $\theta_v$ is the flavor mixing angle (which is set to $\theta_v=0.195$ for this model) and $\Delta_i=\delta m^2/(2E_i)$ is the strength of the one-body term determined by the difference in neutrino squared masses $\delta m^2$ and the energy of each neutrino, $E_i$. 
The strength of the two-body term, $\kappa$, depends on the neutrino density and electroweak couplings, and $\theta_{ij}$ is the angle between the momenta of the $i^{\rm th}$ and $j^{\rm th}$ neutrinos.
The spin operators ${\boldsymbol \sigma}_i$ act in 
the two-dimensional neutrino flavor space ${\boldsymbol\nu}_i = \left(\nu_{i,e} , \nu_{i,\mu} \right)^T$.

For the test-case model simulation presented in Ref.~\cite{Hall:2021rbv}, a monochromatic neutrino beam is assumed, with $E_i=\delta m^2/(4\kappa)$, and with an anisotropic distribution of momentum directions, 
$\theta_{ij}=\arccos (\zeta) \times|i-j|/(N-1)$,
with $\zeta=0.9$.\footnote{i.e.,
$\theta_{12}=\theta_{23}=\theta_{34}=\frac{1}{3}\arccos (\zeta)$, 
$\theta_{13}=\theta_{24}=\frac{2}{3}\arccos (\zeta)$ and 
$\theta_{14}=\arccos (\zeta)$.
}
The time evolution of the system was determined by first-order Trotterization of the evolution operator derived from the Hamiltonian separated into neutrino-pair terms (as opposed to one- and two-body operators)~\cite{Hall:2021rbv}.
One of the observables examined was the probability of the $i^{\rm th}$ neutrino
transforming between flavors
$\nu_e\leftrightarrow\nu_\mu$,
\begin{equation}
    P_{i} (t) = \frac{1}{2} \langle \Psi_t \rvert 1 \mp \sigma^z_i \lvert \Psi_t \rangle \ \ ,
\end{equation}
starting with 
$\lvert \Psi_0\rangle=\lvert \nu_e \nu_e \nu_\mu \nu_\mu\rangle$
and with the sign depending on the initial state of the system, $\nu_{i,e}$ ($-$) or $\nu_{i, \mu}$ ($+$).
The results of those simulations
can be found in Figs.~3 and 4 of their paper~\cite{Hall:2021rbv}.
The evolution of flavor entanglement in the
four-neutrino system was also studied in Ref.~\cite{Hall:2021rbv}. 
The single-neutrino entanglement entropy is given by
\begin{equation}
    S_i(t)=-{\rm Tr}[\rho_i(t)\log_2 (\rho_i(t))]\ \ ,
\end{equation}
where $\rho_i(t)$ is the reduced density matrix for the $i^{\rm th}$ neutrino, with 
$\rho_i(t) = 
{\rm Tr}_{j\ne i}\left[ \ 
\lvert \Psi_t \rangle \langle \Psi_t \rvert \  \right]$.
The concurrence was also studied~\cite{Hall:2021rbv}, and in this work we consider the  logarithmic negativity,
\begin{equation}
    \mathcal{N}_{ij}(t)= \log_2 ||\rho^{\Gamma}_{ij}(t)||_1\ \ ,
\end{equation}
where $\rho_{ij}(t)$ is the two-neutrino
reduced density matrix for the $ij$ neutrino pair, $\Gamma$ indicates the partial transposition of $\rho$, and $||\cdot||_1$ 
is the trace norm.
The logarithmic negativity, related to the concurrence, is an upper bound on the distillable entanglement.
The neutrino Hamiltonian in Eq.~(\ref{eq:nuHami}) is invariant under neutrino exchanges $1\leftrightarrow 4$ and $2\leftrightarrow 3$~\cite{Hall:2021rbv}.
This gives rise to relations between  observables, such as $P_{1}(t)=P_{4}(t)$, $S_{1}(t)=S_{4}(t)$, and $\mathcal{N}_{12}(t)=\mathcal{N}_{34}(t)$.
\begin{figure}[!t]
    \begin{tikzpicture}
    \node(a){\includegraphics[width=0.9\columnwidth]{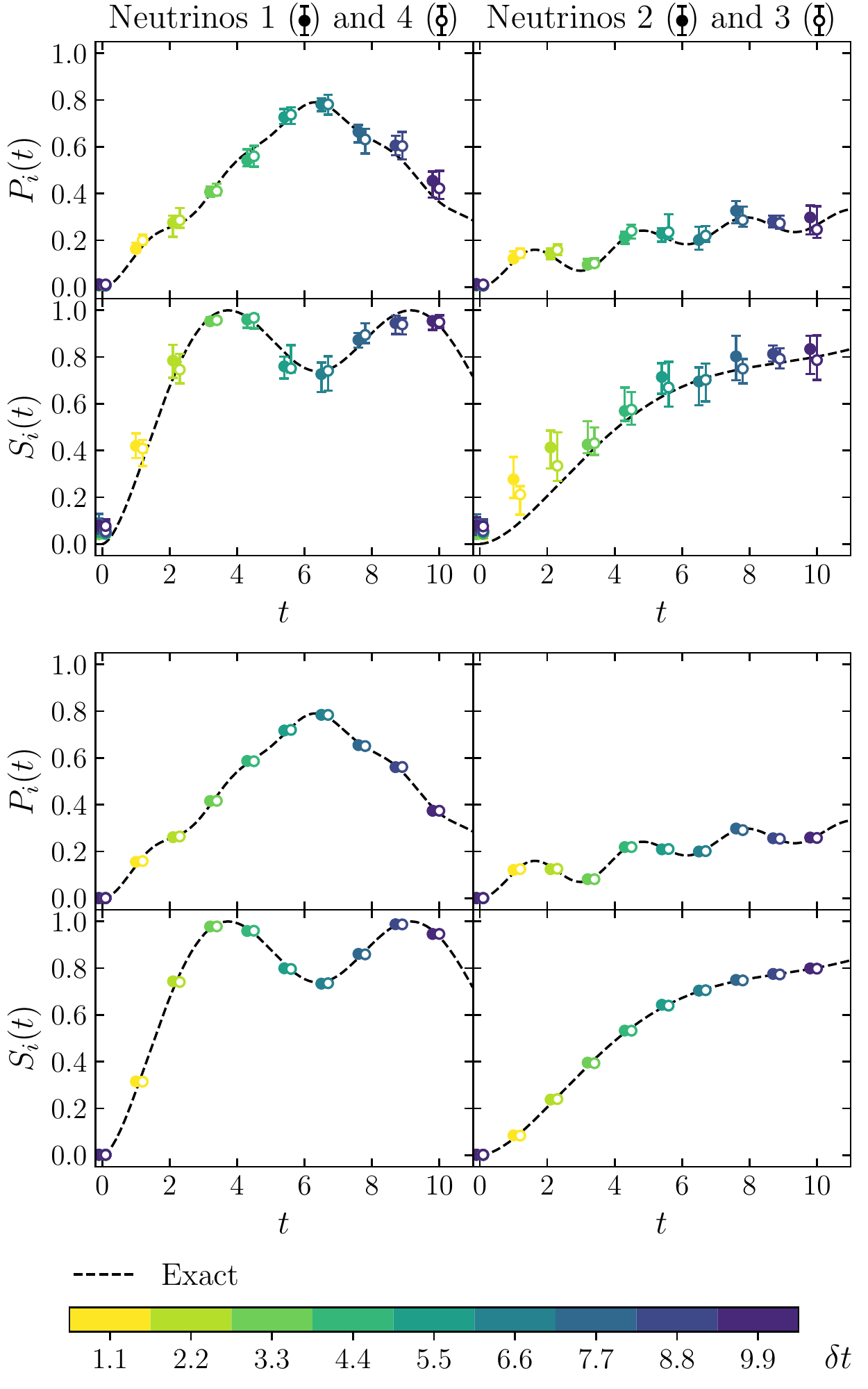}};
    \node at (a.north west) [anchor=north,xshift=1mm,yshift=-1mm]
    {\includegraphics[width=0.04\textwidth]{iconQ1_bf.png}};
    \end{tikzpicture}
	\caption{Probability of flavor transitions (panel rows 1 and 3) and the single-neutrino entanglement entropy (panel rows 2 and 4) for the $1^{\rm st}$ and $4^{\rm th}$ (left panels) and $2^{\rm nd}$ and $3^{\rm rd}$ (right panels) neutrinos as a function of time (in units of $\kappa$). The results are obtained using D-Wave's QA {\tt Advantage} with the maximal number of zoom steps, $K=2$, $N_A=10^3$ anneals, different time shifts $\delta t$, and $n_T=2$ time steps $\{0,\delta t\}$ (they have been shifted slightly along the $x$-axis for clarity). The upper four panels show the raw results, while the lower four panels show the results after two iterations of the procedure described in the main text. The uncertainties correspond to the 68\% confidence intervals determined from $N_{\rm run}=20$ independent runs.
	The dashed curves correspond to the exact theoretical curves.
	}
	\label{fig:Neutrinos_dwave}
\end{figure}
\begin{figure}[!th]
    \begin{tikzpicture}
    \node(a){\includegraphics[width=0.9\columnwidth]{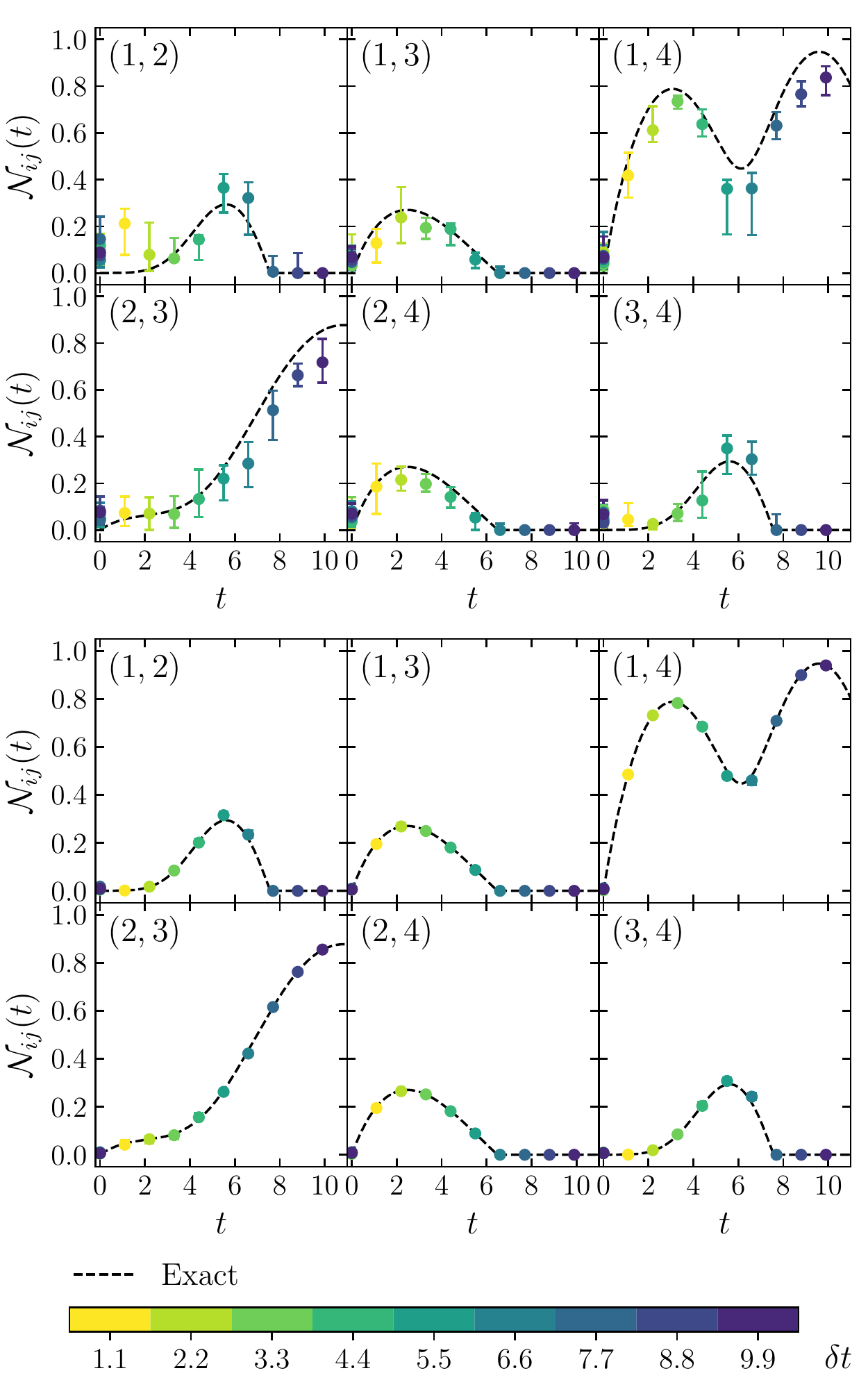}};
    \node at (a.north west) [anchor=north,xshift=1mm,yshift=-1mm]
    {\includegraphics[width=0.04\textwidth]{iconQ1_bf.png}};
    \end{tikzpicture}
	\caption{Logarithmic negativity of pairs of neutrinos as a function of time (in units of $\kappa$). The results are obtained using D-Wave's QA {\tt Advantage} with the maximal number of zoom steps, $K=2$, $N_A=10^3$ anneals, different time shifts $\delta t$, and $n_T=2$ time steps $\{0,\delta t\}$.The upper six panels show the raw results, while the lower six panels show the results after two iterations of the procedure described in the main text. The uncertainties correspond to 68\% confidence intervals determined from $N_{\rm run}=20$ independent runs.
	The dashed curves correspond to the exact theoretical curves.
	}
	\label{fig:Neutrinos_dwave2}
\end{figure}

The results 
of our quantum simulations obtained using {\tt Advantage} are shown in Figs.~\ref{fig:Neutrinos_dwave} and~\ref{fig:Neutrinos_dwave2}.
The implementation of the clock state using the matrix representation of the exact evolution operator between time-slices, without Trotterization into products of unitaries associated with neutrino pairs, eliminates  a significant source of (``theory'') systematic error imposed by circuit-volume limitations of available devices with different architectures, as can be seen by comparing the results shown in Figs.~\ref{fig:Neutrinos_dwave} and~\ref{fig:Neutrinos_dwave2} and the results presented in Figs.~3, 4 and 6 of Ref.~\cite{Hall:2021rbv}.
Further, and equally important, the absence of systematic errors associated 
with device performance in the simulation, dominated by CNOT gates and subsequent mitigation procedures,  improves the accuracy of the simulations of this system
that are possible with {\tt Advantage} compared with other quantum devices.

The uncertainties 
associated with the dynamics of four neutrinos
are considerably larger than for those associated with the single plaquette of SU(3) Yang-Mills lattice gauge theory, discussed in the previous subsection.
This is due to a larger QUBO matrix that is passed to the annealer, and is one indication of the scaling of the capabilities of {\tt Advantage} with increasing system size.  
Adding one more neutrino to the system renders the problem intractable for {\tt Advantage} as the QUBO matrix will not fit onto its QPU.\footnote{
We attempted to study the $N=5$ neutrino system with {\tt Advantage} by setting $K=1$, but the results (and uncertainty estimations) obtained were unreliable.
One of the issues is that $a_{\alpha}$, with $K=1$, only takes two values, $\{-1,0\}$, requiring a large value of $\eta$ to prevent the null solution.} 
Additionally, it can be seen that the uncertainties for the logarithmic negativity in Fig.~\ref{fig:Neutrinos_dwave2} are larger than those of the single-neutrino entanglement entropy in Fig.~\ref{fig:Neutrinos_dwave}, which are in turn larger than the neutrino flavor transition probability, also in Fig.~\ref{fig:Neutrinos_dwave}.
The study of such quantum correlations requires high-precision calculations.  In some cases, the wavefunctions $\lvert \Psi_t \rangle$ are determined with $10^{-1}-10^{-2}$ precision, which is seen to be insufficient. 

%
\begin{figure}[!t]
    \begin{tikzpicture}
    \node(a){\includegraphics[width=0.93\columnwidth]{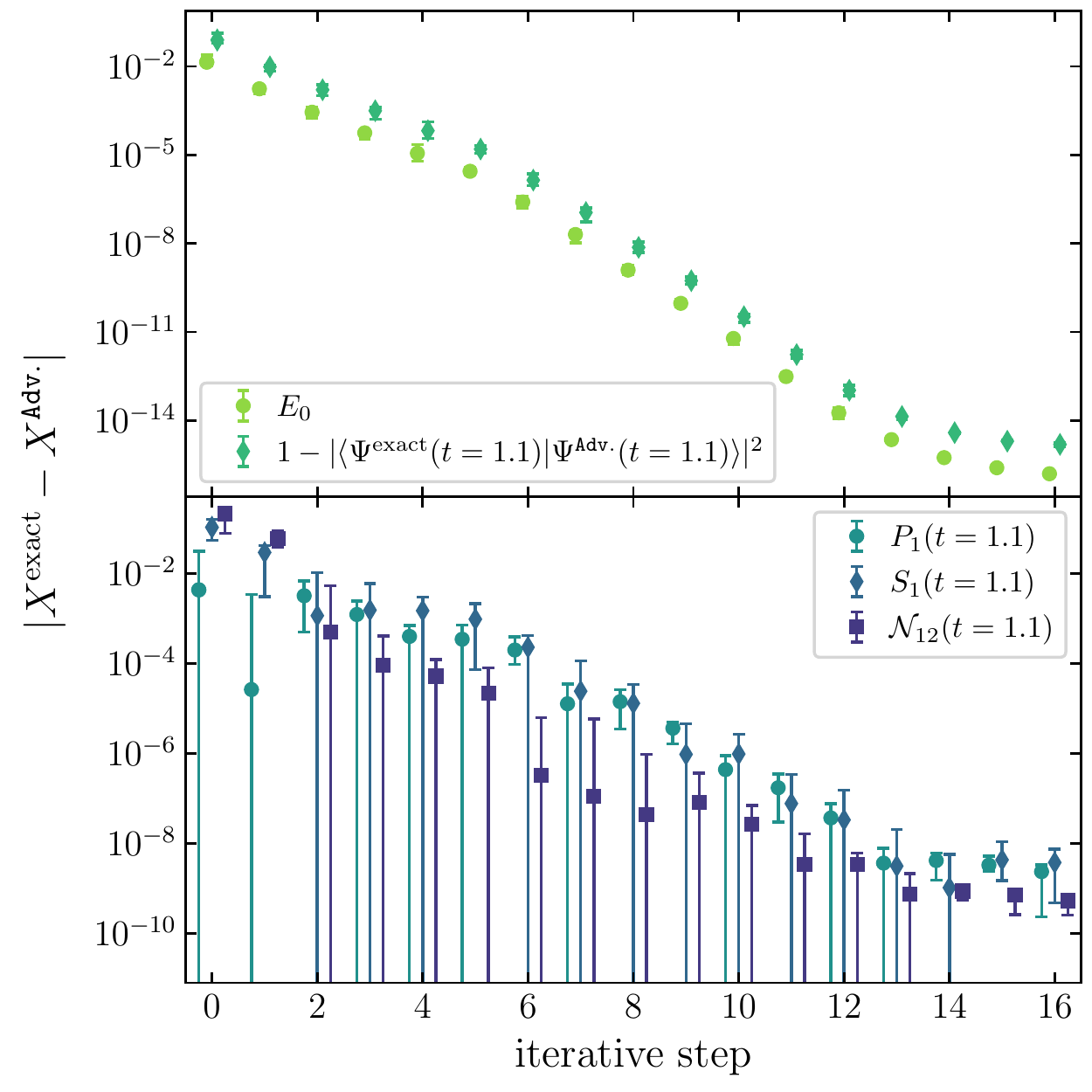}};
    \node at (a.north west) [anchor=north,xshift=3mm,yshift=-1mm]
    {\includegraphics[width=0.04\textwidth]{iconQ1_bf.png}};
    \end{tikzpicture}
	\caption{Convergence of the ground-state energy and wavefunction (upper panel), and the flavor transition probability, single-neutrino entanglement, and logarithmic negativity (lower panel) as a function of the number of steps of the iterative procedure described in the main text. The results are obtained using D-Wave's QA {\tt Advantage} with the maximal number of zoom steps, $K=2$, and $N_A=10^3$ anneals. The uncertainties correspond to 68\% confidence intervals determined from $N_{\rm run}=20$ independent runs.
	}
	\label{fig:Neutrinos_dwave3}
\end{figure}

One way of improving the results (and reduce uncertainties) is to increase the number of anneals $N_A$ by a factor $\xi$.
As this leads to only a $1/\sqrt{\xi}$ reduction in the  uncertainty in energies, 
this (brute force) approach demands excessive computational resources.
A better method for reducing the uncertainties is analogous to the multigrid method used in Sec.~\ref{sec:scalarfield}.
The challenge here is that it is not straightforward to define the Hamiltonian for a smaller system to provide interpolating wavefunctions for the larger system (the four-neutrino case is not continuously connected to the 2,3 (or 5) neutrino systems).
We have found that using the solution from the QA as a starting point for a subsequent anneal, but with $z\neq 0$ (to narrow the window of $a_\alpha$ that the QA can explore), leads to approximately a one-order-of-magnitude reduction in the uncertainty in the energy (while only doubling the number of anneals).
As an example, the results obtained after two steps of this iterative procedure are shown (in panels below the raw results) in Figs.~\ref{fig:Neutrinos_dwave} and~\ref{fig:Neutrinos_dwave2}
and show clear reductions in the uncertainties.

This iterative procedure can be repeated several times until no further improvement is obtained.
For the four-neutrino system, we obtain an ultimate precision of $10^{-16}$ in the energy, and $10^{-8}-10^{-10}$  in the flavor transition probability, single-neutrino entanglement, and logarithmic negativity (as shown in Fig.~\ref{fig:Neutrinos_dwave3} for $t =1.1$, in units of $1/\kappa$).
It appears that the success of this iterative method is due, in part, 
to the ground-state energy of the TEDVP objective function being {\it a priori} known to vanish.

In contrast to the calculations with IBM's superconducting hardware, where scaling to larger problems is limited by qubit and gate fidelity, along with connectivity, for increasing qubit requirements,
the qubit footprint on the annealing devices naively scales exponentially with the number of neutrinos.
This is expected to be mitigated using techniques that have enabled classical computing to provide a series of precision calculations in these model systems.

\section{Conclusions}
\noindent
We have explored the potential of D-Wave's quantum annealers for simulating
some key basic aspects of Standard Model physics.
In particular, the eigenstates and energies of the lowest-lying states of the harmonic oscillator and anharmonic oscillator were 
studied using zooming and a basic coordinate-space multigrid (MG-AQAE).
Deviations in the extracted energies were less than $\sim 10^{-4}$.
These simulations are the basic elements of lattice scalar field theory of importance, for instance, low-energy chiral nuclear physics or high-energy Higgs physics. 
The time evolution of a single plaquette of SU(3) Yang-Mills gauge theory truncated to 
$\{|\mathbf{1}\rangle ,|\mathbf{3}^+\rangle , |\mathbf{6}^+\rangle,|\mathbf{8}\rangle\}$
in the color parity basis,
and of neutrino flavor in
a monochromatic beam with angular dispersion
were also studied through a refinement of the Feynman clock algorithm,  with deviations that can be reduced below $10^{-8}$.
The results of our quantum simulations of the plaquette and neutrino evolution were found to compare favorably with previous quantum simulations performed using IBM's superconducting quantum computers.

Except for the cases in which the Hamiltonian of interest directly maps (or can be efficiently transformed) onto the annealer, which is optimal for transverse-field Ising models (e.g., Ref.~\cite{King:2021}), with the current formulation, the qubit requirements on the device  scale with size of the Hilbert space.
Therefore, the maximum dimensionality of the systems that we have considered in this work 
that can be addressed with {\tt Advantage} remains small. 
For the harmonic and anharmonic oscillators, the maximum number of states in the decimation of the field wavefunction that we could reliably simulate was $n_s=64$.
For time evolution, due to the extra $2 n_T$ factor in the dimensions of the QUBO matrix, this decimation is reduced to 16. 
Further, there is a somewhat unfortunate dependence on the $\eta$ parameter, although this can be partially mitigated by performing a more structured scan, as shown in Ref.~\cite{Teplukhin:2020}.
It seems that paths forward for simulating quantum field theories requires efficiently 
utilizing the innate Ising-Hamiltonian architectures of the annealers, 
including the dynamics of the annealing process.
A  step in this direction has been taken  in Refs.~\cite{Chancellor_2019,Abel:2020ebj},
which utilize the JLP field mapping to qubits but not the quantum Fourier Transform onto conjugate momentum space,
and has the potential to improve the scaling of ground-state preparation.
With the current formulation, the main application can be the preparation of states which are later used in universal quantum computers to perform time evolution, acting as preconditioners to speed up the process of finding ground states using domain decomposition techniques to reduce the size of the problem (for example, by computing the angles to set up the wavefunction for a scalar field theory~\cite{Klco:2019yrb}).
A somewhat different implementation employing Floquet engineering also has promise~\cite{Ciavarella:2022tvc}.

\begin{acknowledgements}
This work was supported in part by the 
U.S.\ Department of Energy,
Office of Science, Office of Nuclear Physics, InQubator for Quantum Simulation (IQuS) under Award Number DOE
(NP) Award DE-SC0020970 (Savage),
and the 
Quantum Science Center (QSC), a National Quantum Information Science Research Center of the U.S.\ Department of Energy (DOE) (Illa).
All calculations performed on D-Wave's QAs were through cloud access~\cite{DwaveLeap}.
We made extensive use of Wolfram {\tt Mathematica}~\cite{Mathematica},
{\tt python}~\cite{python3,Hunter:2007} and {\tt jupyter} notebooks~\cite{PER-GRA:2007} in the {\tt Conda} environment~\cite{anaconda}.

This work was enabled, in part, by
the use of advanced computational, storage, and networking infrastructure provided by the Hyak supercomputer system at the University of Washington.
\end{acknowledgements}

\bibliography{bib}

\onecolumngrid
\clearpage

{
\appendix
\section{The QUBO Matrix}
\label{app:derivation}
\noindent
In this appendix, an outline is presented of the derivation of expressions for the QUBO matrix used in Sec.~\ref{sec:qubo}  to compute eigenstates and energies of a given Hamiltonian,
and in Sec.~\ref{sec:timeevol} to determine the time-evolution of an SU(3) Yang-Mills plaquette and a system of four neutrinos.

Starting from the objective function in Eq.~\eqref{eq:QAE}
for the eigenstates and energies of a given Hamiltonian, 
and using the fixed-point representation for the zoom coefficients in Eq.~\eqref{eq:defa_AQAE_improved_2},
\begin{equation}
    F = \sum_{\alpha\beta} a_{\alpha} a_{\beta} h_{\alpha\beta} = \sum_{\alpha\beta} \left[a^{(z)}_\alpha + \sum_i 2^{i-K-z} (-1)^{\delta_{iK}}q^\alpha_i \right]\left[a^{(z)}_\beta + \sum_j 2^{j-K-z} (-1)^{\delta_{jK}}q^\beta_j \right] h_{\alpha\beta}
    \ \ ,
\end{equation}
which consists of  three different types of terms. 
The first type is the product of $a^{(z)}_\alpha a^{(z)}_\beta$, 
without $q^\alpha_i$ variables.   
It provides a constant off-set to $F$, which 
does not modify the position of the minimum, and thus can be omitted from the QUBO matrix. 
The second type comes from the product of terms with 
$q^\alpha_i$ and $q^\beta_j$, which is of the desired form in  Eq.~\eqref{eq:QUBO_simple}, and
multiplied with a $2^{-2z}$ zooming factor. 
The last type originates from the product of $a^{(z)}_\alpha$ with $q^\beta_j$, which, as it involves a single $q^\beta_j$, 
naively has the potential to be 
problematic for forming a viable QUBO matrix.
The difficulty is averted by using a defining property of binary variables, $q^\beta_j=(q^\beta_j)^2$, 
leading to, for example,
\begin{align}
    \sum_{\alpha\beta} a^{(z)}_\alpha \sum_{j} (-1)^{\delta_{jK}}2^{j-K-z} q^{\beta}_j h_{\alpha\beta} &= \sum_{\beta,j} (-1)^{\delta_{jK}}2^{j-K-z}q^{\beta}_jq^{\beta}_j \underbrace{\left(\sum_{\alpha} a^{(z)}_\alpha h_{\alpha\beta}\right)}_{\alpha\rightarrow \gamma} \nonumber \\
    &= \sum_{\alpha\beta,ij} \delta_{\alpha\beta}\delta_{ij}(-1)^{\delta_{jK}}2^{j-K-z}q^{\alpha}_iq^{\beta}_j \sum_{\gamma} a^{(z)}_\gamma h_{\gamma\beta} 
    \ \ .    
\label{eq:trick}
\end{align}
As $h_{\alpha\beta}$ is symmetric, the two contributions of this form give the same contribution, leading to a QUBO matrix of the form, as given in Eq.~\eqref{eq:QUBOmatrix},
\begin{equation}
    Q_{\alpha,i;\beta,j} = 2^{i+j-2K-2z} (-1)^{\delta_{iK}+\delta_{jK}} h_{\alpha\beta} + 2 \delta_{\alpha\beta} \delta_{ij} 2^{i-K-z} (-1)^{\delta_{iK}} \sum_\gamma a^{(z)}_\gamma h_{\gamma\beta}\ \ .
\end{equation}
In order to reduce this four-index array into a more manageable matrix form, following Ref.~\cite{ARahman:2021ktn}, the indices $\alpha$ and $i$ are combined into $n=K(\alpha-1)+i$.

To construct the QUBO matrix for time evolution using the Feynman clock method, discussed in Sec.~\ref{sec:timeevol}, 
the starting point is to use Eq.~\eqref{eq:QAE_cmplx_2} 
to write the objective function in Eq.~\eqref{eq:QAE_cmplx} 
as a real function, as required for implementation on D-Wave's QA,
\begin{equation}
    F^{(\cbar)} = \sum_{\alpha\beta} \left( a^{\rm Re}_\alpha a^{\rm Re}_\beta \cbar^{\rm Re}_{\alpha\beta} - a^{\rm Re}_\alpha a^{\rm Im}_\beta \cbar^{\rm Im}_{\alpha\beta} + a^{\rm Im}_\alpha a^{\rm Re}_\beta \cbar^{\rm Im}_{\alpha\beta}+ a^{\rm Im}_\alpha a^{\rm Im}_\beta \cbar^{\rm Re}_{\alpha\beta}\right)\ \ .
    \label{eq:objfun_complex}
\end{equation}
The rows and columns of the QUBO matrix are doubled in length to accommodate the real and imaginary parts of each coefficient $a_\alpha = a_\alpha^{\rm Re} + i a_\alpha^{\rm Im}$. 
This is accomplished simply by increasing the range of the $i$ index from $[1,K]$ to $[1,2K]$, 
with $1\leq i \leq K$ used for $a^{\rm Re}_\alpha$ 
and $(K+1)\leq i \leq 2K$ for $a^{\rm Im}_\alpha$. 
Obtaining the expression for the QUBO matrix associated with 
$F^{(\cbar)}$ follows straightforwardly from its derivation given above for the corresponding QUBO matrix for eigenstates and energies.
For example, the first term in Eq.~\eqref{eq:objfun_complex} is analogous to that in Eq.~\eqref{eq:QUBOmatrix},
\begin{equation}
    \sum_{\alpha\beta} a^{\rm Re}_\alpha a^{\rm Re}_\beta \cbar^{\rm Re}_{\alpha\beta} =\sum_{\alpha\beta}\sum_{i,j=1}^K\left[ 2^{i+j-2K-2z} (-1)^{\delta_{iK}+\delta_{jK}} \cbar^{\rm Re}_{\alpha\beta} + 2 \delta_{\alpha\beta} \delta_{ij} 2^{i-K-z} (-1)^{\delta_{iK}} \sum_\gamma a^{{\rm Re},(z)}_\gamma \cbar^{\rm Re}_{\gamma\beta}\right]q^{\alpha}_iq^{\beta}_j\ \ ,
\end{equation}
where the sum over the indices $i,j$ is over the range $[1,K]$.
The last term in Eq.~\eqref{eq:objfun_complex} is also analogous, but with $i,j$ summed over $[K+1,2K]$,
\begin{equation}
    \sum_{\alpha\beta} a^{\rm Im}_\alpha a^{\rm Im}_\beta \cbar^{\rm Re}_{\alpha\beta} =\sum_{\alpha\beta}\sum_{i,j=K+1}^{2K}\left[ 2^{i'+j'-2K-2z} (-1)^{\delta_{i'K}+\delta_{j'K}} \cbar^{\rm Re}_{\alpha\beta} + 2 \delta_{\alpha\beta} \delta_{i'j'} 2^{i'-K-z} (-1)^{\delta_{i'K}} \sum_\gamma a^{{\rm Im},(z)}_\gamma \cbar^{\rm Re}_{\gamma\beta}\right]q^{\alpha}_iq^{\beta}_j\ \ ,
\end{equation}
where $i'$ is defined by $i'\equiv i-K$.

For the second and third terms in Eq.~\eqref{eq:objfun_complex},
we employ the same binary identity, $q^\alpha_i=(q^\alpha_i)^2$, but keep in mind that the indices $i,j$ run over different values. 
For example, for the $a^{\rm Re}_\alpha a^{\rm Im}_\beta$ term,
\begin{align}
    -\sum_{\alpha\beta} a^{\rm Re}_\alpha a^{\rm Im}_\beta \cbar^{\rm Im}_{\alpha\beta} =-\sum_{\alpha\beta}\left[ \sum_{i=1}^{K} \sum_{j=K+1}^{2K} 2^{i+j'-2K-2z} (-1)^{\delta_{iK}+\delta_{j'K}} \cbar^{\rm Im}_{\alpha\beta} \right. & + \sum_{i,j=K+1}^{2K}  \delta_{\alpha\beta} \delta_{i'j'} 2^{j'-K-z} (-1)^{\delta_{j'K}} \sum_\gamma a^{{\rm Re},(z)}_\gamma \cbar^{\rm Im}_{\gamma\beta} \nonumber \\
    &\left. +\sum_{i,j=1}^K \delta_{\alpha\beta} \delta_{ij} 2^{i-K-z} (-1)^{\delta_{iK}} \sum_\gamma a^{{\rm Im},(z)}_\gamma \cbar^{\rm Im}_{\alpha\gamma} \right]q^{\alpha}_iq^{\beta}_j\ \ .   
\end{align}
Collecting together the contributions, 
the following QUBO matrix for the Feynman clock is obtained, 
\begin{equation}
    Q_{\alpha,i;\beta,j} = \begin{cases}
    2^{i+j-2K-2z} (-1)^{\delta_{iK}+\delta_{jK}} \cbar^{\rm Re}_{\alpha\beta} &  \\
    \qquad + \delta_{\alpha\beta} \delta_{ij} 2^{i-K-z} (-1)^{\delta_{iK}} \sum_\gamma \left(2a^{{\rm Re},(z)}_\gamma \cbar^{\rm Re}_{\gamma\beta} - a^{{\rm Im},(z)}_\gamma \cbar^{\rm Im}_{\alpha\gamma} + a^{{\rm Im},(z)}_\gamma \cbar^{\rm Im}_{\gamma\beta}\right) & 1\leq i,j \leq K \, ,\\
    - 2^{i+j'-2K-2z}(-1)^{\delta_{iK} + \delta_{j'K}} \cbar^{\rm Im}_{\alpha\beta} & 1\leq i, j' \leq K \, , \\
    2^{i'+j-2K-2z}(-1)^{\delta_{i'K} + \delta_{jK}} \cbar^{\rm Im}_{\alpha\beta} & 1\leq i',j \leq K \, , \\
    2^{i'+j'-2K-2z}(-1)^{\delta_{i'K} + \delta_{j'K}} \cbar^{\rm Re}_{\alpha\beta} & \\
    \qquad + \delta_{\alpha\beta} \delta_{i'j'} 2^{i'-K-z} (-1)^{\delta_{i'K}} \sum_{\gamma} \left(2a^{{\rm Im},(z)}_\gamma \cbar^{\rm Re}_{\gamma\beta} - a^{{\rm Re},(z)}_\gamma \cbar^{\rm Im}_{\gamma\beta} + a^{{\rm Re},(z)}_\gamma \cbar^{\rm Im}_{\alpha\gamma}\right) & 1\leq i',j' \leq K \, .
\end{cases}
\end{equation}
This can be somewhat simplified because
$\cbar^{\rm Im}_{\alpha\beta}$ is antisymmetric
while 
$\cbar^{\rm Re}_{\alpha\beta}$ is symmetric,
leading to the expressions given in 
Eq.~\eqref{eq:QAE_cmplx_2},
\begin{equation}
    Q_{\alpha,i;\beta,j} = \begin{cases}
    2^{i+j-2K-2z} (-1)^{\delta_{iK}+\delta_{jK}} \cbar^{\rm Re}_{\alpha\beta}+ 2\delta_{\alpha\beta} \delta_{ij} 2^{i-K-z} (-1)^{\delta_{iK}} \sum_\gamma \left(a^{{\rm Re},(z)}_\gamma \cbar^{\rm Re}_{\gamma\beta} + a^{{\rm Im},(z)}_\gamma \cbar^{\rm Im}_{\gamma\beta}\right) & 1\leq i,j \leq K \, , \\
    - 2^{i+j'-2K-2z}(-1)^{\delta_{iK} + \delta_{j'K}} \cbar^{\rm Im}_{\alpha\beta} & 1\leq i, j' \leq K \, , \\
    2^{i'+j-2K-2z}(-1)^{\delta_{i'K} + \delta_{jK}} \cbar^{\rm Im}_{\alpha\beta} & 1\leq i',j \leq K \, , \\
    2^{i'+j'-2K-2z}(-1)^{\delta_{i'K} + \delta_{j'K}} \cbar^{\rm Re}_{\alpha\beta} + 2\delta_{\alpha\beta} \delta_{i'j'} 2^{i'-K-z} (-1)^{\delta_{i'K}} \sum_{\gamma} \left(a^{{\rm Im},(z)}_\gamma \cbar^{\rm Re}_{\gamma\beta}- a^{{\rm Re},(z)}_\gamma \cbar^{\rm Im}_{\gamma\beta}\right) & 1\leq i',j' \leq K \, .
\end{cases}
\end{equation}
%

\section{More on our Results}
\label{app:timings}
\noindent
The (average) time it takes to perform $N_A=10^3$ anneals on both the {\tt neal} simulator as well as on {\tt Advantage} for the HO and AHO Hamiltonian for a single zoom step from Sec.~\ref{sec:scalarfield} is reported in Table~\ref{tab:timings}.
Regarding the time specific to the quantum processor, it takes into account two separate steps.
The first one is the programming time, during which the values of the QUBO matrix are transferred to the quantum processor (same for all problem sizes, $8-15 \; {\rm ms}$).
The next step is the annealing phase, repeated $N_A$ times, which is further decomposed into three steps: the annealing time ($t_A=20\; \mu$s), the readout time (size-dependent, varying between $100-200\; \mu$s), and the delay time (same for all problem sizes, $20.54\; \mu$s), during which the system is allowed to cool and reinitialized.
For {\tt Advantage}, the time spent finding the embedding should also be taken into account (although this part is not computed on the quantum processor, but locally), therefore its contribution is shown in a separate column.
\begin{table}[!b]
\begin{center}
\begin{tabular}{|| c | c | c | c ||} 
 \hline
 Problem size 
 & {\tt neal}
 & {\tt Advantage} (without embedding)
 & {\tt Advantage} (with embedding)\\ [0.5ex] 
 \hline\hline
 $n_s=16$, $K=3$ 
 & $0.7$ s
 & $0.15$ s
 & $10.15$ s \\ 
 $n_s=32$, $K=3$ 
 & $2.0$ s
 & $0.20$ s
 & $100.20$ s \\ 
 $n_s=64$, $K=2$ 
 & $2.6$ s
 & $0.25$ s
 & $200.25$ s \\ 
[0.5ex] 
 \hline
\end{tabular}
\caption{
Comparison of the (average) time (seconds) taken to perform $N_A=10^3$ anneals for a specific problem size (fixed $n_s$ and $K$) and for a single zoom step between the simulator {\tt neal}, run on a 2.40 GHz Intel Core i9-9980HK CPU, and the quantum processor {\tt Advantage} ($t_A=20\; \mu$s), with and without including the time for the embedding.
}
\label{tab:timings}
\end{center}
\end{table}

Although the time required to find the configuration with minimum energy is around an order of magnitude smaller for {\tt Advantage} than {\tt neal} (increasing $t_A$ would reduce this difference), the overhead time spent computing the embedding (not needed for the simulator)  inverts the situation.
As mentioned in Sec.~\ref{subsec:dwave_scalar}, the embedding can be reused for the multiple zoom steps.
This is possible because the required connectivity between the different logic qubits does not change, only the entries of the QUBO matrix change.
More specifically, and looking at Eq.~\eqref{eq:QUBOmatrix}, the additional term obtained from the zooming is added in the diagonal part of the QUBO matrix, where $\alpha=\beta$ and $i=j$ (the same argument can be used for Eq.~\eqref{eq:QAE_cmplx_2}).
Table~\ref{tab:timings} does not take into account the time spent building the QUBO matrix or analyzing the results, since these steps are the same for both cases.

All of the results shown in the main text can be found in HDF5 format~\cite{hdf5} in the file {\tt SSMQA{\_}data.h5}, where each set is labeled by the figure number, with additional metadata to specify the parameters used during its production and the dimension of the array (like the value of $N_{\rm run}$ or the number of zoom levels).
For the results related to Figs.~\ref{fig:SU3_dwave},~\ref{fig:Neutrinos_dwave} and ~\ref{fig:Neutrinos_dwave2}, only the compound states $\lvert \Psi_t\rangle \lvert t \rangle$ are included.

Additionally, in the following tables we provide the values plotted in the figures from the main text.
\begin{table}[!b]
\begin{center}
\begin{tabular}{|| c | c | c | c | c | c | c ||} 
 \hline
 $n$ 
 & $|\delta E^{{\tt Adv.}-16}_n|_{\rm HO}$ 
 & $|\delta E^{{\tt Adv.}-32}_n|_{\rm HO}$
 & $|\delta E^{{\tt Adv.}-64}_n|_{\rm HO}$
 & $|\delta E^{{\tt Adv.}-16}_n|_{\rm AHO}$ 
 & $|\delta E^{{\tt Adv.}-32}_n|_{\rm AHO}$
 & $|\delta E^{{\tt Adv.}-64}_n|_{\rm AHO}$\\ [0.5ex] 
 \hline\hline
 $0$ 
 & $(2.4^{\,+6.1}_{\,-1.5})\times 10^{-6}$ 
 & $(3.9^{\,+7.8}_{\,-2.3})\times 10^{-6}$ 
 & $(1.2^{\,+0.3}_{\,-0.1})\times 10^{-6}$
 & $(3.4^{\,+22}_{\,-3.4})\times 10^{-4}$ 
 & $(1.5^{\,+3.0}_{\,-1.0})\times 10^{-5}$ 
 & $(1.9^{\,+1.1}_{\,-0.2})\times 10^{-6}$ \\ 
 $1$ 
 & $(3.6^{\,+52}_{\,-1.9})\times 10^{-6}$ 
 & $(9.6^{\,+11}_{\,-2.6})\times 10^{-6}$ 
 & $(3.6^{\,+0.2}_{\,-0.2})\times 10^{-6}$
 & $(2.6^{\,+186}_{\,-2.6})\times 10^{-4}$ 
 & $(1.9^{\,+2.4}_{\,-1.1})\times 10^{-5}$ 
 & $(2.7^{\,+0.5}_{\,-0.9})\times 10^{-6}$ \\ 
 $2$ 
 & $(1.9^{\,+5.5}_{\,-1.2})\times 10^{-6}$ 
 & $(9.8^{\,+15}_{\,-3.1})\times 10^{-6}$ 
 & $(2.8^{\,+0.4}_{\,-0.2})\times 10^{-6}$
 & $(0.7^{\,+4487}_{\,-0.7})\times 10^{-5}$ 
 & $(2.2^{\,+2.4}_{\,-1.7})\times 10^{-5}$ 
 & $(8.3^{\,+1.0}_{\,-0.9})\times 10^{-6}$ \\ 
 $3$ 
 & $(0.8^{\,+117}_{\,-0.6})\times 10^{-5}$ 
 & $(9.7^{\,+1.5}_{\,-0.3})\times 10^{-5}$ 
 & $(30^{\,+1.8}_{\,-0.2})\times 10^{-6}$
 & $(1.7^{\,+1.8}_{\,-0.9})\times 10^{-6}$ 
 & $(0.7^{\,+1.8}_{\,-0.8})\times 10^{-5}$ 
 & $(6.0^{\,+1.4}_{\,-2.0})\times 10^{-6}$ \\ 
 $4$ 
 & $(0.5^{\,+744}_{\,-0.5})\times 10^{-5}$ 
 & $(1.6^{\,+1.1}_{\,-0.3})\times 10^{-4}$ 
 & $(8.0^{\,+0.6}_{\,-0.7})\times 10^{-5}$
 & $(0.6^{\,+14}_{\,-0.5})\times 10^{-6}$ 
 & $(6.3^{\,+14}_{\,-4.1})\times 10^{-5}$ 
 & $(5.3^{\,+1.1}_{\,-1.8})\times 10^{-5}$ \\ 
 $5$ 
 & $(2.9^{\,+29}_{\,-1.8})\times 10^{-6}$ 
 & $(4.7^{\,+6.6}_{\,-5.3})\times 10^{-5}$ 
 & $(1.3^{\,+0.5}_{\,-0.5})\times 10^{-4}$
 & $(1.5^{\,+20}_{\,-0.7})\times 10^{-6}$ 
 & $(6.1^{\,+4.2}_{\,-4.9})\times 10^{-4}$ 
 & $(8.1^{\,+1.9}_{\,-0.6})\times 10^{-5}$ \\ 
[0.5ex] 
 \hline
\end{tabular}
\caption{
    Differences between the digitized energies and the corresponding results obtained using D-Wave's {\tt Advantage} for the HO with $m_0=1$, $\phi_{\rm max}=5$, and the AHO with $m_0=1$, $\lambda=32$, $\phi_{\rm max}=2.6$, shown in Figs.~\ref{fig:HO1site_project_dwave} and~\ref{fig:lambphi41site_project_dwave}, with $n_s=\{16,32,64\}$, $K=\{3,3,2\}$, $N_A=10^3$, and the maximum number of zoom steps. The uncertainties correspond to 68\% confidence intervals determined from $N_{\rm run}=20$ independent runs.
}
\label{tab:HO1sitelambphi41site_project_dwave_tab}
\end{center}
\end{table}
The results shown in Figs.~\ref{fig:HO1site_project_dwave} and~\ref{fig:lambphi41site_project_dwave}
are given in Table~\ref{tab:HO1sitelambphi41site_project_dwave_tab}.
\begin{table}[!b]
\begin{center}
\begin{tabular}{|| c | c ||} 
 \hline
 $n$ 
 & $|\delta E^{{\tt Adv.}-32}_n|$\\ [0.5ex] 
 \hline\hline
 $0$ 
 & $(2.5^{\,+5.0}_{\,-2.4})\times 10^{-3}$ \\ 
 $1$ 
 & $(1.2^{\,+16}_{\,-1.1})\times 10^{-3}$ \\ 
[0.5ex] 
 \hline
\end{tabular}
\caption{
    Differences between the digitized energies and the corresponding results obtained using D-Wave's {\tt Advantage} for the AHO with $m^2_0=-4$, $\lambda=1$, $\phi_{\rm max}=9$, shown in Fig.~\ref{fig:muLE0}, with $n_s=32$, $K=3$, $N_A=10^3$, and the maximum number of zoom steps. The uncertainties correspond to 68\% confidence intervals determined from $N_{\rm run}=20$ independent runs.
}
\label{tab:muLE0_tab}
\end{center}
\end{table}
The results shown in Fig.~\ref{fig:muLE0}
are given in Table~\ref{tab:muLE0_tab}.
\begin{table}[!b]
\begin{center}
\begin{tabular}{|| c | c | c ||} 
 \hline
 $n_s$ 
 & ${\tt 2000Q}$
 & ${\tt Advantage}$\\ [0.5ex] 
 \hline\hline
 $4$ 
 & $22^{\,+1}_{\,-0}$
 & $12^{\,+0}_{\,-0}$ \\ 
 $8$ 
 & $83^{\,+1}_{\,-2}$
 & $39^{\,+1}_{\,-1}$ \\
 $16$ 
 & $337^{\,+8}_{\,-9}$
 & $141^{\,+8}_{\,-6}$ \\
 $32$ 
 & $1464^{\,+39}_{\,-109}$
 & $528^{\,+24}_{\,-26}$ \\
 $64$ 
 & $-$
 & $2077^{\,+293}_{\,-254}$ \\
[0.5ex] 
 \hline
\end{tabular}
\caption{
    Number of physical qubits required to map the QUBO matrix for the two available D-Wave's QAs, shown in Fig.~\ref{fig:scaling}, with $K=2$. The uncertainties correspond to 68\% confidence intervals determined from $20$ different embedding with the same problem.
}
\label{tab:scaling_tab}
\end{center}
\end{table}
The results shown in Fig.~\ref{fig:scaling} are given in Table~\ref{tab:scaling_tab}.
\begin{table}[!tbh]
\begin{center}
\begin{tabular}{|| c | c | c | c ||} 
 \hline
 $\delta t $ 
 & $t $ 
 & $|\langle 00 |\hat U_t| 00 \rangle |^2$
 & $\langle \mathcal{\hat H}_E \rangle$ \\ [0.5ex] 
 \hline\hline
 \multirow{3}{*}{$0.2$}
 & $0$ 
 & $0.999988^{\,+10}_{\,-46}$
 & $(0.6^{\,+2.5}_{\,-0.5})\times 10^{-4}$ \\ 
  & $0.2$ 
 & $0.9802^{\,+13}_{\,-3}$
 & $0.0537^{\,+43}_{\,-14}$ \\
  & $0.4$ 
 & $0.9271^{\,+23}_{\,-6}$
 & $0.2018^{\,+97}_{\,-33}$ \\\hline
 \multirow{3}{*}{$0.5$}
 & $0$ 
 & $0.999956^{\,+42}_{\,-64}$
 & $(2.0^{\,+4.2}_{\,-1.8})\times 10^{-4}$ \\ 
  & $0.5$ 
 & $0.8921^{\,+47}_{\,-42}$
 & $0.3030^{\,+95}_{\,-71}$ \\
  & $1.0$ 
 & $0.7139^{\,+66}_{\,-56}$
 & $0.828^{\,+34}_{\,-20}$ \\\hline
 \multirow{3}{*}{$0.7$}
 & $0$ 
 & $0.99995^{\,+3}_{\,-81}$
 & $(0.3^{\,+4.4}_{\,-0.2})\times 10^{-3}$ \\ 
  & $0.7$ 
 & $0.816^{\,+7}_{\,-20}$
 & $0.535^{\,+85}_{\,-14}$ \\
  & $1.4$ 
 & $0.6750^{\,+78}_{\,-51}$
 & $0.903^{\,+45}_{\,-20}$ \\\hline
 \multirow{3}{*}{$0.9$}
 & $0$ 
 & $0.99999^{\,+1}_{\,-13}$
 & $(0.9^{\,+7.9}_{\,-0.8})\times 10^{-4}$ \\ 
  & $0.9$ 
 & $0.7435^{\,+18}_{\,-66}$
 & $0.738^{\,+11}_{\,-8}$ \\
  & $1.8$ 
 & $0.768^{\,+2}_{\,-11}$
 & $0.645^{\,+34}_{\,-9}$ \\\hline
 \multirow{3}{*}{$1.1$}
 & $0$ 
 & $0.999980^{\,+17}_{\,-68}$
 & $(0.9^{\,+4.4}_{\,-0.7})\times 10^{-4}$ \\ 
  & $1.1$ 
 & $0.6949^{\,+91}_{\,-21}$
 & $0.877^{\,+8}_{\,-21}$ \\
  & $2.2$ 
 & $0.9239^{\,+20}_{\,-20}$
 & $0.226^{\,+7}_{\,-10}$ \\\hline
 \multirow{3}{*}{$1.3$}
 & $0$ 
 & $0.99990^{\,+9}_{\,-73}$
 & $(0.4^{\,+4.1}_{\,-0.3})\times 10^{-3}$ \\ 
  & $1.3$ 
 & $0.670^{\,+5}_{\,-26}$
 & $0.926^{\,+93}_{\,-17}$ \\
  & $2.6$ 
 & $0.995^{\,+2}_{\,-13}$
 & $0.025^{\,+76}_{\,-9}$ \\
[0.5ex] 
 \hline
\end{tabular}
\caption{
    The vacuum-to-vacuum probability $|\langle 00 |\hat U_t| 00 \rangle |^2$ and energy in the electric field $\langle \mathcal{\hat H}_E \rangle$ of the one-plaquette system, as shown in Fig.~\ref{fig:SU3_dwave}, with $K=2$ and $N_A=10^3$. The uncertainties correspond to 68\% confidence intervals determined from $N_{\rm run}=20$ independent runs.
}
\label{tab:SU3_dwave_tab}
\end{center}
\end{table}
The results shown in Fig.~\ref{fig:SU3_dwave} are given in Table~\ref{tab:SU3_dwave_tab}.
\begin{table}[!tbh]
\begin{center}
\begin{tabular}{|| c | c | c | c | c | c | c | c | c | c ||} 
 \hline
 $\delta t $ 
 & $t $ 
 & $P_1(t)$
 & $P_2(t)$
 & $P_3(t)$
 & $P_4(t)$
 & $S_1(t)$
 & $S_2(t)$
 & $S_3(t)$
 & $S_4(t)$\\ [0.5ex] 
 \hline\hline
 \multirow{2}{*}{$1.1$}
 & $0$ 
 & $0.0065^{\,+36}_{\,-11}$
 & $0.0073^{\,+48}_{\,-26}$
 & $0.0053^{\,+27}_{\,-34}$
 & $0.0054^{\,+46}_{\,-18}$
 & $0.052^{\,+15}_{\,-8}$
 & $0.054^{\,+40}_{\,-16}$
 & $0.041^{\,+19}_{\,-21}$
 & $0.044^{\,+33}_{\,-10}$\\ 
  & $1.1$ 
 & $0.163^{\,+26}_{\,-15}$
 & $0.121^{\,+31}_{\,-8}$
 & $0.144^{\,+22}_{\,-18}$
 & $0.200^{\,+24}_{\,-19}$
 & $0.420^{\,+54}_{\,-52}$
 & $0.277^{\,+95}_{\,-80}$
 & $0.211^{\,+36}_{\,-86}$
 & $0.408^{\,+38}_{\,-75}$ \\\hline
 \multirow{2}{*}{$2.2$}
 & $0$ 
 & $0.0092^{\,+50}_{\,-31}$
 & $0.0077^{\,+27}_{\,-21}$
 & $0.0062^{\,+28}_{\,-33}$
 & $0.0070^{\,+15}_{\,-27}$
 & $0.071^{\,+25}_{\,-23}$
 & $0.057^{\,+23}_{\,-17}$
 & $0.051^{\,+23}_{\,-25}$
 & $0.056^{\,+6}_{\,-19}$\\
  & $2.2$ 
 & $0.276^{\,+30}_{\,-61}$
 & $0.143^{\,+22}_{\,-23}$
 & $0.159^{\,+23}_{\,-23}$
 & $0.285^{\,+53}_{\,-32}$
 & $0.785^{\,+68}_{\,-74}$
 & $0.413^{\,+72}_{\,-90}$
 & $0.334^{\,+143}_{\,-64}$
 & $0.746^{\,+67}_{\,-59}$ \\\hline
 \multirow{2}{*}{$3.3$}
 & $0$ 
 & $0.0054^{\,+31}_{\,-25}$
 & $0.0057^{\,+37}_{\,-21}$
 & $0.0057^{\,+36}_{\,-24}$
 & $0.0067^{\,+29}_{\,-36}$
 & $0.041^{\,+7}_{\,-18}$
 & $0.042^{\,+33}_{\,-10}$
 & $0.049^{\,+19}_{\,-21}$
 & $0.051^{\,+25}_{\,-20}$\\
  & $3.3$ 
 & $0.407^{\,+21}_{\,-21}$
 & $0.095^{\,+25}_{\,-13}$
 & $0.101^{\,+21}_{\,-12}$
 & $0.411^{\,+31}_{\,-20}$
 & $0.953^{\,+18}_{\,-15}$
 & $0.426^{\,+100}_{\,-37}$
 & $0.431^{\,+68}_{\,-50}$
 & $0.958^{\,+14}_{\,-10}$ \\\hline
 \multirow{2}{*}{$4.4$}
 & $0$ 
 & $0.0078^{\,+32}_{\,-29}$
 & $0.0075^{\,+15}_{\,-39}$
 & $0.0056^{\,+14}_{\,-18}$
 & $0.0058^{\,+8}_{\,-22}$
 & $0.058^{\,+19}_{\,-16}$
 & $0.059^{\,+8}_{\,-27}$
 & $0.043^{\,+10}_{\,-14}$
 & $0.048^{\,+4}_{\,-14}$\\
  & $4.4$ 
 & $0.542^{\,+47}_{\,-26}$
 & $0.212^{\,+24}_{\,-27}$
 & $0.240^{\,+26}_{\,-32}$
 & $0.559^{\,+45}_{\,-44}$
 & $0.961^{\,+17}_{\,-36}$
 & $0.569^{\,+100}_{\,-42}$
 & $0.575^{\,+75}_{\,-65}$
 & $0.968^{\,+13}_{\,-46}$ \\\hline
 \multirow{2}{*}{$5.5$}
 & $0$ 
 & $0.0088^{\,+71}_{\,-41}$
 & $0.0092^{\,+37}_{\,-44}$
 & $0.0060^{\,+52}_{\,-24}$
 & $0.0066^{\,+28}_{\,-30}$
 & $0.051^{\,+55}_{\,-14}$
 & $0.058^{\,+26}_{\,-26}$
 & $0.051^{\,+32}_{\,-24}$
 & $0.043^{\,+25}_{\,-12}$\\
  & $5.5$ 
 & $0.727^{\,+36}_{\,-29}$
 & $0.229^{\,+23}_{\,-35}$
 & $0.234^{\,+76}_{\,-29}$
 & $0.737^{\,+32}_{\,-40}$
 & $0.760^{\,+41}_{\,-51}$
 & $0.715^{\,+59}_{\,-73}$
 & $0.670^{\,+110}_{\,-83}$
 & $0.752^{\,+99}_{\,-18}$ \\\hline
 \multirow{2}{*}{$6.6$}
 & $0$ 
 & $0.0085^{\,+46}_{\,-35}$
 & $0.0086^{\,+55}_{\,-41}$
 & $0.0092^{\,+37}_{\,-26}$
 & $0.0092^{\,+38}_{\,-33}$
 & $0.064^{\,+19}_{\,-23}$
 & $0.066^{\,+26}_{\,-28}$
 & $0.059^{\,+20}_{\,-15}$
 & $0.067^{\,+27}_{\,-20}$\\ 
  & $6.6$ 
 & $0.782^{\,+26}_{\,-29}$
 & $0.201^{\,+56}_{\,-42}$
 & $0.220^{\,+41}_{\,-26}$
 & $0.782^{\,+41}_{\,-44}$
 & $0.727^{\,+50}_{\,-77}$
 & $0.695^{\,+61}_{\,-101}$
 & $0.703^{\,+69}_{\,-93}$
 & $0.741^{\,+62}_{\,-85}$ \\\hline
 \multirow{2}{*}{$7.7$}
 & $0$ 
 & $0.0123^{\,+73}_{\,-60}$
 & $0.0133^{\,+69}_{\,-56}$
 & $0.0066^{\,+53}_{\,-35}$
 & $0.0083^{\,+22}_{\,-43}$
 & $0.087^{\,+41}_{\,-33}$
 & $0.082^{\,+46}_{\,-27}$
 & $0.054^{\,+39}_{\,-26}$
 & $0.051^{\,+23}_{\,-14}$\\
  & $7.7$ 
 & $0.665^{\,+29}_{\,-46}$
 & $0.326^{\,+42}_{\,-53}$
 & $0.286^{\,+57}_{\,-28}$
 & $0.632^{\,+45}_{\,-60}$
 & $0.873^{\,+30}_{\,-33}$
 & $0.803^{\,+87}_{\,-103}$
 & $0.750^{\,+41}_{\,-63}$
 & $0.895^{\,+51}_{\,-34}$ \\\hline
 \multirow{2}{*}{$8.8$}
 & $0$ 
 & $0.009^{\,+13}_{\,-4}$
 & $0.0083^{\,+67}_{\,-34}$
 & $0.0081^{\,+69}_{\,-49}$
 & $0.0065^{\,+68}_{\,-31}$
 & $0.065^{\,+45}_{\,-25}$
 & $0.059^{\,+24}_{\,-16}$
 & $0.055^{\,+49}_{\,-27}$
 & $0.053^{\,+39}_{\,-24}$\\
  & $8.8$ 
 & $0.606^{\,+41}_{\,-42}$
 & $0.280^{\,+25}_{\,-24}$
 & $0.273^{\,+33}_{\,-22}$
 & $0.603^{\,+61}_{\,-57}$
 & $0.945^{\,+32}_{\,-47}$
 & $0.813^{\,+37}_{\,-32}$
 & $0.793^{\,+44}_{\,-41}$
 & $0.939^{\,+27}_{\,-42}$ \\\hline
 \multirow{2}{*}{$9.9$}
 & $0$ 
 & $0.0104^{\,+49}_{\,-28}$
 & $0.0106^{\,+66}_{\,-29}$
 & $0.0106^{\,+24}_{\,-43}$
 & $0.0106^{\,+45}_{\,-40}$
 & $0.078^{\,+23}_{\,-20}$
 & $0.079^{\,+34}_{\,-25}$
 & $0.075^{\,+16}_{\,-24}$
 & $0.077^{\,+28}_{\,-27}$\\
  & $9.9$ 
 & $0.455^{\,+39}_{\,-72}$
 & $0.298^{\,+51}_{\,-73}$
 & $0.246^{\,+99}_{\,-36}$
 & $0.422^{\,+76}_{\,-45}$
 & $0.954^{\,+16}_{\,-38}$
 & $0.834^{\,+57}_{\,-107}$
 & $0.787^{\,+106}_{\,-85}$
 & $0.949^{\,+30}_{\,-24}$ \\
[0.5ex] 
 \hline
\end{tabular}
\caption{
    The (raw) probability of flavor transitions $P_i(t)$ and single-neutrino entanglement entropy $S_i(t)$, as shown in Fig.~\ref{fig:Neutrinos_dwave}, with $K=2$ and $N_A=10^3$. The uncertainties correspond to 68\% confidence intervals determined from $N_{\rm run}=20$ independent runs.
}
\label{tab:Neutrinos_dwave_tab}
\end{center}
\end{table}
\begin{table}[!tbh]
\begin{center}
\begin{tabular}{|| c | c | c | c | c | c | c | c | c | c ||} 
 \hline
 $\delta t $ 
 & $t $ 
 & $P_1(t)$
 & $P_2(t)$
 & $P_3(t)$
 & $P_4(t)$
 & $S_1(t)$
 & $S_2(t)$
 & $S_3(t)$
 & $S_4(t)$\\ [0.5ex] 
 \hline\hline
 \multirow{2}{*}{$1.1$}
 & $0$ 
 & $0.00013 ^{\,+11}_{\,-7}$
 & $0.00015 ^{\,+ 6}_{\,- 7}$
 & $0.00010 ^{\,+ 9}_{\,- 3}$
 & $0.00011 ^{\,+ 8}_{\,- 4}$
 & $0.0018 ^{\,+ 12}_{\,- 9}$
 & $0.0020 ^{\,+ 8}_{\,- 8}$
 & $0.0013 ^{\,+ 11}_{\,- 4}$
 & $0.0013 ^{\,+ 10}_{\,- 4}$\\ 
  & $1.1$ 
 & $0.1553 ^{\,+ 35}_{\,- 27}$
 & $0.1205 ^{\,+ 25}_{\,- 29}$
 & $0.1242 ^{\,+ 44}_{\,- 24}$
 & $0.1586 ^{\,+ 39}_{\,- 36}$
 & $0.3154 ^{\,+ 92}_{\,- 146}$
 & $0.0843 ^{\,+ 102}_{\,- 94}$
 & $0.0832 ^{\,+ 100}_{\,- 75}$
 & $0.3143 ^{\,+ 69}_{\,- 82}$ \\\hline
 \multirow{2}{*}{$2.2$}
 & $0$ 
 & $0.00013 ^{\,+ 4}_{\,- 5}$
 & $0.00012 ^{\,+ 13}_{\,- 6}$
 & $0.00008 ^{\,+ 4}_{\,- 5}$
 & $0.00011 ^{\,+ 8}_{\,- 4}$
 & $0.0016 ^{\,+ 5}_{\,- 7}$
 & $0.0015 ^{\,+ 11}_{\,- 7}$
 & $0.0011 ^{\,+ 5}_{\,- 6}$
 & $0.0014 ^{\,+ 7}_{\,- 5}$\\
  & $2.2$ 
 & $0.2610 ^{\,+ 47}_{\,- 42}$
 & $0.1240 ^{\,+ 23}_{\,- 53}$
 & $0.1256 ^{\,+ 29}_{\,- 54}$
 & $0.2639 ^{\,+ 32}_{\,- 60}$
 & $0.7437 ^{\,+ 132}_{\,- 88}$
 & $0.2377 ^{\,+ 32}_{\,- 96}$
 & $0.2397 ^{\,+ 128}_{\,- 84}$
 & $0.7408 ^{\,+ 76}_{\,- 67}$ \\\hline
 \multirow{2}{*}{$3.3$}
 & $0$ 
 & $0.00007 ^{\,+7}_{\,-4}$
 & $0.00007 ^{\,+ 9}_{\,- 4}$
 & $0.00009 ^{\,+ 5}_{\,- 4}$
 & $0.00008 ^{\,+ 3}_{\,- 3}$
 & $0.0009 ^{\,+ 8}_{\,- 4}$
 & $0.0010 ^{\,+ 9}_{\,- 5}$
 & $0.0012 ^{\,+ 8}_{\,- 4}$
 & $0.0010 ^{\,+ 4}_{\,- 3}$\\
  & $3.3$ 
 & $0.4159 ^{\,+ 31}_{\,- 14}$
 & $0.0811 ^{\,+ 13}_{\,- 34}$
 & $0.0810 ^{\,+ 25}_{\,- 24}$
 & $0.4164 ^{\,+ 22}_{\,- 22}$
 & $0.9785 ^{\,+ 16}_{\,- 11}$
 & $0.3958 ^{\,+ 70}_{\,- 119}$
 & $0.3933 ^{\,+ 102}_{\,- 85}$
 & $0.9788 ^{\,+ 16}_{\,- 14}$ \\\hline
 \multirow{2}{*}{$4.4$}
 & $0$ 
 & $0.00013 ^{\,+ 5}_{\,- 3}$
 & $0.00012 ^{\,+ 10}_{\,- 4}$
 & $0.00010 ^{\,+ 5}_{\,- 3}$
 & $0.00010 ^{\,+ 9}_{\,- 4}$
 & $0.0018 ^{\,+ 5}_{\,- 8}$
 & $0.0016 ^{\,+ 11}_{\,- 6}$
 & $0.0012 ^{\,+ 6}_{\,- 3}$
 & $0.0014 ^{\,+ 9}_{\,- 5}$\\
  & $4.4$ 
 & $0.5871 ^{\,+ 72}_{\,- 51}$
 & $0.2185 ^{\,+ 38}_{\,- 47}$
 & $0.2185 ^{\,+ 45}_{\,- 58}$
 & $0.5859 ^{\,+ 48}_{\,- 67}$
 & $0.9593 ^{\,+ 31}_{\,- 28}$
 & $0.5328 ^{\,+ 100}_{\,- 83}$
 & $0.5325 ^{\,+ 159}_{\,- 73}$
 & $0.9603 ^{\,+ 43}_{\,- 46}$ \\\hline
 \multirow{2}{*}{$5.5$}
 & $0$ 
 & $0.00010 ^{\,+ 6}_{\,- 4}$
 & $0.00011 ^{\,+ 6}_{\,- 3}$
 & $0.00009 ^{\,+ 6}_{\,- 3}$
 & $0.00014 ^{\,+ 7}_{\,- 4}$
 & $0.0013 ^{\,+ 5}_{\,- 4}$
 & $0.0014 ^{\,+ 3}_{\,- 4}$
 & $0.0013 ^{\,+ 8}_{\,- 4}$
 & $0.0017 ^{\,+ 4}_{\,- 8}$\\
  & $5.5$ 
 & $0.7174 ^{\,+ 35}_{\,- 61}$
 & $0.2088 ^{\,+ 45}_{\,- 29}$
 & $0.2101 ^{\,+ 17}_{\,- 48}$
 & $0.7202 ^{\,+ 50}_{\,- 33}$
 & $0.7996 ^{\,+ 53}_{\,- 34}$
 & $0.6434 ^{\,+ 80}_{\,- 93}$
 & $0.6396 ^{\,+ 57}_{\,- 72}$
 & $0.7967 ^{\,+ 72}_{\,- 69}$ \\\hline
 \multirow{2}{*}{$6.6$}
 & $0$ 
 & $0.00011 ^{\,+ 9}_{\,- 6}$
 & $0.00013 ^{\,+ 7}_{\,- 6}$
 & $0.00011 ^{\,+ 9}_{\,- 3}$
 & $0.00013 ^{\,+ 7}_{\,- 7}$
 & $0.0016 ^{\,+ 7}_{\,- 8}$
 & $0.0016 ^{\,+ 7}_{\,- 7}$
 & $0.0013 ^{\,+ 7}_{\,- 5}$
 & $0.0017 ^{\,+ 10}_{\,- 8}$\\ 
  & $6.6$ 
 & $0.7846 ^{\,+ 36}_{\,- 28}$
 & $0.1996 ^{\,+ 17}_{\,- 48}$
 & $0.2012 ^{\,+ 57}_{\,- 41}$
 & $0.7843 ^{\,+ 49}_{\,- 49}$
 & $0.7339 ^{\,+ 78}_{\,- 61}$
 & $0.7041 ^{\,+ 87}_{\,- 94}$
 & $0.7057 ^{\,+ 117}_{\,- 80}$
 & $0.7356 ^{\,+ 105}_{\,- 65}$ \\\hline
 \multirow{2}{*}{$7.7$}
 & $0$ 
 & $0.00014 ^{\,+ 4}_{\,- 6}$
 & $0.00014 ^{\,+ 4}_{\,- 7}$
 & $0.00006 ^{\,+ 7}_{\,- 2}$
 & $0.00010 ^{\,+ 3}_{\,- 6}$
 & $0.0017 ^{\,+ 6}_{\,- 6}$
 & $0.0018 ^{\,+ 4}_{\,- 8}$
 & $0.0009 ^{\,+ 7}_{\,- 3}$
 & $0.0013 ^{\,+ 4}_{\,- 7}$\\
  & $7.7$ 
 & $0.6550 ^{\,+ 42}_{\,- 48}$
 & $0.2978 ^{\,+ 37}_{\,- 44}$
 & $0.2911 ^{\,+ 54}_{\,- 48}$
 & $0.6510 ^{\,+ 44}_{\,- 70}$
 & $0.8612 ^{\,+ 60}_{\,- 52}$
 & $0.7497 ^{\,+ 88}_{\,- 89}$
 & $0.7477 ^{\,+ 56}_{\,- 93}$
 & $0.8592 ^{\,+ 62}_{\,- 66}$ \\\hline
 \multirow{2}{*}{$8.8$}
 & $0$ 
 & $0.00009 ^{\,+ 10}_{\,- 2}$
 & $0.00012 ^{\,+ 5}_{\,- 4}$
 & $0.00011 ^{\,+ 7}_{\,- 5}$
 & $0.00010 ^{\,+ 8}_{\,- 5}$
 & $0.0012 ^{\,+ 9}_{\,- 4}$
 & $0.0015 ^{\,+ 7}_{\,- 5}$
 & $0.0013 ^{\,+ 9}_{\,- 6}$
 & $0.0012 ^{\,+ 11}_{\,- 5}$\\
  & $8.8$ 
 & $0.5607 ^{\,+ 58}_{\,- 53}$
 & $0.2562 ^{\,+ 40}_{\,- 52}$
 & $0.2540 ^{\,+ 48}_{\,- 29}$
 & $0.5614 ^{\,+ 68}_{\,- 46}$
 & $0.9881 ^{\,+ 15}_{\,- 36}$
 & $0.7755 ^{\,+ 63}_{\,- 90}$
 & $0.7732 ^{\,+ 75}_{\,- 82}$
 & $0.9874 ^{\,+ 18}_{\,- 30}$ \\\hline
 \multirow{2}{*}{$9.9$}
 & $0$ 
 & $0.00011 ^{\,+ 6}_{\,- 7}$
 & $0.00011 ^{\,+ 8}_{\,- 6}$
 & $0.00009 ^{\,+ 6}_{\,- 4}$
 & $0.00008 ^{\,+ 11}_{\,- 4}$
 & $0.0015 ^{\,+ 8}_{\,- 8}$
 & $0.0015 ^{\,+ 8}_{\,- 8}$
 & $0.0012 ^{\,+ 5}_{\,- 5}$
 & $0.0011 ^{\,+ 16}_{\,- 4}$\\
  & $9.9$ 
 & $0.3740 ^{\,+ 36}_{\,- 34}$
 & $0.2593 ^{\,+ 40}_{\,- 41}$
 & $0.2576 ^{\,+ 70}_{\,- 45}$
 & $0.3740 ^{\,+ 29}_{\,- 36}$
 & $0.9464 ^{\,+ 37}_{\,- 23}$
 & $0.7991 ^{\,+ 143}_{\,- 47}$
 & $0.7980 ^{\,+ 136}_{\,- 38}$
 & $0.9466 ^{\,+ 27}_{\,- 22}$ \\
[0.5ex] 
 \hline
\end{tabular}
\caption{
    The probability of flavor transitions $P_i(t)$ and single-neutrino entanglement entropy $S_i(t)$, as shown in Fig.~\ref{fig:Neutrinos_dwave}, with $K=2$ and $N_A=10^3$, after two steps of the iterative procedure. The uncertainties correspond to 68\% confidence intervals determined from $N_{\rm run}=20$ independent runs.
}
\label{tab:Neutrinos_dwave_tab_2}
\end{center}
\end{table}
The  results shown in Fig.~\ref{fig:Neutrinos_dwave} are given in Table~\ref{tab:Neutrinos_dwave_tab} (raw results) and~\ref{tab:Neutrinos_dwave_tab_2} (after two steps of the iterative procedure), and the 
results shown in Fig.~\ref{fig:Neutrinos_dwave2} are given in Table~\ref{tab:Neutrinos_dwave_tab2} (raw results) and~\ref{tab:Neutrinos_dwave_tab2_2} (after two steps of the iterative procedure).
\begin{table}[!tbh]
\begin{center}
\begin{tabular}{|| c | c | c | c | c | c | c | c ||} 
 \hline
 $\delta t $ 
 & $t $ 
 & $\mathcal{N}_{12}(t)$
 & $\mathcal{N}_{13}(t)$
 & $\mathcal{N}_{14}(t)$
 & $\mathcal{N}_{23}(t)$
 & $\mathcal{N}_{24}(t)$
 & $\mathcal{N}_{34}(t)$\\ [0.5ex] 
 \hline\hline
 \multirow{2}{*}{$1.1$}
 & $0$ 
 & $0.107^{\,+32}_{\,-29}$
 & $0.063^{\,+44}_{\,-42}$
 & $0.053^{\,+52}_{\,-33}$
 & $0.048^{\,+44}_{\,-32}$
 & $0.078^{\,+62}_{\,-47}$
 & $0.054^{\,+13}_{\,-20}$\\ 
  & $1.1$ 
 & $0.212^{\,+64}_{\,-134}$
 & $0.128^{\,+61}_{\,-83}$
 & $0.417^{\,+98}_{\,-95}$
 & $0.073^{\,+70}_{\,-56}$
 & $0.186^{\,+99}_{\,-116}$
 & $0.045^{\,+70}_{\,-27}$\\\hline
 \multirow{2}{*}{$2.2$}
 & $0$ 
 & $0.103^{\,+67}_{\,-23}$
 & $0.068^{\,+97}_{\,-37}$
 & $0.088^{\,+39}_{\,-49}$
 & $0.048^{\,+49}_{\,-30}$
 & $0.046^{\,+46}_{\,-37}$
 & $0.028^{\,+46}_{\,-17}$\\
  & $2.2$ 
 & $0.078^{\,+137}_{\,-69}$
 & $0.238^{\,+130}_{\,-110}$
 & $0.611^{\,+103}_{\,-51}$
 & $0.071^{\,+69}_{\,-70}$
 & $0.215^{\,+56}_{\,-47}$
 & $0.026^{\,+10}_{\,-22}$\\\hline
 \multirow{2}{*}{$3.3$}
 & $0$ 
 & $0.056^{\,+34}_{\,-26}$
 & $0.033^{\,+45}_{\,-16}$
 & $0.038^{\,+48}_{\,-27}$
 & $0.049^{\,+44}_{\,-35}$
 & $0.064^{\,+78}_{\,-43}$
 & $0.085^{\,+44}_{\,-36}$\\
  & $3.3$ 
 & $0.063^{\,+87}_{\,-12}$
 & $0.194^{\,+43}_{\,-47}$
 & $0.735^{\,+24}_{\,-31}$
 & $0.068^{\,+76}_{\,-59}$
 & $0.197^{\,+43}_{\,-33}$
 & $0.071^{\,+42}_{\,-32}$\\\hline
 \multirow{2}{*}{$4.4$}
 & $0$ 
 & $0.120^{\,+31}_{\,-38}$
 & $0.057^{\,+36}_{\,-21}$
 & $0.045^{\,+22}_{\,-12}$
 & $0.066^{\,+9}_{\,-28}$
 & $0.036^{\,+25}_{\,-23}$
 & $0.052^{\,+56}_{\,-25}$\\
  & $4.4$ 
 & $0.143^{\,+19}_{\,-88}$
 & $0.188^{\,+23}_{\,-68}$
 & $0.637^{\,+64}_{\,-52}$
 & $0.133^{\,+127}_{\,-77}$
 & $0.142^{\,+42}_{\,-46}$
 & $0.126^{\,+124}_{\,-74}$\\\hline
 \multirow{2}{*}{$5.5$}
 & $0$ 
 & $0.089^{\,+57}_{\,-33}$
 & $0.062^{\,+56}_{\,-47}$
 & $0.059^{\,+62}_{\,-38}$
 & $0.051^{\,+44}_{\,-39}$
 & $0.064^{\,+54}_{\,-43}$
 & $0.042^{\,+30}_{\,-29}$\\
  & $5.5$ 
 & $0.365^{\,+60}_{\,-106}$
 & $0.058^{\,+29}_{\,-36}$
 & $0.361^{\,+38}_{\,-196}$
 & $0.220^{\,+57}_{\,-93}$
 & $0.053^{\,+19}_{\,-53}$
 & $0.349^{\,+55}_{\,-109}$\\\hline
 \multirow{2}{*}{$6.6$}
 & $0$ 
 & $0.056^{\,+67}_{\,-32}$
 & $0.062^{\,+29}_{\,-18}$
 & $0.074^{\,+102}_{\,-53}$
 & $0.078^{\,+39}_{\,-47}$
 & $0.081^{\,+42}_{\,-41}$
 & $0.075^{\,+49}_{\,-53}$\\ 
  & $6.6$ 
 & $0.321^{\,+66}_{\,-157}$
 & $0.000^{\,+27}_{\,-0}$
 & $0.363^{\,+66}_{\,-200}$
 & $0.285^{\,+92}_{\,-102}$
 & $0.000^{\,+28}_{\,-0}$
 & $0.303^{\,+75}_{\,-65}$\\\hline
 \multirow{2}{*}{$7.7$}
 & $0$ 
 & $0.146^{\,+96}_{\,-43}$
 & $0.047^{\,+29}_{\,-34}$
 & $0.061^{\,+33}_{\,-44}$
 & $0.042^{\,+ 58}_{\,-19}$
 & $0.055^{\,+40}_{\,-23}$
 & $0.040^{\,+36}_{\,-29}$\\
  & $7.7$ 
 & $0.006^{\,+68}_{\,-6}$
 & $0$
 & $0.631^{\,+57}_{\,-58}$
 & $0.513^{\,+84}_{\,-128}$
 & $0$
 & $0.000^{\,+67}_{\,-0}$\\\hline
 \multirow{2}{*}{$8.8$}
 & $0$ 
 & $0.076^{\,+49}_{\,-33}$
 & $0.056^{\,+48}_{\,-36}$
 & $0.072^{\,+33}_{\,-35}$
 & $0.083^{\,+60}_{\,-59}$
 & $0.065^{\,+46}_{\,-32}$
 & $0.031^{\,+43}_{\,-18}$\\
  & $8.8$ 
 & $0.000^{\,+86}_{\,-0}$
 & $0$
 & $0.766^{\,+56}_{\,-52}$
 & $0.663^{\,+50}_{\,-46}$
 & $0$
 & $0$\\\hline
 \multirow{2}{*}{$9.9$}
 & $0$ 
 & $0.088^{\,+45}_{\,-43}$
 & $0.069^{\,+45}_{\,-43}$
 & $0.067^{\,+90}_{\,-27}$
 & $0.076^{\,+20}_{\,-49}$
 & $0.071^{\,+43}_{\,-28}$
 & $0.068^{\,+61}_{\,-50}$\\
  & $9.9$ 
 & $0$
 & $0$
 & $0.837^{\,+47}_{\,-76}$
 & $0.718^{\,+100}_{\,-87}$
 & $0.000^{\,+28}_{\,-0}$
 & $0$\\
[0.5ex] 
 \hline
\end{tabular}
\caption{
    The (raw) logarithmic negativity for the different neutrino pairs $\mathcal{N}_{ij}(t)$, as shown in Fig.~\ref{fig:Neutrinos_dwave2}, with $K=2$ and $N_A=10^3$. The uncertainties correspond to 68\% confidence intervals determined from $N_{\rm run}=20$ independent runs.
}
\label{tab:Neutrinos_dwave_tab2}
\end{center}
\end{table}
\begin{table}[!tbh]
\begin{center}
\begin{tabular}{|| c | c | c | c | c | c | c | c ||} 
 \hline
 $\delta t $ 
 & $t $ 
 & $\mathcal{N}_{12}(t)$
 & $\mathcal{N}_{13}(t)$
 & $\mathcal{N}_{14}(t)$
 & $\mathcal{N}_{23}(t)$
 & $\mathcal{N}_{24}(t)$
 & $\mathcal{N}_{34}(t)$\\ [0.5ex] 
 \hline\hline
 \multirow{2}{*}{$1.1$}
 & $0$ 
 & $0.0164 ^{\,+ 66}_{\,- 70}$
 & $0.0085 ^{\,+ 76}_{\,- 50}$
 & $0.0068 ^{\,+ 40}_{\,- 41}$
 & $0.0073 ^{\,+ 49}_{\,- 25}$
 & $0.0069 ^{\,+ 83}_{\,- 53}$
 & $0.0063 ^{\,+ 39}_{\,- 33}$\\ 
  & $1.1$ 
 & $0.0015 ^{\,+ 48}_{\,- 13}$
 & $0.1948 ^{\,+ 133}_{\,- 114}$
 & $0.4851 ^{\,+ 66}_{\,- 109}$
 & $0.0420 ^{\,+ 181}_{\,- 124}$
 & $0.1945 ^{\,+ 150}_{\,-134}$
 & $0.0007 ^{\,+ 6}_{\,- 7}$\\\hline
 \multirow{2}{*}{$2.2$}
 & $0$ 
 & $0.0089 ^{\,+ 99}_{\,- 30}$
 & $0.0053 ^{\,+ 51}_{\,- 24}$
 & $0.0076 ^{\,+ 53}_{\,- 35}$
 & $0.0053 ^{\,+ 62}_{\,- 17}$
 & $0.0077 ^{\,+ 97}_{\,- 29}$
 & $0.0070 ^{\,+ 73}_{\,- 17}$\\
  & $2.2$ 
 & $0.0173 ^{\,+ 36}_{\,- 17}$
 & $0.2683 ^{\,+ 143}_{\,- 75}$
 & $0.7316 ^{\,+ 41}_{\,- 61}$
 & $0.0649 ^{\,+ 136}_{\,- 166}$
 & $0.2650 ^{\,+ 61}_{\,- 81}$
 & $0.0190 ^{\,+ 17}_{\,- 34}$\\\hline
 \multirow{2}{*}{$3.3$}
 & $0$ 
 & $0.0051 ^{\,+ 46}_{\,- 23}$
 & $0.0058 ^{\,+ 44}_{\,- 22}$
 & $0.0027 ^{\,+ 59}_{\,- 10}$
 & $0.0064 ^{\,+ 49}_{\,- 38}$
 & $0.0036 ^{\,+ 47}_{\,- 19}$
 & $0.0089 ^{\,+ 74}_{\,- 43}$\\
  & $3.3$ 
 & $0.0847 ^{\,+ 53}_{\,- 40}$
 & $0.2495 ^{\,+ 110}_{\,- 48}$
 & $0.7827 ^{\,+ 86}_{\,- 56}$
 & $0.0816 ^{\,+ 65}_{\,- 169}$
 & $0.2515 ^{\,+ 68}_{\,- 53}$
 & $0.0851 ^{\,+ 37}_{\,- 33}$\\\hline
 \multirow{2}{*}{$4.4$}
 & $0$ 
 & $0.0154 ^{\,+ 72}_{\,- 23}$
 & $0.0076 ^{\,+ 68}_{\,- 37}$
 & $0.0076 ^{\,+ 43}_{\,- 42}$
 & $0.0071 ^{\,+ 36}_{\,- 25}$
 & $0.0075 ^{\,+ 39}_{\,- 38}$
 & $0.0065 ^{\,+ 73}_{\,- 41}$\\
  & $4.4$ 
 & $0.2013 ^{\,+ 127}_{\,- 91}$
 & $0.1805 ^{\,+ 93}_{\,- 56}$
 & $0.6847 ^{\,+ 68}_{\,- 37}$
 & $0.1563 ^{\,+ 163}_{\,- 113}$
 & $0.1816 ^{\,+ 66}_{\,- 66}$
 & $0.2038 ^{\,+ 164}_{\,- 103}$\\\hline
 \multirow{2}{*}{$5.5$}
 & $0$ 
 & $0.0066 ^{\,+ 25}_{\,- 20}$
 & $0.0088 ^{\,+ 26}_{\,- 34}$
 & $0.0101 ^{\,+ 38}_{\,- 57}$
 & $0.0054 ^{\,+ 62}_{\,- 29}$
 & $0.0105 ^{\,+ 91}_{\,- 50}$
 & $0.0074 ^{\,+ 76}_{\,- 35}$\\
  & $5.5$ 
 & $0.3155 ^{\,+ 137}_{\,- 97}$
 & $0.0869 ^{\,+ 25}_{\,- 38}$
 & $0.4790 ^{\,+ 114}_{\,- 81}$
 & $0.2623 ^{\,+ 54}_{\,- 83}$
 & $0.0888 ^{\,+ 18}_{\,- 54}$
 & $0.3076 ^{\,+ 147}_{\,- 101}$\\\hline
 \multirow{2}{*}{$6.6$}
 & $0$ 
 & $0.0082 ^{\,+ 56}_{\,- 52}$
 & $0.0081 ^{\,+ 64}_{\,- 44}$
 & $0.0107 ^{\,+ 95}_{\,- 60}$
 & $0.0105 ^{\,+ 25}_{\,- 43}$
 & $0.0101 ^{\,+ 44}_{\,- 61}$
 & $0.0070 ^{\,+ 79}_{\,- 31}$\\ 
  & $6.6$ 
 & $0.2342 ^{\,+ 173}_{\,- 125}$
 & $0$
 & $0.4603 ^{\,+ 115}_{\,- 182}$
 & $0.4220 ^{\,+ 59}_{\,- 81}$
 & $0$
 & $0.2423 ^{\,+ 142}_{\,- 126}$\\\hline
 \multirow{2}{*}{$7.7$}
 & $0$ 
 & $0.0174 ^{\,+ 39}_{\,- 55}$
 & $0.0066 ^{\,+ 54}_{\,- 37}$
 & $0.0077 ^{\,+ 40}_{\,- 49}$
 & $0.0076 ^{\,+ 54}_{\,- 25}$
 & $0.0091 ^{\,+ 34}_{\,- 69}$
 & $0.0073 ^{\,+ 38}_{\,- 25}$\\
  & $7.7$ 
 & $0$
 & $0$
 & $0.7085 ^{\,+ 89}_{\,- 42}$
 & $0.6158 ^{\,+ 61}_{\,- 87}$
 & $0$
 & $0$\\\hline
 \multirow{2}{*}{$8.8$}
 & $0$ 
 & $0.0088 ^{\,+ 27}_{\,- 63}$
 & $0.0066 ^{\,+ 69}_{\,- 31}$
 & $0.0082 ^{\,+ 34}_{\,- 37}$
 & $0.0065 ^{\,+ 48}_{\,- 34}$
 & $0.0076 ^{\,+ 60}_{\,- 44}$
 & $0.0070 ^{\,+ 65}_{\,- 37}$\\
  & $8.8$ 
 & $0$
 & $0$
 & $0.8992 ^{\,+ 20}_{\,- 25}$
 & $0.7624 ^{\,+ 42}_{\,- 47}$
 & $0$
 & $0$\\\hline
 \multirow{2}{*}{$9.9$}
 & $0$ 
 & $0.0098 ^{\,+ 119}_{\,- 55}$
 & $0.0054 ^{\,+ 32}_{\,- 26}$
 & $0.0073 ^{\,+ 26}_{\,- 41}$
 & $0.0065 ^{\,+ 51}_{\,- 22}$
 & $0.0082 ^{\,+ 33}_{\,- 21}$
 & $0.0071 ^{\,+ 64}_{\,- 27}$\\
  & $9.9$ 
 & $0$
 & $0$
 & $0.9397 ^{\,+ 20}_{\,- 34}$
 & $0.8554 ^{\,+ 66}_{\,- 48}$
 & $0$
 & $0$\\
[0.5ex] 
 \hline
\end{tabular}
\caption{
    The logarithmic negativity for the different neutrino pairs $\mathcal{N}_{ij}(t)$, as shown in Fig.~\ref{fig:Neutrinos_dwave2}, with $K=2$ and $N_A=10^3$, after two steps of the iterative procedure. The uncertainties correspond to 68\% confidence intervals determined from $N_{\rm run}=20$ independent runs.
}
\label{tab:Neutrinos_dwave_tab2_2}
\end{center}
\end{table}
}

\let\addcontentsline\oldaddcontentsline
\onecolumngrid
\clearpage
\newpage




\beginsupplement

\title{Supplemental Material\\ Basic Elements for Simulations of Standard Model Physics with Quantum Annealers: Multigrid and Clock States
}

\maketitle
\onecolumngrid

\vspace{-1cm}
\tableofcontents

\section{D-Wave quantum annealer processor}

\begin{figure}[!htb]
\centering
\includegraphics[width=0.45\columnwidth]{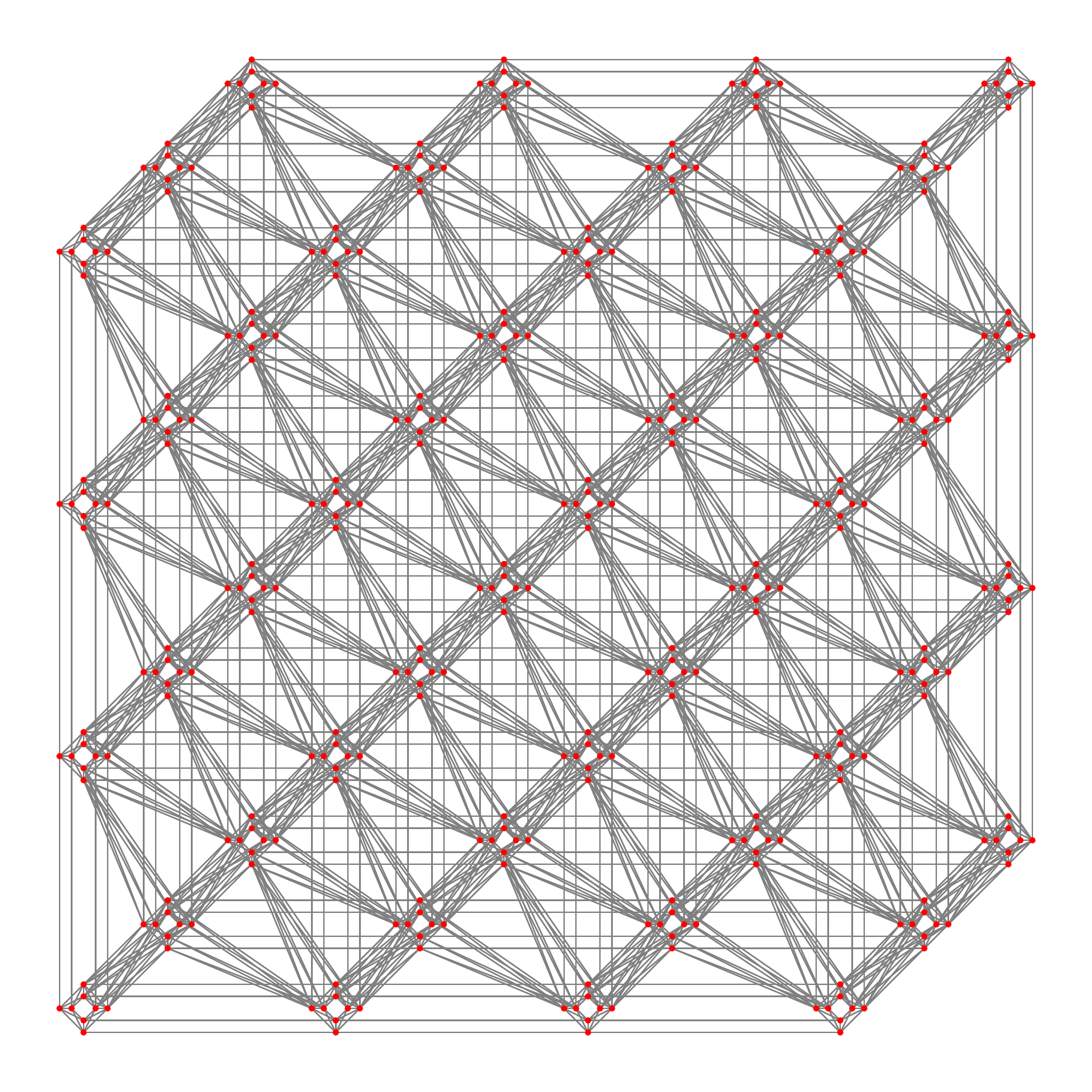}
\includegraphics[width=0.45\columnwidth]{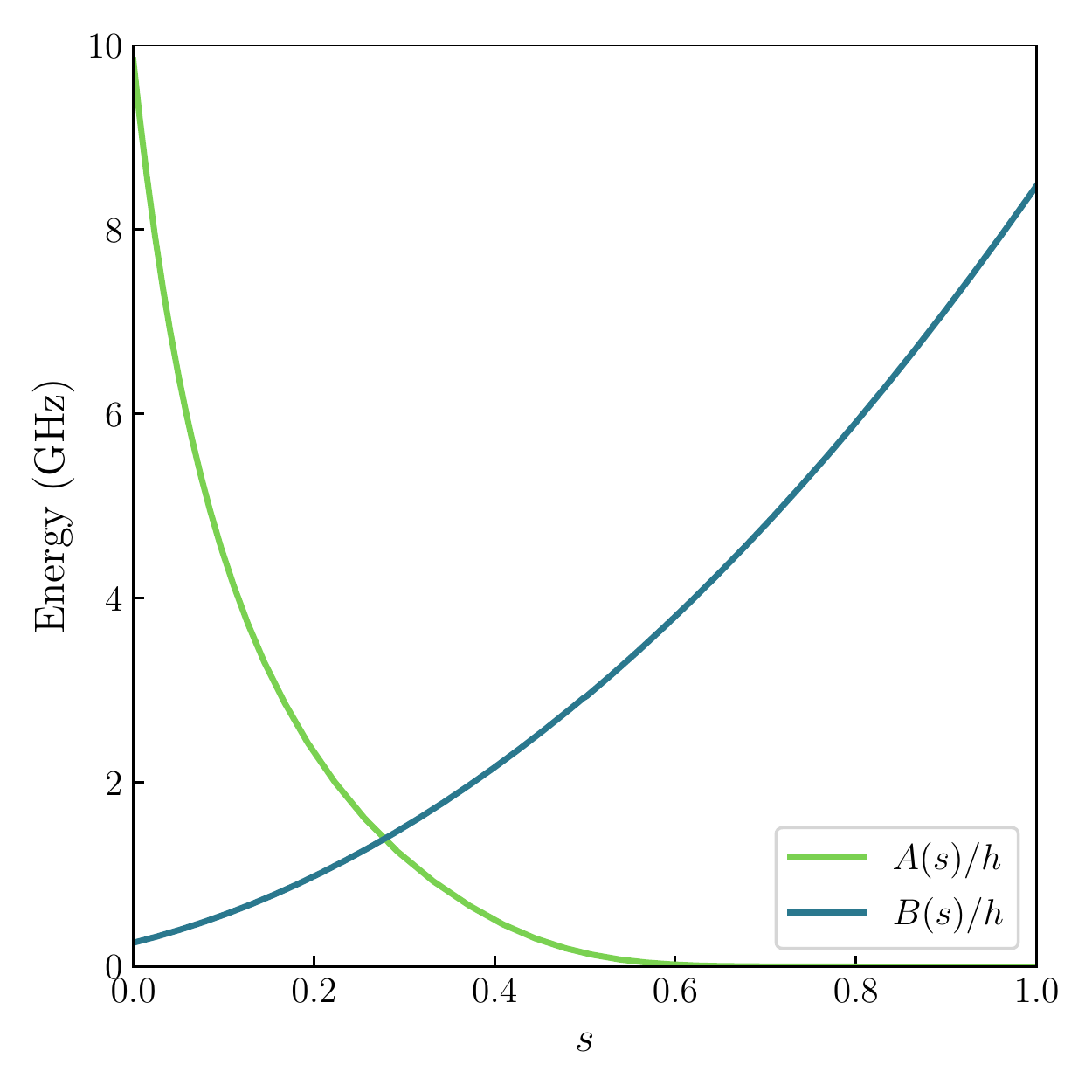}
\caption{(Left) A sketch of the {\tt Pegasus} topology of the quantum annealer {\tt Advantage} (the full processor is not shown for clarity), where each qubit is represented by a red point, and the connections between qubits by the gray lines. (Right) The default annealing schedule of {\tt Advantage} defined via the functions $A(s)$ and $B(s)$ (divided by Planck's constant $h$) in Eq.~(\ref{eq:Hsform}), as a function of the annealing parameter $s$.
}
\label{fig:adv}
\end{figure}

We briefly describe the quantum processor {\tt Advantage}, 
outlining relevant information from the 
detailed description that can be found on the 
D-Wave website~\cite{DWaveAdv} and reference therein.  
{\tt Advantage} consists of 5627 superconducting flux (rf-SQUID) qubits, with 15 connections per qubit via compound Josephson junctions rf-SQUID couplers~\cite{PhysRevB.80.052506}, as shown schematically in the left panel in Fig.~\ref{fig:adv}, operating at a temperature $T=15.4(1)$ mK. 
The time-dependent Hamiltonian that drives the annealing process is:
\begin{equation}
    \hat{\mathcal{H}}(s)=-\frac{A(s)}{2}\sum_i\sigma^{(x)}_i+\frac{B(s)}{2}\left(\sum_i h_i\sigma^{(z)}_i+\sum_{ij} J_{ij}\sigma^{(z)}_i\sigma^{(z)}_j\right)
    \ \ ,
\label{eq:Hsform}
\end{equation}
where $A(s)$ and $B(s)$ are time-dependent functions.
The right panel of Fig.~\ref{fig:adv} shows the employed annealing schedule as a function of the parameter $s$, which, by default, has a linear dependence on time, $s=t/t_A$, with $t_A$ being the total annealing time.
The Pauli matrices $\sigma_i$ act on the $i^{\rm th}$ qubit, and the coefficients $h_i$ and $J_{ij}$ are specified by the user.
In our case, it is simpler to formulate the optimization problem with the QUBO form, but there is a one-to-one correspondence between the QUBO matrix $Q_{ij}$ and the Ising coefficients $h_i$ and $J_{ij}$ (the derivation of the QUBO matrix for our problems can be found in Appendix A of the main text). 
The transformation between these two forms is performed by the D-Wave API~\cite{DwaveNeal}.

\section{Hamiltonian construction}
\noindent
In this section, we provide some of our {\tt Mathematica} scripts~\cite{Mathematica} relevant to the systems studied in this work.

\subsection{Harmonic and Anharmonic Oscillators}
\noindent
The Hamiltonians for digitized harmonic (HO) and anharmonic oscillators (AHO)~\cite{Jordan:2011ci,Klco:2018zqz} are defined once the parameters $n_s$ ({\tt ns}), $\phi_{\rm max}$ ({\tt phimax}), $m_0^2$ ({\tt m02}), and $\lambda$ ({\tt lambda}) are fixed, as described in the main text. 
For example, for the HO case with $n_s=32$, $\phi_{\rm max}=5$, and $m_0^2=1$, the Hamiltonian matrix for this system can be generated by:
\begin{tcolorbox}[breakable, size=fbox, boxrule=1pt, pad at break*=1mm,colback=cellbackgroundmma, colframe=cellborder,fontupper=\footnotesize]
\begin{mmaCell}[leftmargin=3.5em,labelsep=.5em,moredefined={phimax, ns,m02,lambda},morefunctionlocal={i}]{Input} 
ns = 32;
phimax = 5;
m02 = 1;
lambda = 0;
\end{mmaCell}
\begin{mmaCell}[leftmargin=3.5em,labelsep=.5em,moredefined={phimax, ns},morefunctionlocal={i}]{Input}
deltaphi = 2*phimax/(ns-1);
philist = Table[-1+2/(ns-1)*i, \{i,0,ns-1\}]*phimax;
\end{mmaCell}
\begin{mmaCell}[leftmargin=3.5em,labelsep=.5em,moredefined={phimax, ns},morefunctionlocal={i}]{Input}
kmax = \(\pi\)/phimax;
klist = Table[i-(ns-1)/2, \{i,0,ns-1\}]*2*kmax/ns;
\end{mmaCell}
\begin{mmaCell}[leftmargin=3.5em,labelsep=.5em,moredefined={phimax, ns},morefunctionlocal={i}]{Input}
phi2 = DiagonalMatrix[philist^2]/2;
phi4 = DiagonalMatrix[philist^4]/4!;
\end{mmaCell}
\begin{mmaCell}[leftmargin=3.5em,labelsep=.5em,moredefined={phimax, ns},morefunctionlocal={i,j}]{Input}
ftMat = Table[Exp[I*philist[[i]]*klist[[j]]], \{i,1,ns\}, \{j,1,ns\}];
pi2 = ftMat.DiagonalMatrix[klist^2].ConjugateTranspose[ftMat]*deltaphi^2/ns;
\end{mmaCell}
\begin{mmaCell}[leftmargin=3.5em,labelsep=.5em,moredefined={m02,lambda}]{Input}
hamiltonian = pi2 + m02*phi2 + lambda*phi4;
Export["HO_phimax5_nq5_ham.dat",hamiltonian]
\end{mmaCell}
\end{tcolorbox}

\subsection{One plaquette in SU(3) lattice gauge theory}
\noindent
The Hamiltonian matrix describing a one-plaquette system in SU(3) Yang-Mills lattice gauge theory~\cite{Ciavarella:2021nmj} in the color-parity basis that is truncated to $\{|\mathbf{1}\rangle ,|\mathbf{3}^+\rangle , |\mathbf{6}^+\rangle,|\mathbf{8}\rangle\}$, 
and with $g=1$, is 
\begin{tcolorbox}[breakable, size=fbox, boxrule=1pt, pad at break*=1mm,colback=cellbackgroundmma, colframe=cellborder,fontupper=\footnotesize]
\begin{mmaCell}[leftmargin=3.5em,labelsep=.5em]{Input}
hamiltonian = 1/2*\{\{0,0,0,0\},\{0,16/3,0,0\},\{0,0,40/3,0\},\{0,0,0,12\}\}
+1/2*(6*IdentityMatrix[4]+\{\{0,\(\sqrt{2}\),0,0\},\{\(\sqrt{2}\),1,1,\(\sqrt{2}\)\},\{0,1,0,\(\sqrt{2}\)\},\{0,\(\sqrt{2}\),\(\sqrt{2}\),0\}\});
Export["hamiltonian_SU3.dat",hamiltonian]
\end{mmaCell}
\end{tcolorbox}

\subsection{Coherent flavor evolution in a four-neutrino system}
\noindent
The Hamiltonian 
describing  coherent flavor evolution in a simplified four-neutrino system~\cite{Hall:2021rbv}
is specified by the mixing angle,
which is fixed to $\theta_v=0.195$, and the distribution of momentum directions, fixed to  $\theta_{ij}=\arccos(0.9) \times|i-j|/(4-1)$, 
and can be generated by:
\begin{tcolorbox}[breakable, size=fbox, boxrule=1pt, pad at break*=1mm,colback=cellbackgroundmma, colframe=cellborder,fontupper=\footnotesize]
\begin{mmaCell}[leftmargin=3.5em,labelsep=.5em,moredefined={PauliMatrix}]{Input}
pauliVec = \{PauliMatrix[1],PauliMatrix[2],PauliMatrix[3]\};
bfield = \{Sin[2*0.195],0,-Cos[2*0.195]\};
op1bodymat = bfield.pauliVec;
\end{mmaCell}
\begin{mmaCell}[leftmargin=3.5em,labelsep=.5em,moredefined={KroneckerProduct},morefunctionlocal={i,j,k}]{Input}
op1body = 
  Sum[KroneckerProduct[If[i == 1, op1bodymat, IdentityMatrix[2]], 
    If[i == 2, op1bodymat, IdentityMatrix[2]], 
    If[i == 3, op1bodymat, IdentityMatrix[2]], 
    If[i == 4, op1bodymat, IdentityMatrix[2]]], \{i,1,4\}];
op2body = 
  Sum[(1 - Cos[ArcCos[0.9]*Abs[i-j]/(4-1)])*
    Sum[KroneckerProduct[If[i == 1, pauliVec[[k]], IdentityMatrix[2]],
      If[i == 2 || j == 2, pauliVec[[k]], IdentityMatrix[2]], 
      If[i == 3 || j == 3, pauliVec[[k]], IdentityMatrix[2]], 
      If[j == 4, pauliVec[[k]], IdentityMatrix[2]]], \{k,1,3\}], \{i,1,3\}, \{j,i+1,4\}];
\end{mmaCell}
\begin{mmaCell}[leftmargin=3.6em,labelsep=.5em]{Input}
hamiltonian = op1body + op2body;
Export["hamiltonian_Neutrinos.dat",hamiltonian]
\end{mmaCell}
\end{tcolorbox}

\section{Annealer job script}
\noindent
In this section, we provide {\tt python} scripts~\cite{python3} used to send jobs to the D-Wave annealer (or simulator), which are similar to the ones used in Ref.~\cite{ARahman:2021ktn}. 
The packages that are required to run the following scripts are:
\begin{tcolorbox}[breakable, size=fbox, boxrule=1pt, pad at break*=1mm,colback=cellbackground, colframe=cellborder,fontupper=\footnotesize]
\prompt{In}{incolor}{1}{\boxspacing}
\begin{Verbatim}[commandchars=\\\{\}]
\PY{k+kn}{from} \PY{n+nn}{dwave}\PY{n+nn}{.}\PY{n+nn}{system} \PY{k+kn}{import} \PY{n}{DWaveSampler}\PY{p}{,} \PY{n}{FixedEmbeddingComposite}
\PY{k+kn}{from} \PY{n+nn}{neal}         \PY{k+kn}{import} \PY{n}{SimulatedAnnealingSampler}
\PY{k+kn}{from} \PY{n+nn}{minorminer} \PY{k+kn}{import} \PY{n}{find\PYZus{}embedding}
\PY{k+kn}{from} \PY{n+nn}{collections}  \PY{k+kn}{import} \PY{n}{defaultdict}
\PY{k+kn}{import} \PY{n+nn}{numpy} \PY{k}{as} \PY{n+nn}{np}
\PY{k+kn}{import} \PY{n+nn}{scipy}
\PY{k+kn}{import} \PY{n+nn}{itertools}
\PY{k+kn}{import} \PY{n+nn}{time}
\end{Verbatim}
\end{tcolorbox}

\subsection{Ground-state and excited-state energy extraction}
\noindent
The main parts of the workflow are performed by two functions. The first is responsible for finding the embedding of the QUBO matrix onto the D-Wave {\tt Advantage} annealer ({\tt energy\_embedding()}),
with input variables: the Hamiltonian as a matrix ({\tt listham}), the dimension of matrix ({\tt dimH}), the value of $K$ ({\tt Kval}), the value of $\eta$ ({\tt etaval}),
the parameter {\tt tokenval} (the authentication token needed for access),   
the list of eigenvectors $|\Psi_n\rangle$ required to be projected out ({\tt veclist}, non-zero only when extracting excited states), and the value of $\mu$ ({\tt muval}).
\begin{tcolorbox}[breakable, size=fbox, boxrule=1pt, pad at break*=1mm,colback=cellbackground, colframe=cellborder,fontupper=\footnotesize]
\prompt{In}{incolor}{2}{\boxspacing}
\begin{Verbatim}[commandchars=\\\{\}]
\PY{k}{def} \PY{n+nf}{energy\PYZus{}embedding}\PY{p}{(}\PY{n}{dimH}\PY{p}{,} \PY{n}{listham}\PY{p}{,} \PY{n}{Kval}\PY{p}{,} \PY{n}{etaval}\PY{p}{,} \PY{n}{tokenval}\PY{p}{,} \PY{n}{veclist}\PY{p}{,} \PY{n}{muval}\PY{p}{)}\PY{p}{:}

    \PY{n}{delta} \PY{o}{=} \PY{n}{np}\PY{o}{.}\PY{n}{zeros}\PY{p}{(}\PY{p}{(}\PY{n}{Kval}\PY{o}{+}\PY{l+m+mi}{1}\PY{p}{,}\PY{n}{Kval}\PY{o}{+}\PY{l+m+mi}{1}\PY{p}{)}\PY{p}{,}\PY{n+nb}{int}\PY{p}{)}
    \PY{n}{i} \PY{o}{=} \PY{n}{np}\PY{o}{.}\PY{n}{arange}\PY{p}{(}\PY{n}{Kval}\PY{o}{+}\PY{l+m+mi}{1}\PY{p}{)}
    \PY{n}{delta}\PY{p}{[}\PY{n}{i}\PY{p}{,}\PY{n}{i}\PY{p}{]} \PY{o}{=} \PY{l+m+mi}{1}
    
    \PY{n}{Hlambda} \PY{o}{=} \PY{n}{defaultdict}\PY{p}{(}\PY{n+nb}{float}\PY{p}{)}

    \PY{k}{for} \PY{n}{num} \PY{o+ow}{in} \PY{n+nb}{range}\PY{p}{(}\PY{l+m+mi}{1}\PY{p}{,} \PY{n}{dimH}\PY{o}{+}\PY{l+m+mi}{1}\PY{p}{)}\PY{p}{:}
        \PY{n}{Hlambda}\PY{p}{[}\PY{n}{num}\PY{p}{,}\PY{n}{num}\PY{p}{]} \PY{o}{\PYZhy{}}\PY{o}{=} \PY{n}{etaval}
        
    \PY{k}{for} \PY{n}{num1} \PY{o+ow}{in} \PY{n+nb}{range}\PY{p}{(}\PY{l+m+mi}{1}\PY{p}{,} \PY{n}{dimH}\PY{o}{+}\PY{l+m+mi}{1}\PY{p}{)}\PY{p}{:}
        \PY{k}{for} \PY{n}{num2} \PY{o+ow}{in} \PY{n+nb}{range}\PY{p}{(}\PY{l+m+mi}{1}\PY{p}{,} \PY{n}{dimH}\PY{o}{+}\PY{l+m+mi}{1}\PY{p}{)}\PY{p}{:}
            \PY{n}{Hlambda}\PY{p}{[}\PY{n}{num1}\PY{p}{,}\PY{n}{num2}\PY{p}{]} \PY{o}{+}\PY{o}{=} \PY{n}{listham}\PY{p}{[}\PY{n}{num1}\PY{o}{\PYZhy{}}\PY{l+m+mi}{1}\PY{p}{,}\PY{n}{num2}\PY{o}{\PYZhy{}}\PY{l+m+mi}{1}\PY{p}{]}
        
    \PY{k}{for} \PY{n}{vecs} \PY{o+ow}{in} \PY{n}{veclist}\PY{p}{:}
        \PY{k}{for} \PY{n}{xval} \PY{o+ow}{in} \PY{n+nb}{range}\PY{p}{(}\PY{l+m+mi}{1}\PY{p}{,} \PY{n}{dimH}\PY{o}{+}\PY{l+m+mi}{1}\PY{p}{)}\PY{p}{:}
            \PY{k}{for} \PY{n}{yval} \PY{o+ow}{in} \PY{n+nb}{range}\PY{p}{(}\PY{l+m+mi}{1}\PY{p}{,} \PY{n}{dimH}\PY{o}{+}\PY{l+m+mi}{1}\PY{p}{)}\PY{p}{:}
                \PY{n}{Hlambda}\PY{p}{[}\PY{n}{xval}\PY{p}{,}\PY{n}{yval}\PY{p}{]} \PY{o}{+}\PY{o}{=} \PY{n}{muval}\PY{o}{*}\PY{n}{np}\PY{o}{.}\PY{n}{outer}\PY{p}{(}\PY{n}{vecs}\PY{p}{,}\PY{n}{vecs}\PY{p}{)}\PY{p}{[}\PY{n}{xval}\PY{o}{\PYZhy{}}\PY{l+m+mi}{1}\PY{p}{,}\PY{n}{yval}\PY{o}{\PYZhy{}}\PY{l+m+mi}{1}\PY{p}{]}

    \PY{n}{acenter} \PY{o}{=} \PY{p}{[}\PY{l+m+mi}{0}\PY{p}{]}\PY{o}{*}\PY{n}{dimH}
    \PY{n}{zoom} \PY{o}{=} \PY{l+m+mi}{0}
    \PY{n}{Q} \PY{o}{=} \PY{n}{defaultdict}\PY{p}{(}\PY{n+nb}{float}\PY{p}{)}
    \PY{k}{for} \PY{n}{alpha} \PY{o+ow}{in} \PY{n+nb}{range}\PY{p}{(}\PY{l+m+mi}{1}\PY{p}{,}\PY{n}{dimH}\PY{o}{+}\PY{l+m+mi}{1}\PY{p}{)}\PY{p}{:}
        \PY{k}{for} \PY{n}{beta} \PY{o+ow}{in} \PY{n+nb}{range}\PY{p}{(}\PY{l+m+mi}{1}\PY{p}{,}\PY{n}{dimH}\PY{o}{+}\PY{l+m+mi}{1}\PY{p}{)}\PY{p}{:}
            \PY{k}{for} \PY{n}{i}\PY{p}{,}\PY{n}{j} \PY{o+ow}{in} \PY{n}{itertools}\PY{o}{.}\PY{n}{product}\PY{p}{(}\PY{n+nb}{range}\PY{p}{(}\PY{l+m+mi}{1}\PY{p}{,}\PY{n}{Kval}\PY{o}{+}\PY{l+m+mi}{1}\PY{p}{)}\PY{p}{,}\PY{n+nb}{range}\PY{p}{(}\PY{l+m+mi}{1}\PY{p}{,}\PY{n}{Kval}\PY{o}{+}\PY{l+m+mi}{1}\PY{p}{)}\PY{p}{)}\PY{p}{:}
                \PY{n}{n} \PY{o}{=} \PY{n}{Kval}\PY{o}{*}\PY{p}{(}\PY{n}{alpha}\PY{o}{\PYZhy{}}\PY{l+m+mi}{1}\PY{p}{)} \PY{o}{+} \PY{n}{i}
                \PY{n}{m} \PY{o}{=} \PY{n}{Kval}\PY{o}{*}\PY{p}{(}\PY{n}{beta}\PY{o}{\PYZhy{}}\PY{l+m+mi}{1}\PY{p}{)} \PY{o}{+} \PY{n}{j}
                \PY{n}{Q}\PY{p}{[}\PY{n}{n}\PY{p}{,}\PY{n}{m}\PY{p}{]} \PY{o}{+}\PY{o}{=} \PY{n}{Hlambda}\PY{p}{[}\PY{n}{alpha}\PY{p}{,}\PY{n}{beta}\PY{p}{]}\PY{o}{*}\PY{l+m+mi}{2}\PY{o}{*}\PY{o}{*}\PY{p}{(}\PY{n}{i}\PY{o}{+}\PY{n}{j}\PY{o}{\PYZhy{}}\PY{l+m+mi}{2}\PY{o}{*}\PY{n}{Kval}\PY{o}{\PYZhy{}}\PY{l+m+mi}{2}\PY{o}{*}\PY{n}{zoom}\PY{p}{)}\PY{o}{*}\PY{p}{(}\PY{o}{\PYZhy{}}\PY{l+m+mi}{1}\PY{p}{)}\PY{o}{*}\PY{o}{*}\PY{p}{(}\PY{n}{delta}\PY{p}{[}\PY{n}{i}\PY{p}{,}\PY{n}{Kval}\PY{p}{]}\PY{o}{+}\PY{n}{delta}\PY{p}{[}\PY{n}{j}\PY{p}{,}\PY{n}{Kval}\PY{p}{]}\PY{p}{)}
                \PY{k}{if} \PY{n}{alpha}\PY{o}{==}\PY{n}{beta} \PY{o+ow}{and} \PY{n}{i}\PY{o}{==}\PY{n}{j}\PY{p}{:}
                    \PY{k}{for} \PY{n}{loop} \PY{o+ow}{in} \PY{n+nb}{range}\PY{p}{(}\PY{l+m+mi}{1}\PY{p}{,}\PY{n}{dimH}\PY{o}{+}\PY{l+m+mi}{1}\PY{p}{)}\PY{p}{:}
                        \PY{n}{Q}\PY{p}{[}\PY{n}{n}\PY{p}{,}\PY{n}{m}\PY{p}{]} \PY{o}{+}\PY{o}{=} \PY{l+m+mi}{2}\PY{o}{*}\PY{n}{acenter}\PY{p}{[}\PY{p}{(}\PY{n}{loop}\PY{o}{\PYZhy{}}\PY{l+m+mi}{1}\PY{p}{)}\PY{o}{*}\PY{l+m+mi}{2}\PY{p}{]}\PY{o}{*}\PY{l+m+mi}{2}\PY{o}{*}\PY{o}{*}\PY{p}{(}\PY{n}{i}\PY{o}{\PYZhy{}}\PY{n}{Kval}\PY{o}{\PYZhy{}}\PY{n}{zoom}\PY{p}{)}\PY{o}{*}\PY{n}{Hlambda}\PY{p}{[}\PY{n}{loop}\PY{p}{,}\PY{n}{beta}\PY{p}{]}\PY{o}{*}\PY{p}{(}\PY{o}{\PYZhy{}}\PY{l+m+mi}{1}\PY{p}{)}\PY{o}{*}\PY{o}{*}\PY{p}{(}\PY{n}{delta}\PY{p}{[}\PY{n}{i}\PY{p}{,}\PY{n}{Kval}\PY{p}{]}\PY{p}{)}

    \PY{n}{start} \PY{o}{=} \PY{n}{time}\PY{o}{.}\PY{n}{time}\PY{p}{(}\PY{p}{)}
    \PY{n}{sampler} \PY{o}{=} \PY{n}{DWaveSampler}\PY{p}{(}\PY{n}{solver}\PY{o}{=}\PY{p}{\PYZob{}}\PY{l+s+s1}{\PYZsq{}}\PY{l+s+s1}{topology\PYZus{}\PYZus{}type\PYZus{}\PYZus{}eq}\PY{l+s+s1}{\PYZsq{}}\PY{p}{:}\PY{l+s+s1}{\PYZsq{}}\PY{l+s+s1}{pegasus}\PY{l+s+s1}{\PYZsq{}}\PY{p}{\PYZcb{}}\PY{p}{,}\PY{n}{token}\PY{o}{=}\PY{n}{tokenval}\PY{p}{)}
    \PY{n}{embedding} \PY{o}{=} \PY{n}{find\PYZus{}embedding}\PY{p}{(}\PY{n}{Q}\PY{p}{,} \PY{n}{sampler}\PY{o}{.}\PY{n}{edgelist}\PY{p}{)}
    \PY{n}{end} \PY{o}{=} \PY{n}{time}\PY{o}{.}\PY{n}{time}\PY{p}{(}\PY{p}{)}
    \PY{n+nb}{print}\PY{p}{(}\PY{l+s+sa}{f}\PY{l+s+s2}{\PYZdq{}}\PY{l+s+s2}{Number of logical qubits: }\PY{l+s+si}{\PYZob{}}\PY{n+nb}{len}\PY{p}{(}\PY{n}{embedding}\PY{o}{.}\PY{n}{keys}\PY{p}{(}\PY{p}{)}\PY{p}{)}\PY{l+s+si}{\PYZcb{}}\PY{l+s+s2}{\PYZdq{}}\PY{p}{)}
    \PY{n+nb}{print}\PY{p}{(}\PY{l+s+sa}{f}\PY{l+s+s2}{\PYZdq{}}\PY{l+s+s2}{Number of physical qubits used in embedding: }\PY{l+s+si}{\PYZob{}}\PY{n+nb}{sum}\PY{p}{(}\PY{n+nb}{len}\PY{p}{(}\PY{n}{chain}\PY{p}{)} \PY{k}{for} \PY{n}{chain} \PY{o+ow}{in} \PY{n}{embedding}\PY{o}{.}\PY{n}{values}\PY{p}{(}\PY{p}{)}\PY{p}{)}\PY{l+s+si}{\PYZcb{}}\PY{l+s+s2}{\PYZdq{}}\PY{p}{)}
    \PY{n+nb}{print}\PY{p}{(}\PY{l+s+sa}{f}\PY{l+s+s2}{\PYZdq{}}\PY{l+s+s2}{Runtime of computing the embedding is }\PY{l+s+si}{\PYZob{}}\PY{n}{end} \PY{o}{\PYZhy{}} \PY{n}{start}\PY{l+s+si}{\PYZcb{}}\PY{l+s+s2}{\PYZdq{}}\PY{p}{)}
    \PY{k}{return}\PY{p}{(}\PY{n}{embedding}\PY{p}{)}
\end{Verbatim}
\end{tcolorbox}
The above function includes projections against lower-lying states in the spectrum through the use of a chemical potential, specifically the line (in the above, cell [2])
\begin{eqnarray}
{\tt Hlambda[xval,yval] += muval*np.outer(vecs,vecs)[xval-1,yval-1] }
\ \ .
\end{eqnarray}

The second function is responsible for finding the ground-state energy (and its corresponding wavefunction) for a specific Hamiltonian ({\tt energy\_multigrid\_project()}). Most of its parameters are similar to those of the previous function, and the new ones are the number of anneals $N_A$ ({\tt shots}), the maximum number of zoom levels $z^{\rm max}+1$ ({\tt zoommax}), the starting points $a_{\alpha}$ for an anneal if the multigrid method is used ({\tt acenterval}), and its starting zoom level $z^{\rm init}$ ({\tt zoomint}). If the flag {\tt quantum} is set to ``0'', the simulator {\tt neal} is used, but if {\tt quantum} is set to ``1'', the quantum processor {\tt Advantage} is used, requiring the authentication token ({\tt tokenval}), a label for the job submitted to the queue ({\tt jobid}), the embedding computed with the previous function ({\tt embedding}), and the value of $\omega$ ({\tt chainval}) to fix the chain strength $c_s=\omega \,{\rm max}(|Q|)$. The annealing time $t_A$ is fixed at 20 $\mu$s. 
The outputs of this function are two arrays, one for the ground-state energy and another for its wavefunction at each zoom level.
\begin{tcolorbox}[breakable, size=fbox, boxrule=1pt, pad at break*=1mm,colback=cellbackground, colframe=cellborder,fontupper=\footnotesize]
\prompt{In}{incolor}{3}{\boxspacing}
\begin{Verbatim}[commandchars=\\\{\}]
\PY{k}{def} \PY{n+nf}{energy\PYZus{}multigrid\PYZus{}project}\PY{p}{(}\PY{n}{dimH}\PY{p}{,} \PY{n}{listham}\PY{p}{,} \PY{n}{Kval}\PY{p}{,} \PY{n}{etaval}\PY{p}{,} \PY{n}{shots}\PY{p}{,} \PY{n}{zoommax}\PY{p}{,} \PY{n}{quantum}\PY{p}{,} \PY{n}{tokenval}\PY{p}{,} \PY{n}{jobid}\PY{p}{,} \PY{n}{embedding}\PY{p}{,}
                             \PY{n}{chainval}\PY{p}{,} \PY{n}{acenterval}\PY{p}{,} \PY{n}{zoomint}\PY{p}{,} \PY{n}{veclist}\PY{p}{,} \PY{n}{muval}\PY{p}{)}\PY{p}{:}
    
    \PY{n}{delta} \PY{o}{=} \PY{n}{np}\PY{o}{.}\PY{n}{zeros}\PY{p}{(}\PY{p}{(}\PY{n}{Kval}\PY{o}{+}\PY{l+m+mi}{1}\PY{p}{,}\PY{n}{Kval}\PY{o}{+}\PY{l+m+mi}{1}\PY{p}{)}\PY{p}{,}\PY{n+nb}{int}\PY{p}{)}
    \PY{n}{i} \PY{o}{=} \PY{n}{np}\PY{o}{.}\PY{n}{arange}\PY{p}{(}\PY{n}{Kval}\PY{o}{+}\PY{l+m+mi}{1}\PY{p}{)}
    \PY{n}{delta}\PY{p}{[}\PY{n}{i}\PY{p}{,}\PY{n}{i}\PY{p}{]} \PY{o}{=} \PY{l+m+mi}{1}
    
    \PY{n}{Hlambda} \PY{o}{=} \PY{n}{defaultdict}\PY{p}{(}\PY{n+nb}{float}\PY{p}{)}

    \PY{k}{for} \PY{n}{num} \PY{o+ow}{in} \PY{n+nb}{range}\PY{p}{(}\PY{l+m+mi}{1}\PY{p}{,} \PY{n}{dimH}\PY{o}{+}\PY{l+m+mi}{1}\PY{p}{)}\PY{p}{:}
        \PY{n}{Hlambda}\PY{p}{[}\PY{n}{num}\PY{p}{,}\PY{n}{num}\PY{p}{]} \PY{o}{\PYZhy{}}\PY{o}{=} \PY{n}{etaval}

    \PY{k}{for} \PY{n}{num1} \PY{o+ow}{in} \PY{n+nb}{range}\PY{p}{(}\PY{l+m+mi}{1}\PY{p}{,} \PY{n}{dimH}\PY{o}{+}\PY{l+m+mi}{1}\PY{p}{)}\PY{p}{:}
        \PY{k}{for} \PY{n}{num2} \PY{o+ow}{in} \PY{n+nb}{range}\PY{p}{(}\PY{l+m+mi}{1}\PY{p}{,} \PY{n}{dimH}\PY{o}{+}\PY{l+m+mi}{1}\PY{p}{)}\PY{p}{:}
            \PY{n}{Hlambda}\PY{p}{[}\PY{n}{num1}\PY{p}{,}\PY{n}{num2}\PY{p}{]} \PY{o}{+}\PY{o}{=} \PY{n}{listham}\PY{p}{[}\PY{n}{num1}\PY{o}{\PYZhy{}}\PY{l+m+mi}{1}\PY{p}{,}\PY{n}{num2}\PY{o}{\PYZhy{}}\PY{l+m+mi}{1}\PY{p}{]}
        
    \PY{k}{for} \PY{n}{vecs} \PY{o+ow}{in} \PY{n}{veclist}\PY{p}{:}
        \PY{k}{for} \PY{n}{xval} \PY{o+ow}{in} \PY{n+nb}{range}\PY{p}{(}\PY{l+m+mi}{1}\PY{p}{,} \PY{n}{dimH}\PY{o}{+}\PY{l+m+mi}{1}\PY{p}{)}\PY{p}{:}
            \PY{k}{for} \PY{n}{yval} \PY{o+ow}{in} \PY{n+nb}{range}\PY{p}{(}\PY{l+m+mi}{1}\PY{p}{,} \PY{n}{dimH}\PY{o}{+}\PY{l+m+mi}{1}\PY{p}{)}\PY{p}{:}
                \PY{n}{Hlambda}\PY{p}{[}\PY{n}{xval}\PY{p}{,}\PY{n}{yval}\PY{p}{]} \PY{o}{+}\PY{o}{=} \PY{n}{muval}\PY{o}{*}\PY{n}{np}\PY{o}{.}\PY{n}{outer}\PY{p}{(}\PY{n}{vecs}\PY{p}{,}\PY{n}{vecs}\PY{p}{)}\PY{p}{[}\PY{n}{xval}\PY{o}{\PYZhy{}}\PY{l+m+mi}{1}\PY{p}{,}\PY{n}{yval}\PY{o}{\PYZhy{}}\PY{l+m+mi}{1}\PY{p}{]}

    \PY{n}{acenter} \PY{o}{=} \PY{n}{acenterval}
    
    \PY{n}{listen} \PY{o}{=} \PY{p}{[}\PY{p}{]}
    \PY{n}{listvec} \PY{o}{=} \PY{p}{[}\PY{p}{]}
    \PY{n}{listchain} \PY{o}{=} \PY{p}{[}\PY{p}{]}
    \PY{k}{for} \PY{n}{zoom} \PY{o+ow}{in} \PY{n+nb}{range}\PY{p}{(}\PY{n}{zoomint}\PY{p}{,}\PY{n}{zoommax}\PY{p}{)}\PY{p}{:}
        \PY{n}{Q} \PY{o}{=} \PY{n}{defaultdict}\PY{p}{(}\PY{n+nb}{float}\PY{p}{)}
        \PY{k}{for} \PY{n}{alpha} \PY{o+ow}{in} \PY{n+nb}{range}\PY{p}{(}\PY{l+m+mi}{1}\PY{p}{,}\PY{n}{dimH}\PY{o}{+}\PY{l+m+mi}{1}\PY{p}{)}\PY{p}{:}
            \PY{k}{for} \PY{n}{beta} \PY{o+ow}{in} \PY{n+nb}{range}\PY{p}{(}\PY{l+m+mi}{1}\PY{p}{,}\PY{n}{dimH}\PY{o}{+}\PY{l+m+mi}{1}\PY{p}{)}\PY{p}{:}
                \PY{k}{for} \PY{n}{i}\PY{p}{,}\PY{n}{j} \PY{o+ow}{in} \PY{n}{itertools}\PY{o}{.}\PY{n}{product}\PY{p}{(}\PY{n+nb}{range}\PY{p}{(}\PY{l+m+mi}{1}\PY{p}{,}\PY{n}{Kval}\PY{o}{+}\PY{l+m+mi}{1}\PY{p}{)}\PY{p}{,}\PY{n+nb}{range}\PY{p}{(}\PY{l+m+mi}{1}\PY{p}{,}\PY{n}{Kval}\PY{o}{+}\PY{l+m+mi}{1}\PY{p}{)}\PY{p}{)}\PY{p}{:}
                    \PY{n}{n} \PY{o}{=} \PY{n}{Kval}\PY{o}{*}\PY{p}{(}\PY{n}{alpha}\PY{o}{\PYZhy{}}\PY{l+m+mi}{1}\PY{p}{)} \PY{o}{+} \PY{n}{i}
                    \PY{n}{m} \PY{o}{=} \PY{n}{Kval}\PY{o}{*}\PY{p}{(}\PY{n}{beta}\PY{o}{\PYZhy{}}\PY{l+m+mi}{1}\PY{p}{)} \PY{o}{+} \PY{n}{j}
                    \PY{n}{Q}\PY{p}{[}\PY{n}{n}\PY{p}{,}\PY{n}{m}\PY{p}{]} \PY{o}{+}\PY{o}{=} \PY{n}{Hlambda}\PY{p}{[}\PY{n}{alpha}\PY{p}{,}\PY{n}{beta}\PY{p}{]}\PY{o}{*}\PY{l+m+mi}{2}\PY{o}{*}\PY{o}{*}\PY{p}{(}\PY{n}{i}\PY{o}{+}\PY{n}{j}\PY{o}{\PYZhy{}}\PY{l+m+mi}{2}\PY{o}{*}\PY{n}{Kval}\PY{o}{\PYZhy{}}\PY{l+m+mi}{2}\PY{o}{*}\PY{n}{zoom}\PY{p}{)}\PY{o}{*}\PY{p}{(}\PY{o}{\PYZhy{}}\PY{l+m+mi}{1}\PY{p}{)}\PY{o}{*}\PY{o}{*}\PY{p}{(}\PY{n}{delta}\PY{p}{[}\PY{n}{i}\PY{p}{,}\PY{n}{Kval}\PY{p}{]}\PY{o}{+}\PY{n}{delta}\PY{p}{[}\PY{n}{j}\PY{p}{,}\PY{n}{Kval}\PY{p}{]}\PY{p}{)}
                    \PY{k}{if} \PY{n}{alpha}\PY{o}{==}\PY{n}{beta} \PY{o+ow}{and} \PY{n}{n}\PY{o}{==}\PY{n}{m}\PY{p}{:}
                        \PY{k}{for} \PY{n}{loop} \PY{o+ow}{in} \PY{n+nb}{range}\PY{p}{(}\PY{l+m+mi}{1}\PY{p}{,}\PY{n}{dimH}\PY{o}{+}\PY{l+m+mi}{1}\PY{p}{)}\PY{p}{:}
                            \PY{n}{Q}\PY{p}{[}\PY{n}{n}\PY{p}{,}\PY{n}{m}\PY{p}{]} \PY{o}{+}\PY{o}{=} \PY{l+m+mi}{2}\PY{o}{*}\PY{n}{acenter}\PY{p}{[}\PY{p}{(}\PY{n}{loop}\PY{o}{\PYZhy{}}\PY{l+m+mi}{1}\PY{p}{)}\PY{o}{*}\PY{l+m+mi}{2}\PY{p}{]}\PY{o}{*}\PY{l+m+mi}{2}\PY{o}{*}\PY{o}{*}\PY{p}{(}\PY{n}{i}\PY{o}{\PYZhy{}}\PY{n}{Kval}\PY{o}{\PYZhy{}}\PY{n}{zoom}\PY{p}{)}\PY{o}{*}\PY{n}{Hlambda}\PY{p}{[}\PY{n}{loop}\PY{p}{,}\PY{n}{beta}\PY{p}{]}\PY{o}{*}\PY{p}{(}\PY{o}{\PYZhy{}}\PY{l+m+mi}{1}\PY{p}{)}\PY{o}{*}\PY{o}{*}\PY{p}{(}\PY{n}{delta}\PY{p}{[}\PY{n}{i}\PY{p}{,}\PY{n}{Kval}\PY{p}{]}\PY{p}{)}
        

        \PY{k}{if} \PY{n}{quantum}\PY{o}{==}\PY{l+m+mi}{1}\PY{p}{:}
            \PY{n}{Q\PYZus{}values} \PY{o}{=} \PY{n}{Q}\PY{o}{.}\PY{n}{values}\PY{p}{(}\PY{p}{)}
            \PY{n}{chainstrength} \PY{o}{=} \PY{n}{chainval}\PY{o}{*}\PY{n+nb}{max}\PY{p}{(}\PY{n+nb}{max}\PY{p}{(}\PY{n}{Q\PYZus{}values}\PY{p}{)}\PY{p}{,}\PY{n+nb}{abs}\PY{p}{(}\PY{n+nb}{min}\PY{p}{(}\PY{n}{Q\PYZus{}values}\PY{p}{)}\PY{p}{)}\PY{p}{)}
            \PY{n}{sampler} \PY{o}{=} \PY{n}{FixedEmbeddingComposite}\PY{p}{(}\PY{n}{DWaveSampler}\PY{p}{(}\PY{n}{solver}\PY{o}{=}\PY{p}{\PYZob{}}\PY{l+s+s1}{\PYZsq{}}\PY{l+s+s1}{topology\PYZus{}\PYZus{}type\PYZus{}\PYZus{}eq}\PY{l+s+s1}{\PYZsq{}}\PY{p}{:}\PY{l+s+s1}{\PYZsq{}}\PY{l+s+s1}{pegasus}\PY{l+s+s1}{\PYZsq{}}\PY{p}{\PYZcb{}}\PY{p}{,}\PY{n}{token}\PY{o}{=}\PY{n}{tokenval}\PY{p}{)}\PY{p}{,}
                                              \PY{n}{embedding}\PY{p}{)}
            \PY{n}{sampleset} \PY{o}{=} \PY{n}{sampler}\PY{o}{.}\PY{n}{sample\PYZus{}qubo}\PY{p}{(}\PY{n}{Q}\PY{p}{,}\PY{n}{num\PYZus{}reads}\PY{o}{=}\PY{n}{shots}\PY{p}{,}\PY{n}{chain\PYZus{}strength}\PY{o}{=}\PY{n}{chainstrength}\PY{p}{,}\PY{n}{annealing\PYZus{}time}\PY{o}{=}\PY{l+m+mi}{20}\PY{p}{,}
                                            \PY{n}{label}\PY{o}{=}\PY{n}{jobid}\PY{p}{)}
            \PY{n}{rawoutput} \PY{o}{=} \PY{n}{sampleset}\PY{o}{.}\PY{n}{aggregate}\PY{p}{(}\PY{p}{)}
        \PY{k}{else}\PY{p}{:}
            \PY{n}{sampler} \PY{o}{=} \PY{n}{SimulatedAnnealingSampler}\PY{p}{(}\PY{p}{)}
            \PY{n}{sampleset} \PY{o}{=} \PY{n}{sampler}\PY{o}{.}\PY{n}{sample\PYZus{}qubo}\PY{p}{(}\PY{n}{Q}\PY{p}{,}\PY{n}{num\PYZus{}reads}\PY{o}{=}\PY{n}{shots}\PY{p}{)}
            \PY{n}{rawoutput} \PY{o}{=} \PY{n}{sampleset}\PY{o}{.}\PY{n}{aggregate}\PY{p}{(}\PY{p}{)}

        \PY{n}{minimumevalue} \PY{o}{=} \PY{l+m+mf}{100.0}
        \PY{n}{minimuma} \PY{o}{=} \PY{p}{[}\PY{p}{]}
        \PY{n}{minimumunita} \PY{o}{=} \PY{p}{[}\PY{p}{]}
        \PY{n}{warning} \PY{o}{=} \PY{l+m+mi}{0}
        \PY{n}{chaincount} \PY{o}{=} \PY{l+m+mi}{0}

        \PY{k}{for} \PY{n}{irow} \PY{o+ow}{in} \PY{n+nb}{range}\PY{p}{(}\PY{n+nb}{len}\PY{p}{(}\PY{n}{rawoutput}\PY{o}{.}\PY{n}{record}\PY{p}{)}\PY{p}{)}\PY{p}{:}
            \PY{k}{if} \PY{n}{quantum}\PY{o}{==}\PY{l+m+mi}{1}\PY{p}{:}
                \PY{n}{chain} \PY{o}{=} \PY{n}{rawoutput}\PY{o}{.}\PY{n}{record}\PY{p}{[}\PY{n}{irow}\PY{p}{]}\PY{p}{[}\PY{l+m+mi}{3}\PY{p}{]}
            \PY{n}{numoc} \PY{o}{=} \PY{n}{rawoutput}\PY{o}{.}\PY{n}{record}\PY{p}{[}\PY{n}{irow}\PY{p}{]}\PY{p}{[}\PY{l+m+mi}{2}\PY{p}{]}
            \PY{n}{a} \PY{o}{=} \PY{p}{[}\PY{p}{]}
            \PY{k}{for} \PY{n}{alphaminus1} \PY{o+ow}{in} \PY{n+nb}{range}\PY{p}{(}\PY{n}{dimH}\PY{p}{)}\PY{p}{:}
                \PY{n}{a}\PY{o}{.}\PY{n}{append}\PY{p}{(}\PY{l+m+mi}{0}\PY{p}{)}
                \PY{k}{for} \PY{n}{kminus1} \PY{o+ow}{in} \PY{n+nb}{range}\PY{p}{(}\PY{n}{Kval}\PY{o}{\PYZhy{}}\PY{l+m+mi}{1}\PY{p}{)}\PY{p}{:}
                    \PY{n}{i} \PY{o}{=} \PY{n}{Kval}\PY{o}{*}\PY{n}{alphaminus1} \PY{o}{+} \PY{n}{kminus1}
                    \PY{n}{a}\PY{p}{[}\PY{n}{alphaminus1}\PY{p}{]} \PY{o}{+}\PY{o}{=} \PY{l+m+mi}{2}\PY{o}{*}\PY{o}{*}\PY{p}{(}\PY{l+m+mi}{1}\PY{o}{+}\PY{n}{kminus1}\PY{o}{\PYZhy{}}\PY{n}{Kval}\PY{o}{\PYZhy{}}\PY{n}{zoom}\PY{p}{)}\PY{o}{*}\PY{n}{rawoutput}\PY{o}{.}\PY{n}{record}\PY{p}{[}\PY{n}{irow}\PY{p}{]}\PY{p}{[}\PY{l+m+mi}{0}\PY{p}{]}\PY{p}{[}\PY{n}{i}\PY{p}{]}
                \PY{n}{i} \PY{o}{=} \PY{n}{myK}\PY{o}{*}\PY{n}{alphaminus1} \PY{o}{+} \PY{n}{myK} \PY{o}{\PYZhy{}} \PY{l+m+mi}{1}
                \PY{n}{a}\PY{p}{[}\PY{n}{alphaminus1}\PY{p}{]} \PY{o}{+}\PY{o}{=} \PY{n}{acenter}\PY{p}{[}\PY{n}{alphaminus1}\PY{p}{]}\PY{o}{\PYZhy{}}\PY{l+m+mi}{2}\PY{o}{*}\PY{o}{*}\PY{p}{(}\PY{o}{\PYZhy{}}\PY{n}{zoom}\PY{p}{)}\PY{o}{*}\PY{n}{rawoutput}\PY{o}{.}\PY{n}{record}\PY{p}{[}\PY{n}{irow}\PY{p}{]}\PY{p}{[}\PY{l+m+mi}{0}\PY{p}{]}\PY{p}{[}\PY{n}{i}\PY{p}{]}
            \PY{n}{anorm} \PY{o}{=} \PY{n}{np}\PY{o}{.}\PY{n}{sqrt}\PY{p}{(}\PY{n+nb}{sum}\PY{p}{(}\PY{n}{a}\PY{p}{[}\PY{n}{i}\PY{p}{]}\PY{o}{*}\PY{o}{*}\PY{l+m+mi}{2} \PY{k}{for} \PY{n}{i} \PY{o+ow}{in} \PY{n+nb}{range}\PY{p}{(}\PY{n}{dimH}\PY{p}{)}\PY{p}{)}\PY{p}{)}
            \PY{k}{if} \PY{n}{anorm}\PY{o}{\PYZlt{}}\PY{l+m+mf}{1.0e\PYZhy{}6}\PY{p}{:}
                \PY{n}{warning} \PY{o}{+}\PY{o}{=} \PY{n}{numoc}
            \PY{k}{else}\PY{p}{:}
                \PY{n}{unita} \PY{o}{=} \PY{p}{[}\PY{n}{a}\PY{p}{[}\PY{n}{i}\PY{p}{]}\PY{o}{/}\PY{n}{anorm} \PY{k}{for} \PY{n}{i} \PY{o+ow}{in} \PY{n+nb}{range}\PY{p}{(}\PY{n}{dimH}\PY{p}{)}\PY{p}{]}
                \PY{n}{evalue} \PY{o}{=} \PY{n}{np}\PY{o}{.}\PY{n}{matmul}\PY{p}{(}\PY{n}{unita}\PY{p}{,}\PY{n}{np}\PY{o}{.}\PY{n}{matmul}\PY{p}{(}\PY{n}{listham}\PY{p}{,}\PY{n}{unita}\PY{p}{)}\PY{p}{)}
                \PY{k}{if} \PY{n}{quantum}\PY{o}{==}\PY{l+m+mi}{1}\PY{p}{:}
                    \PY{k}{if} \PY{n}{chain}\PY{o}{\PYZgt{}}\PY{l+m+mf}{1.0e\PYZhy{}6}\PY{p}{:}
                        \PY{n}{chaincount} \PY{o}{+}\PY{o}{=} \PY{l+m+mi}{1}
                \PY{n}{minimumevalue} \PY{o}{=} \PY{n+nb}{min}\PY{p}{(}\PY{n}{evalue}\PY{p}{,}\PY{n}{minimumevalue}\PY{p}{)}
                \PY{k}{if} \PY{n}{evalue}\PY{o}{==}\PY{n}{minimumevalue}\PY{p}{:}
                    \PY{n}{minimuma} \PY{o}{=} \PY{n}{a}
                    \PY{n}{minimumunita} \PY{o}{=} \PY{n}{unita}
                    \PY{k}{if} \PY{n}{quantum}\PY{o}{==}\PY{l+m+mi}{1}\PY{p}{:}
                        \PY{n}{minimumchain} \PY{o}{=} \PY{n}{chain}
            
        \PY{n}{acenter} \PY{o}{=} \PY{n}{minimuma}

        \PY{n}{listen}\PY{o}{.}\PY{n}{append}\PY{p}{(}\PY{n}{np}\PY{o}{.}\PY{n}{matmul}\PY{p}{(}\PY{n}{minimumunita}\PY{p}{,}\PY{n}{np}\PY{o}{.}\PY{n}{matmul}\PY{p}{(}\PY{n}{listham}\PY{p}{,}\PY{n}{minimumunita}\PY{p}{)}\PY{p}{)}\PY{p}{)}
        \PY{n}{listvec}\PY{o}{.}\PY{n}{append}\PY{p}{(}\PY{n}{minimumunita}\PY{p}{)}

    \PY{k}{return}\PY{p}{(}\PY{n}{listen}\PY{p}{,}\PY{n}{listvec}\PY{p}{)}
\end{Verbatim}
\end{tcolorbox}

As an example of how these functions are used, we show the computation of the ground state and first-excited state of the HO with $n_s=16$, and then, with the corresponding interpolated wavefunctions, compute the same states for $n_s=32$ using the MG-AQAE solver.
First, the Hamiltonian for $n_s=16$ is loaded, 
and the embedding on the device determined:
\begin{tcolorbox}[breakable, size=fbox, boxrule=1pt, pad at break*=1mm,colback=cellbackground, colframe=cellborder,fontupper=\footnotesize]
\prompt{In}{incolor}{4}{\boxspacing}
\begin{Verbatim}[commandchars=\\\{\}]
\PY{n}{listham16} \PY{o}{=} \PY{n}{np}\PY{o}{.}\PY{n}{genfromtxt}\PY{p}{(}\PY{l+s+s1}{\PYZsq{}}\PY{l+s+s1}{HO\PYZus{}phimax5\PYZus{}nq4\PYZus{}ham.dat}\PY{l+s+s1}{\PYZsq{}}\PY{p}{)}
\PY{n}{embedding16} \PY{o}{=} \PY{n}{energy\PYZus{}embedding}\PY{p}{(}\PY{l+m+mi}{16}\PY{p}{,} \PY{n}{listham16}\PY{p}{,} \PY{l+m+mi}{3}\PY{p}{,} \PY{l+m+mf}{0.51}\PY{p}{,} \PY{n}{tokenval}\PY{p}{,} \PY{p}{[}\PY{p}{[}\PY{l+m+mi}{0}\PY{p}{]}\PY{o}{*}\PY{l+m+mi}{16}\PY{p}{]}\PY{p}{,} \PY{l+m+mi}{0}\PY{p}{)}
\end{Verbatim}
\end{tcolorbox}
Then, the ground-state energy and wavefunction 
are determined.
This is repeated $N_{\rm run}=20$ times as a way to quantify the systematic uncertainty in the results obtained from different calls to the machine.
\begin{tcolorbox}[breakable, size=fbox, boxrule=1pt, pad at break*=1mm,colback=cellbackground, colframe=cellborder,fontupper=\footnotesize]
\prompt{In}{incolor}{5}{\boxspacing}
\begin{Verbatim}[commandchars=\\\{\}]
\PY{n}{res16\PYZus{}results\PYZus{}n0} \PY{o}{=} \PY{p}{[}\PY{p}{]}
\PY{k}{for} \PY{n}{val} \PY{o+ow}{in} \PY{n+nb}{range}\PY{p}{(}\PY{l+m+mi}{20}\PY{p}{)}\PY{p}{:}
    \PY{n}{res16} \PY{o}{=} \PY{n}{energy\PYZus{}multigrid\PYZus{}project}\PY{p}{(}\PY{l+m+mi}{16}\PY{p}{,} \PY{n}{listham16}\PY{p}{,} \PY{l+m+mi}{3}\PY{p}{,} \PY{l+m+mf}{0.51}\PY{p}{,} \PY{l+m+mi}{1000}\PY{p}{,} \PY{l+m+mi}{15}\PY{p}{,} \PY{l+m+mi}{1}\PY{p}{,} \PY{n}{tokenval}\PY{p}{,} \PY{l+s+s1}{\PYZsq{}}\PY{l+s+s1}{16\PYZus{}n0}\PY{l+s+s1}{\PYZsq{}}\PY{p}{,} \PY{n}{embedding16}\PY{p}{,} \PY{l+m+mf}{0.2}\PY{p}{,} \PY{p}{[}\PY{l+m+mi}{0}\PY{p}{]}\PY{o}{*}\PY{l+m+mi}{16}\PY{p}{,} 
                                     \PY{l+m+mi}{0}\PY{p}{,}\PY{p}{[}\PY{p}{[}\PY{l+m+mi}{0}\PY{p}{]}\PY{o}{*}\PY{l+m+mi}{16}\PY{p}{]}\PY{p}{,} \PY{l+m+mi}{0}\PY{p}{)}
    \PY{n}{res16\PYZus{}results\PYZus{}n0}\PY{o}{.}\PY{n}{append}\PY{p}{(}\PY{n}{res16}\PY{p}{)}
\end{Verbatim}
\end{tcolorbox}
The energies and wavefunctions obtained at each zoom step are saved in different files, and the wavefunction associated with the corresponding minimum energy with respect to the $N_{\rm run}=20$ samples is saved in a file.
\begin{tcolorbox}[breakable, size=fbox, boxrule=1pt, pad at break*=1mm,colback=cellbackground, colframe=cellborder,fontupper=\footnotesize]
\prompt{In}{incolor}{6}{\boxspacing}
\begin{Verbatim}[commandchars=\\\{\}]
\PY{k}{for} \PY{n}{zoom} \PY{o+ow}{in} \PY{n+nb}{range}\PY{p}{(}\PY{l+m+mi}{15}\PY{p}{)}\PY{p}{:}
    \PY{n}{energy} \PY{o}{=} \PY{p}{[}\PY{n}{i}\PY{p}{[}\PY{l+m+mi}{0}\PY{p}{]}\PY{p}{[}\PY{n}{zoom}\PY{p}{]} \PY{k}{for} \PY{n}{i} \PY{o+ow}{in} \PY{n}{res16\PYZus{}results\PYZus{}n0}\PY{p}{]}
    \PY{n}{np}\PY{o}{.}\PY{n}{savetxt}\PY{p}{(}\PY{l+s+s1}{\PYZsq{}}\PY{l+s+s1}{HO\PYZus{}phimax5\PYZus{}nq4\PYZus{}eta0.51\PYZus{}z}\PY{l+s+s1}{\PYZsq{}}\PY{o}{+}\PY{n+nb}{str}\PY{p}{(}\PY{n}{zoom}\PY{p}{)}\PY{o}{+}\PY{l+s+s1}{\PYZsq{}}\PY{l+s+s1}{\PYZus{}K3\PYZus{}shots1000\PYZus{}state0\PYZus{}en.dat}\PY{l+s+s1}{\PYZsq{}}\PY{p}{,}\PY{n}{energy}\PY{p}{)}
    \PY{n}{vectors} \PY{o}{=} \PY{p}{[}\PY{n}{i}\PY{p}{[}\PY{l+m+mi}{1}\PY{p}{]}\PY{p}{[}\PY{n}{zoom}\PY{p}{]} \PY{k}{for} \PY{n}{i} \PY{o+ow}{in} \PY{n}{res16\PYZus{}results\PYZus{}n0}\PY{p}{]}
    \PY{n}{np}\PY{o}{.}\PY{n}{savetxt}\PY{p}{(}\PY{l+s+s1}{\PYZsq{}}\PY{l+s+s1}{HO\PYZus{}phimax5\PYZus{}nq4\PYZus{}eta0.51\PYZus{}z}\PY{l+s+s1}{\PYZsq{}}\PY{o}{+}\PY{n+nb}{str}\PY{p}{(}\PY{n}{zoom}\PY{p}{)}\PY{o}{+}\PY{l+s+s1}{\PYZsq{}}\PY{l+s+s1}{\PYZus{}K3\PYZus{}shots1000\PYZus{}state0\PYZus{}vec.dat}\PY{l+s+s1}{\PYZsq{}}\PY{p}{,}\PY{n}{vectors}\PY{p}{)}
\end{Verbatim}
\end{tcolorbox}
\begin{tcolorbox}[breakable, size=fbox, boxrule=1pt, pad at break*=1mm,colback=cellbackground, colframe=cellborder,fontupper=\footnotesize]
\prompt{In}{incolor}{7}{\boxspacing}
\begin{Verbatim}[commandchars=\\\{\}]
\PY{n}{listvecs} \PY{o}{=} \PY{p}{[}\PY{p}{]}
\PY{n}{energy} \PY{o}{=} \PY{p}{[}\PY{n}{i}\PY{p}{[}\PY{l+m+mi}{0}\PY{p}{]}\PY{p}{[}\PY{l+m+mi}{14}\PY{p}{]} \PY{k}{for} \PY{n}{i} \PY{o+ow}{in} \PY{n}{res16\PYZus{}results\PYZus{}n0}\PY{p}{]}
\PY{n}{locmin} \PY{o}{=} \PY{n}{energy}\PY{o}{.}\PY{n}{index}\PY{p}{(}\PY{n+nb}{min}\PY{p}{(}\PY{n}{energy}\PY{p}{)}\PY{p}{)}
\PY{n}{vectors} \PY{o}{=} \PY{p}{[}\PY{n}{i}\PY{p}{[}\PY{l+m+mi}{1}\PY{p}{]}\PY{p}{[}\PY{l+m+mi}{14}\PY{p}{]} \PY{k}{for} \PY{n}{i} \PY{o+ow}{in} \PY{n}{res16\PYZus{}results\PYZus{}n0}\PY{p}{]}
\PY{n}{eigenvector} \PY{o}{=} \PY{n}{vectors}\PY{p}{[}\PY{n}{locmin}\PY{p}{]}
\PY{n}{listvecs}\PY{o}{.}\PY{n}{append}\PY{p}{(}\PY{n}{eigenvector}\PY{p}{)}
\PY{n}{np}\PY{o}{.}\PY{n}{savetxt}\PY{p}{(}\PY{l+s+s1}{\PYZsq{}}\PY{l+s+s1}{HO\PYZus{}phimax5\PYZus{}nq4\PYZus{}1.dat}\PY{l+s+s1}{\PYZsq{}}\PY{p}{,} \PY{n}{listvecs}\PY{p}{)}
\end{Verbatim}
\end{tcolorbox}
The ground-state wavefunction is loaded,
and an offset in the ground-state energy by  $\mu=10$ is included so that the first-excited state can be determined.
\begin{tcolorbox}[breakable, size=fbox, boxrule=1pt, pad at break*=1mm,colback=cellbackground, colframe=cellborder,fontupper=\footnotesize]
\prompt{In}{incolor}{8}{\boxspacing}
\begin{Verbatim}[commandchars=\\\{\}]
\PY{n}{listvecs} \PY{o}{=} \PY{n}{np}\PY{o}{.}\PY{n}{loadtxt}\PY{p}{(}\PY{l+s+s1}{\PYZsq{}}\PY{l+s+s1}{HO\PYZus{}phimax5\PYZus{}nq4\PYZus{}1.dat}\PY{l+s+s1}{\PYZsq{}}\PY{p}{,}\PY{n}{ndmin}\PY{o}{=}\PY{l+m+mi}{2}\PY{p}{)}
\PY{n}{embedding16\PYZus{}proj1} \PY{o}{=} \PY{n}{energy\PYZus{}embedding}\PY{p}{(}\PY{l+m+mi}{16}\PY{p}{,} \PY{n}{listham16}\PY{p}{,} \PY{l+m+mi}{3}\PY{p}{,} \PY{l+m+mf}{1.51}\PY{p}{,} \PY{n}{tokenval}\PY{p}{,} \PY{n}{listvecs}\PY{p}{,} \PY{l+m+mi}{10}\PY{p}{)}
\end{Verbatim}
\end{tcolorbox}
Again, $N_{\rm run}=20$ samples are obtained for the first-excited state energy and wavefunction, with the results saved in separated files, along with the wavefunction with the minimum energy.
\begin{tcolorbox}[breakable, size=fbox, boxrule=1pt, pad at break*=1mm,colback=cellbackground, colframe=cellborder,fontupper=\footnotesize]
\prompt{In}{incolor}{9}{\boxspacing}
\begin{Verbatim}[commandchars=\\\{\}]
\PY{n}{res16\PYZus{}results\PYZus{}n1} \PY{o}{=} \PY{p}{[}\PY{p}{]}
\PY{k}{for} \PY{n}{val} \PY{o+ow}{in} \PY{n+nb}{range}\PY{p}{(}\PY{l+m+mi}{20}\PY{p}{)}\PY{p}{:}
    \PY{n}{res16} \PY{o}{=} \PY{n}{energy\PYZus{}multigrid\PYZus{}project}\PY{p}{(}\PY{l+m+mi}{16}\PY{p}{,} \PY{n}{listham16}\PY{p}{,} \PY{l+m+mi}{3}\PY{p}{,} \PY{l+m+mf}{1.51}\PY{p}{,} \PY{l+m+mi}{1000}\PY{p}{,} \PY{l+m+mi}{15}\PY{p}{,} \PY{l+m+mi}{1}\PY{p}{,} \PY{n}{tokenval}\PY{p}{,} \PY{l+s+s1}{\PYZsq{}}\PY{l+s+s1}{16\PYZus{}n1}\PY{l+s+s1}{\PYZsq{}}\PY{p}{,} \PY{n}{embedding16\PYZus{}proj1}\PY{p}{,} \PY{l+m+mf}{0.2}\PY{p}{,}
                                     \PY{p}{[}\PY{l+m+mi}{0}\PY{p}{]}\PY{o}{*}\PY{l+m+mi}{16}\PY{p}{,} \PY{l+m+mi}{0}\PY{p}{,} \PY{n}{listvecs}\PY{p}{,} \PY{l+m+mi}{10}\PY{p}{)}
    \PY{n}{res16\PYZus{}results\PYZus{}n1}\PY{o}{.}\PY{n}{append}\PY{p}{(}\PY{n}{res16}\PY{p}{)}
\end{Verbatim}
\end{tcolorbox}
\begin{tcolorbox}[breakable, size=fbox, boxrule=1pt, pad at break*=1mm,colback=cellbackground, colframe=cellborder,fontupper=\footnotesize]
\prompt{In}{incolor}{10}{\boxspacing}
\begin{Verbatim}[commandchars=\\\{\}]
\PY{k}{for} \PY{n}{zoom} \PY{o+ow}{in} \PY{n+nb}{range}\PY{p}{(}\PY{l+m+mi}{15}\PY{p}{)}\PY{p}{:}
    \PY{n}{energy} \PY{o}{=} \PY{p}{[}\PY{n}{i}\PY{p}{[}\PY{l+m+mi}{0}\PY{p}{]}\PY{p}{[}\PY{n}{zoom}\PY{p}{]} \PY{k}{for} \PY{n}{i} \PY{o+ow}{in} \PY{n}{res16\PYZus{}results\PYZus{}n1}\PY{p}{]}
    \PY{n}{np}\PY{o}{.}\PY{n}{savetxt}\PY{p}{(}\PY{l+s+s1}{\PYZsq{}}\PY{l+s+s1}{HO\PYZus{}phimax5\PYZus{}nq4\PYZus{}eta1.51\PYZus{}z}\PY{l+s+s1}{\PYZsq{}}\PY{o}{+}\PY{n+nb}{str}\PY{p}{(}\PY{n}{zoom}\PY{p}{)}\PY{o}{+}\PY{l+s+s1}{\PYZsq{}}\PY{l+s+s1}{\PYZus{}K3\PYZus{}shots1000\PYZus{}state1\PYZus{}en.dat}\PY{l+s+s1}{\PYZsq{}}\PY{p}{,}\PY{n}{energy}\PY{p}{)}
    \PY{n}{vectors} \PY{o}{=} \PY{p}{[}\PY{n}{i}\PY{p}{[}\PY{l+m+mi}{1}\PY{p}{]}\PY{p}{[}\PY{n}{zoom}\PY{p}{]} \PY{k}{for} \PY{n}{i} \PY{o+ow}{in} \PY{n}{res16\PYZus{}results\PYZus{}n1}\PY{p}{]}
    \PY{n}{np}\PY{o}{.}\PY{n}{savetxt}\PY{p}{(}\PY{l+s+s1}{\PYZsq{}}\PY{l+s+s1}{HO\PYZus{}phimax5\PYZus{}nq4\PYZus{}eta1.51\PYZus{}z}\PY{l+s+s1}{\PYZsq{}}\PY{o}{+}\PY{n+nb}{str}\PY{p}{(}\PY{n}{zoom}\PY{p}{)}\PY{o}{+}\PY{l+s+s1}{\PYZsq{}}\PY{l+s+s1}{\PYZus{}K3\PYZus{}shots1000\PYZus{}state1\PYZus{}vec.dat}\PY{l+s+s1}{\PYZsq{}}\PY{p}{,}\PY{n}{vectors}\PY{p}{)}
\end{Verbatim}
\end{tcolorbox}
\begin{tcolorbox}[breakable, size=fbox, boxrule=1pt, pad at break*=1mm,colback=cellbackground, colframe=cellborder,fontupper=\footnotesize]
\prompt{In}{incolor}{11}{\boxspacing}
\begin{Verbatim}[commandchars=\\\{\}]
\PY{n}{listvecs} \PY{o}{=} \PY{n}{np}\PY{o}{.}\PY{n}{loadtxt}\PY{p}{(}\PY{l+s+s1}{\PYZsq{}}\PY{l+s+s1}{HO\PYZus{}phimax5\PYZus{}nq4\PYZus{}1.dat}\PY{l+s+s1}{\PYZsq{}}\PY{p}{,}\PY{n}{ndmin}\PY{o}{=}\PY{l+m+mi}{2}\PY{p}{)}
\PY{n}{energy} \PY{o}{=} \PY{p}{[}\PY{n}{i}\PY{p}{[}\PY{l+m+mi}{0}\PY{p}{]}\PY{p}{[}\PY{l+m+mi}{14}\PY{p}{]} \PY{k}{for} \PY{n}{i} \PY{o+ow}{in} \PY{n}{res16\PYZus{}results\PYZus{}n1}\PY{p}{]}
\PY{n}{locmin} \PY{o}{=} \PY{n}{energy}\PY{o}{.}\PY{n}{index}\PY{p}{(}\PY{n+nb}{min}\PY{p}{(}\PY{n}{energy}\PY{p}{)}\PY{p}{)}
\PY{n}{vectors} \PY{o}{=} \PY{p}{[}\PY{n}{i}\PY{p}{[}\PY{l+m+mi}{1}\PY{p}{]}\PY{p}{[}\PY{l+m+mi}{14}\PY{p}{]} \PY{k}{for} \PY{n}{i} \PY{o+ow}{in} \PY{n}{res16\PYZus{}results\PYZus{}n1}\PY{p}{]}
\PY{n}{eigenvector} \PY{o}{=} \PY{n}{vectors}\PY{p}{[}\PY{n}{locmin}\PY{p}{]}
\PY{n}{listvecs} \PY{o}{=} \PY{n}{np}\PY{o}{.}\PY{n}{append}\PY{p}{(}\PY{n}{listvecs}\PY{p}{,}\PY{p}{[}\PY{n}{eigenvector}\PY{p}{]}\PY{p}{,}\PY{n}{axis}\PY{o}{=}\PY{l+m+mi}{0}\PY{p}{)}
\PY{n}{np}\PY{o}{.}\PY{n}{savetxt}\PY{p}{(}\PY{l+s+s1}{\PYZsq{}}\PY{l+s+s1}{HO\PYZus{}phimax5\PYZus{}nq4\PYZus{}2.dat}\PY{l+s+s1}{\PYZsq{}}\PY{p}{,} \PY{n}{listvecs}\PY{p}{)}
\end{Verbatim}
\end{tcolorbox}

To prepare for calculations in the 
 $n_s=32$ system, the $n_s=16$ wavefunction for the ground state is loaded, 
 and an interpolation from 16 points to 32 points is performed to provide starting values for the  $a_{\alpha}$ coefficients.
\begin{tcolorbox}[breakable, size=fbox, boxrule=1pt, pad at break*=1mm,colback=cellbackground, colframe=cellborder,fontupper=\footnotesize]
\prompt{In}{incolor}{12}{\boxspacing}
\begin{Verbatim}[commandchars=\\\{\}]
\PY{n}{listvecs} \PY{o}{=} \PY{p}{[}\PY{p}{]}
\PY{n}{acenterval\PYZus{}file} \PY{o}{=} \PY{n}{np}\PY{o}{.}\PY{n}{loadtxt}\PY{p}{(}\PY{l+s+s1}{\PYZsq{}}\PY{l+s+s1}{HO\PYZus{}phimax5\PYZus{}nq4\PYZus{}1.dat}\PY{l+s+s1}{\PYZsq{}}\PY{p}{,}\PY{n}{ndmin}\PY{o}{=}\PY{l+m+mi}{2}\PY{p}{)}
\end{Verbatim}
\end{tcolorbox}
\begin{tcolorbox}[breakable, size=fbox, boxrule=1pt, pad at break*=1mm,colback=cellbackground, colframe=cellborder,fontupper=\footnotesize]
\prompt{In}{incolor}{13}{\boxspacing}
\begin{Verbatim}[commandchars=\\\{\}]
\PY{n}{x16} \PY{o}{=} \PY{n}{np}\PY{o}{.}\PY{n}{arange}\PY{p}{(}\PY{o}{\PYZhy{}}\PY{l+m+mi}{5}\PY{p}{,}\PY{l+m+mi}{5}\PY{o}{+}\PY{l+m+mi}{9}\PY{o}{/}\PY{l+m+mi}{15}\PY{p}{,}\PY{l+m+mi}{10}\PY{o}{/}\PY{l+m+mi}{15}\PY{p}{)}
\PY{n}{tck16} \PY{o}{=} \PY{n}{scipy}\PY{o}{.}\PY{n}{interpolate}\PY{o}{.}\PY{n}{splrep}\PY{p}{(}\PY{n}{x16}\PY{p}{,} \PY{n}{acenterval\PYZus{}file}\PY{p}{[}\PY{l+m+mi}{-1}\PY{p}{]}\PY{p}{,} \PY{n}{s}\PY{o}{=}\PY{l+m+mi}{0}\PY{p}{)}
\PY{n}{x32} \PY{o}{=} \PY{n}{np}\PY{o}{.}\PY{n}{arange}\PY{p}{(}\PY{o}{\PYZhy{}}\PY{l+m+mi}{5}\PY{p}{,}\PY{l+m+mi}{5}\PY{o}{+}\PY{l+m+mi}{10}\PY{o}{/}\PY{l+m+mi}{31}\PY{p}{,}\PY{l+m+mi}{10}\PY{o}{/}\PY{l+m+mi}{31}\PY{p}{)}
\PY{n}{y32} \PY{o}{=} \PY{n}{scipy}\PY{o}{.}\PY{n}{interpolate}\PY{o}{.}\PY{n}{splev}\PY{p}{(}\PY{n}{x32}\PY{p}{,} \PY{n}{tck16}\PY{p}{,} \PY{n}{der}\PY{o}{=}\PY{l+m+mi}{0}\PY{p}{)}
\PY{n}{y32norm} \PY{o}{=} \PY{n}{np}\PY{o}{.}\PY{n}{sqrt}\PY{p}{(}\PY{n+nb}{sum}\PY{p}{(}\PY{n}{y32}\PY{p}{[}\PY{n}{i}\PY{p}{]}\PY{o}{*}\PY{o}{*}\PY{l+m+mi}{2} \PY{k}{for} \PY{n}{i} \PY{o+ow}{in} \PY{n+nb}{range}\PY{p}{(}\PY{l+m+mi}{32}\PY{p}{)}\PY{p}{)}\PY{p}{)}
\PY{n}{unity32} \PY{o}{=} \PY{p}{[}\PY{n}{y32}\PY{p}{[}\PY{n}{i}\PY{p}{]}\PY{o}{/}\PY{n}{y32norm} \PY{k}{for} \PY{n}{i} \PY{o+ow}{in} \PY{n+nb}{range}\PY{p}{(}\PY{l+m+mi}{32}\PY{p}{)}\PY{p}{]}
\end{Verbatim}
\end{tcolorbox}
The embedding for the $n_s=32$ system is determined:
\begin{tcolorbox}[breakable, size=fbox, boxrule=1pt, pad at break*=1mm,colback=cellbackground, colframe=cellborder,fontupper=\footnotesize]
\prompt{In}{incolor}{14}{\boxspacing}
\begin{Verbatim}[commandchars=\\\{\}]
\PY{n}{listham32} \PY{o}{=} \PY{n}{np}\PY{o}{.}\PY{n}{genfromtxt}\PY{p}{(}\PY{l+s+s1}{\PYZsq{}}\PY{l+s+s1}{HO\PYZus{}phimax5\PYZus{}nq5\PYZus{}ham.dat}\PY{l+s+s1}{\PYZsq{}}\PY{p}{)}
\PY{n}{embedding32} \PY{o}{=} \PY{n}{energy\PYZus{}embedding}\PY{p}{(}\PY{l+m+mi}{32}\PY{p}{,} \PY{n}{listham32}\PY{p}{,} \PY{l+m+mi}{3}\PY{p}{,} \PY{l+m+mf}{0.51}\PY{p}{,} \PY{n}{tokenval}\PY{p}{,} \PY{p}{[}\PY{p}{[}\PY{l+m+mi}{0}\PY{p}{]}\PY{o}{*}\PY{l+m+mi}{32}\PY{p}{]}\PY{p}{,} \PY{l+m+mi}{0}\PY{p}{)}
\end{Verbatim}
\end{tcolorbox}
As in the case of $n_s=16$, $N_{\rm run}=20$ samples are obtained for the ground-state energy and wavefunction of the $n_s=32$ system. 
Notice that  $z^{\rm init}=8$ is used to narrow the window of  values that $a_{\alpha}$ can take, and $z^{\rm max}+1=15+8$, so that 14 levels of zoom are still applied. 
The next steps are similar to the $n_s=16$ case.
\begin{tcolorbox}[breakable, size=fbox, boxrule=1pt, pad at break*=1mm,colback=cellbackground, colframe=cellborder,fontupper=\footnotesize]
\prompt{In}{incolor}{15}{\boxspacing}
\begin{Verbatim}[commandchars=\\\{\}]
\PY{n}{res32\PYZus{}results\PYZus{}n0} \PY{o}{=} \PY{p}{[}\PY{p}{]}
\PY{k}{for} \PY{n}{val} \PY{o+ow}{in} \PY{n}{np}\PY{o}{.}\PY{n}{arange}\PY{p}{(}\PY{l+m+mi}{20}\PY{p}{)}\PY{p}{:}
    \PY{n}{res32} \PY{o}{=} \PY{n}{energy\PYZus{}multigrid\PYZus{}project}\PY{p}{(}\PY{l+m+mi}{32}\PY{p}{,} \PY{n}{listham32}\PY{p}{,} \PY{l+m+mi}{3}\PY{p}{,} \PY{l+m+mf}{0.51}\PY{p}{,} \PY{l+m+mi}{1000}\PY{p}{,} \PY{l+m+mi}{15}\PY{o}{+}\PY{l+m+mi}{8}\PY{p}{,} \PY{l+m+mi}{1}\PY{p}{,} \PY{n}{tokenval}\PY{p}{,} \PY{l+s+s1}{\PYZsq{}}\PY{l+s+s1}{32\PYZus{}n0}\PY{l+s+s1}{\PYZsq{}}\PY{p}{,} \PY{n}{embedding32}\PY{p}{,} \PY{l+m+mf}{0.2}\PY{p}{,} 
                                     \PY{n}{unity32}\PY{p}{,} \PY{l+m+mi}{8}\PY{p}{,} \PY{p}{[}\PY{p}{[}\PY{l+m+mi}{0}\PY{p}{]}\PY{o}{*}\PY{l+m+mi}{16}\PY{p}{]}\PY{p}{,} \PY{l+m+mi}{0}\PY{p}{)}
    \PY{n}{res32\PYZus{}results\PYZus{}n0}\PY{o}{.}\PY{n}{append}\PY{p}{(}\PY{n}{res32}\PY{p}{)}
\end{Verbatim}
\end{tcolorbox}
\begin{tcolorbox}[breakable, size=fbox, boxrule=1pt, pad at break*=1mm,colback=cellbackground, colframe=cellborder,fontupper=\footnotesize]
\prompt{In}{incolor}{16}{\boxspacing}
\begin{Verbatim}[commandchars=\\\{\}]
\PY{k}{for} \PY{n}{zoom} \PY{o+ow}{in} \PY{n+nb}{range}\PY{p}{(}\PY{l+m+mi}{15}\PY{p}{)}\PY{p}{:}
    \PY{n}{energy} \PY{o}{=} \PY{p}{[}\PY{n}{i}\PY{p}{[}\PY{l+m+mi}{0}\PY{p}{]}\PY{p}{[}\PY{n}{zoom}\PY{p}{]} \PY{k}{for} \PY{n}{i} \PY{o+ow}{in} \PY{n}{res32\PYZus{}results\PYZus{}n0}\PY{p}{]}
    \PY{n}{np}\PY{o}{.}\PY{n}{savetxt}\PY{p}{(}\PY{l+s+s1}{\PYZsq{}}\PY{l+s+s1}{HO\PYZus{}phimax5\PYZus{}nq5\PYZus{}eta0.51\PYZus{}z}\PY{l+s+s1}{\PYZsq{}}\PY{o}{+}\PY{n+nb}{str}\PY{p}{(}\PY{n}{zoom}\PY{p}{)}\PY{o}{+}\PY{l+s+s1}{\PYZsq{}}\PY{l+s+s1}{\PYZus{}K3\PYZus{}shots1000\PYZus{}multigrid\PYZus{}zinit8\PYZus{}state0\PYZus{}en.dat}\PY{l+s+s1}{\PYZsq{}}\PY{p}{,}\PY{n}{energy}\PY{p}{)}
    \PY{n}{vectors} \PY{o}{=} \PY{p}{[}\PY{n}{i}\PY{p}{[}\PY{l+m+mi}{1}\PY{p}{]}\PY{p}{[}\PY{n}{zoom}\PY{p}{]} \PY{k}{for} \PY{n}{i} \PY{o+ow}{in} \PY{n}{res32\PYZus{}results\PYZus{}n0}\PY{p}{]}
    \PY{n}{np}\PY{o}{.}\PY{n}{savetxt}\PY{p}{(}\PY{l+s+s1}{\PYZsq{}}\PY{l+s+s1}{HO\PYZus{}phimax5\PYZus{}nq5\PYZus{}eta0.51\PYZus{}z}\PY{l+s+s1}{\PYZsq{}}\PY{o}{+}\PY{n+nb}{str}\PY{p}{(}\PY{n}{zoom}\PY{p}{)}\PY{o}{+}\PY{l+s+s1}{\PYZsq{}}\PY{l+s+s1}{\PYZus{}K3\PYZus{}shots1000\PYZus{}multigrid\PYZus{}zinit8\PYZus{}state0\PYZus{}vec.dat}\PY{l+s+s1}{\PYZsq{}}\PY{p}{,}\PY{n}{vectors}\PY{p}{)}
\end{Verbatim}
\end{tcolorbox}
\begin{tcolorbox}[breakable, size=fbox, boxrule=1pt, pad at break*=1mm,colback=cellbackground, colframe=cellborder,fontupper=\footnotesize]
\prompt{In}{incolor}{17}{\boxspacing}
\begin{Verbatim}[commandchars=\\\{\}]
\PY{n}{listvecs} \PY{o}{=} \PY{p}{[}\PY{p}{]}
\PY{n}{energy} \PY{o}{=} \PY{p}{[}\PY{n}{i}\PY{p}{[}\PY{l+m+mi}{0}\PY{p}{]}\PY{p}{[}\PY{l+m+mi}{14}\PY{p}{]} \PY{k}{for} \PY{n}{i} \PY{o+ow}{in} \PY{n}{res32\PYZus{}results\PYZus{}n0}\PY{p}{]}
\PY{n}{locmin} \PY{o}{=} \PY{n}{energy}\PY{o}{.}\PY{n}{index}\PY{p}{(}\PY{n+nb}{min}\PY{p}{(}\PY{n}{energy}\PY{p}{)}\PY{p}{)}
\PY{n}{vectors} \PY{o}{=} \PY{p}{[}\PY{n}{i}\PY{p}{[}\PY{l+m+mi}{1}\PY{p}{]}\PY{p}{[}\PY{l+m+mi}{14}\PY{p}{]} \PY{k}{for} \PY{n}{i} \PY{o+ow}{in} \PY{n}{res32\PYZus{}results\PYZus{}n0}\PY{p}{]}
\PY{n}{eigenvector} \PY{o}{=} \PY{n}{vectors}\PY{p}{[}\PY{n}{locmin}\PY{p}{]}
\PY{n}{listvecs}\PY{o}{.}\PY{n}{append}\PY{p}{(}\PY{n}{eigenvector}\PY{p}{)}
\PY{n}{np}\PY{o}{.}\PY{n}{savetxt}\PY{p}{(}\PY{l+s+s1}{\PYZsq{}}\PY{l+s+s1}{HO\PYZus{}phimax5\PYZus{}nq5\PYZus{}1.dat}\PY{l+s+s1}{\PYZsq{}}\PY{p}{,} \PY{n}{listvecs}\PY{p}{)}
\end{Verbatim}
\end{tcolorbox}
The same process is repeated for the first-excited state:
\begin{tcolorbox}[breakable, size=fbox, boxrule=1pt, pad at break*=1mm,colback=cellbackground, colframe=cellborder,fontupper=\footnotesize]
\prompt{In}{incolor}{18}{\boxspacing}
\begin{Verbatim}[commandchars=\\\{\}]
\PY{n}{listvecs} \PY{o}{=} \PY{n}{np}\PY{o}{.}\PY{n}{loadtxt}\PY{p}{(}\PY{l+s+s1}{\PYZsq{}}\PY{l+s+s1}{HO\PYZus{}phimax5\PYZus{}nq5\PYZus{}1.dat}\PY{l+s+s1}{\PYZsq{}}\PY{p}{,}\PY{n}{ndmin}\PY{o}{=}\PY{l+m+mi}{2}\PY{p}{)}
\PY{n}{acenterval\PYZus{}file} \PY{o}{=} \PY{n}{np}\PY{o}{.}\PY{n}{loadtxt}\PY{p}{(}\PY{l+s+s1}{\PYZsq{}}\PY{l+s+s1}{HO\PYZus{}phimax5\PYZus{}nq4\PYZus{}2.dat}\PY{l+s+s1}{\PYZsq{}}\PY{p}{,}\PY{n}{ndmin}\PY{o}{=}\PY{l+m+mi}{2}\PY{p}{)}
\end{Verbatim}
\end{tcolorbox}
\begin{tcolorbox}[breakable, size=fbox, boxrule=1pt, pad at break*=1mm,colback=cellbackground, colframe=cellborder,fontupper=\footnotesize]
\prompt{In}{incolor}{19}{\boxspacing}
\begin{Verbatim}[commandchars=\\\{\}]
\PY{n}{x16} \PY{o}{=} \PY{n}{np}\PY{o}{.}\PY{n}{arange}\PY{p}{(}\PY{o}{\PYZhy{}}\PY{l+m+mi}{5}\PY{p}{,}\PY{l+m+mi}{5}\PY{o}{+}\PY{l+m+mi}{9}\PY{o}{/}\PY{l+m+mi}{15}\PY{p}{,}\PY{l+m+mi}{10}\PY{o}{/}\PY{l+m+mi}{15}\PY{p}{)}
\PY{n}{tck16} \PY{o}{=} \PY{n}{scipy}\PY{o}{.}\PY{n}{interpolate}\PY{o}{.}\PY{n}{splrep}\PY{p}{(}\PY{n}{x16}\PY{p}{,} \PY{n}{acenterval\PYZus{}file}\PY{p}{[}\PY{l+m+mi}{-1}\PY{p}{]}\PY{p}{,} \PY{n}{s}\PY{o}{=}\PY{l+m+mi}{0}\PY{p}{)}
\PY{n}{x32} \PY{o}{=} \PY{n}{np}\PY{o}{.}\PY{n}{arange}\PY{p}{(}\PY{o}{\PYZhy{}}\PY{l+m+mi}{5}\PY{p}{,}\PY{l+m+mi}{5}\PY{o}{+}\PY{l+m+mi}{10}\PY{o}{/}\PY{l+m+mi}{31}\PY{p}{,}\PY{l+m+mi}{10}\PY{o}{/}\PY{l+m+mi}{31}\PY{p}{)}
\PY{n}{y32} \PY{o}{=} \PY{n}{scipy}\PY{o}{.}\PY{n}{interpolate}\PY{o}{.}\PY{n}{splev}\PY{p}{(}\PY{n}{x32}\PY{p}{,} \PY{n}{tck16}\PY{p}{,} \PY{n}{der}\PY{o}{=}\PY{l+m+mi}{0}\PY{p}{)}
\PY{n}{y32norm} \PY{o}{=} \PY{n}{np}\PY{o}{.}\PY{n}{sqrt}\PY{p}{(}\PY{n+nb}{sum}\PY{p}{(}\PY{n}{y32}\PY{p}{[}\PY{n}{i}\PY{p}{]}\PY{o}{*}\PY{o}{*}\PY{l+m+mi}{2} \PY{k}{for} \PY{n}{i} \PY{o+ow}{in} \PY{n+nb}{range}\PY{p}{(}\PY{l+m+mi}{32}\PY{p}{)}\PY{p}{)}\PY{p}{)}
\PY{n}{unity32\PYZus{}proj1} \PY{o}{=} \PY{p}{[}\PY{n}{y32}\PY{p}{[}\PY{n}{i}\PY{p}{]}\PY{o}{/}\PY{n}{y32norm} \PY{k}{for} \PY{n}{i} \PY{o+ow}{in} \PY{n+nb}{range}\PY{p}{(}\PY{l+m+mi}{32}\PY{p}{)}\PY{p}{]}
\end{Verbatim}
\end{tcolorbox}
\begin{tcolorbox}[breakable, size=fbox, boxrule=1pt, pad at break*=1mm,colback=cellbackground, colframe=cellborder,fontupper=\footnotesize]
\prompt{In}{incolor}{20}{\boxspacing}
\begin{Verbatim}[commandchars=\\\{\}]
\PY{n}{embedding32\PYZus{}proj1} \PY{o}{=} \PY{n}{energy\PYZus{}embedding}\PY{p}{(}\PY{l+m+mi}{32}\PY{p}{,} \PY{n}{listham32}\PY{p}{,} \PY{l+m+mi}{3}\PY{p}{,} \PY{l+m+mi}{1.51}\PY{p}{,} \PY{n}{tokenval}\PY{p}{,} \PY{n}{listvecs}\PY{p}{,} \PY{l+m+mi}{10}\PY{p}{)}
\end{Verbatim}
\end{tcolorbox}
\begin{tcolorbox}[breakable, size=fbox, boxrule=1pt, pad at break*=1mm,colback=cellbackground, colframe=cellborder,fontupper=\footnotesize]
\prompt{In}{incolor}{21}{\boxspacing}
\begin{Verbatim}[commandchars=\\\{\}]
\PY{n}{res32\PYZus{}results\PYZus{}n1} \PY{o}{=} \PY{p}{[}\PY{p}{]}
\PY{k}{for} \PY{n}{val} \PY{o+ow}{in} \PY{n}{np}\PY{o}{.}\PY{n}{arange}\PY{p}{(}\PY{l+m+mi}{20}\PY{p}{)}\PY{p}{:}
    \PY{n}{res32} \PY{o}{=} \PY{n}{energy\PYZus{}multigrid\PYZus{}project}\PY{p}{(}\PY{l+m+mi}{32}\PY{p}{,} \PY{n}{listham32}\PY{p}{,} \PY{l+m+mi}{3}\PY{p}{,} \PY{l+m+mf}{1.51}\PY{p}{,} \PY{l+m+mi}{1000}\PY{p}{,} \PY{l+m+mi}{15}\PY{o}{+}\PY{l+m+mi}{8}\PY{p}{,} \PY{l+m+mi}{1}\PY{p}{,} \PY{n}{tokenval}\PY{p}{,} \PY{l+s+s1}{\PYZsq{}}\PY{l+s+s1}{32\PYZus{}n1}\PY{l+s+s1}{\PYZsq{}}\PY{p}{,} \PY{n}{embedding32\PYZus{}proj1}\PY{p}{,} \PY{l+m+mf}{0.2}\PY{p}{,}
                                     \PY{n}{unity32\PYZus{}proj1}\PY{p}{,} \PY{l+m+mi}{8}\PY{p}{,} \PY{n}{listvecs}\PY{p}{,} \PY{l+m+mi}{10}\PY{p}{)}
    \PY{n}{res32\PYZus{}results\PYZus{}n1}\PY{o}{.}\PY{n}{append}\PY{p}{(}\PY{n}{res32}\PY{p}{)}
\end{Verbatim}
\end{tcolorbox}
\begin{tcolorbox}[breakable, size=fbox, boxrule=1pt, pad at break*=1mm,colback=cellbackground, colframe=cellborder,fontupper=\footnotesize]
\prompt{In}{incolor}{22}{\boxspacing}
\begin{Verbatim}[commandchars=\\\{\}]
\PY{k}{for} \PY{n}{zoom} \PY{o+ow}{in} \PY{n+nb}{range}\PY{p}{(}\PY{l+m+mi}{15}\PY{p}{)}\PY{p}{:}
    \PY{n}{energy} \PY{o}{=} \PY{p}{[}\PY{n}{i}\PY{p}{[}\PY{l+m+mi}{0}\PY{p}{]}\PY{p}{[}\PY{n}{zoom}\PY{p}{]} \PY{k}{for} \PY{n}{i} \PY{o+ow}{in} \PY{n}{res32\PYZus{}results\PYZus{}n1}\PY{p}{]}
    \PY{n}{np}\PY{o}{.}\PY{n}{savetxt}\PY{p}{(}\PY{l+s+s1}{\PYZsq{}}\PY{l+s+s1}{HO\PYZus{}phimax5\PYZus{}nq5\PYZus{}eta1.51\PYZus{}z}\PY{l+s+s1}{\PYZsq{}}\PY{o}{+}\PY{n+nb}{str}\PY{p}{(}\PY{n}{zoom}\PY{p}{)}\PY{o}{+}\PY{l+s+s1}{\PYZsq{}}\PY{l+s+s1}{\PYZus{}K3\PYZus{}shots1000\PYZus{}multigrid\PYZus{}zinit8\PYZus{}state1\PYZus{}en.dat}\PY{l+s+s1}{\PYZsq{}}\PY{p}{,}\PY{n}{energy}\PY{p}{)}
    \PY{n}{vectors} \PY{o}{=} \PY{p}{[}\PY{n}{i}\PY{p}{[}\PY{l+m+mi}{1}\PY{p}{]}\PY{p}{[}\PY{n}{zoom}\PY{p}{]} \PY{k}{for} \PY{n}{i} \PY{o+ow}{in} \PY{n}{res32\PYZus{}results\PYZus{}n1}\PY{p}{]}
    \PY{n}{np}\PY{o}{.}\PY{n}{savetxt}\PY{p}{(}\PY{l+s+s1}{\PYZsq{}}\PY{l+s+s1}{HO\PYZus{}phimax5\PYZus{}nq5\PYZus{}eta1.51\PYZus{}z}\PY{l+s+s1}{\PYZsq{}}\PY{o}{+}\PY{n+nb}{str}\PY{p}{(}\PY{n}{zoom}\PY{p}{)}\PY{o}{+}\PY{l+s+s1}{\PYZsq{}}\PY{l+s+s1}{\PYZus{}K3\PYZus{}shots1000\PYZus{}multigrid\PYZus{}zinit8\PYZus{}state1\PYZus{}vec.dat}\PY{l+s+s1}{\PYZsq{}}\PY{p}{,}\PY{n}{vectors}\PY{p}{)}
\end{Verbatim}
\end{tcolorbox}
\begin{tcolorbox}[breakable, size=fbox, boxrule=1pt, pad at break*=1mm,colback=cellbackground, colframe=cellborder,fontupper=\footnotesize]
\prompt{In}{incolor}{23}{\boxspacing}
\begin{Verbatim}[commandchars=\\\{\}]
\PY{n}{listvecs} \PY{o}{=} \PY{n}{np}\PY{o}{.}\PY{n}{loadtxt}\PY{p}{(}\PY{l+s+s1}{\PYZsq{}}\PY{l+s+s1}{HO\PYZus{}phimax5\PYZus{}nq5\PYZus{}1.dat}\PY{l+s+s1}{\PYZsq{}}\PY{p}{,}\PY{n}{ndmin}\PY{o}{=}\PY{l+m+mi}{2}\PY{p}{)}
\PY{n}{energy} \PY{o}{=} \PY{p}{[}\PY{n}{i}\PY{p}{[}\PY{l+m+mi}{0}\PY{p}{]}\PY{p}{[}\PY{l+m+mi}{14}\PY{p}{]} \PY{k}{for} \PY{n}{i} \PY{o+ow}{in} \PY{n}{res32\PYZus{}results\PYZus{}n1}\PY{p}{]}
\PY{n}{locmin} \PY{o}{=} \PY{n}{energy}\PY{o}{.}\PY{n}{index}\PY{p}{(}\PY{n+nb}{min}\PY{p}{(}\PY{n}{energy}\PY{p}{)}\PY{p}{)}
\PY{n}{vectors} \PY{o}{=} \PY{p}{[}\PY{n}{i}\PY{p}{[}\PY{l+m+mi}{1}\PY{p}{]}\PY{p}{[}\PY{l+m+mi}{14}\PY{p}{]} \PY{k}{for} \PY{n}{i} \PY{o+ow}{in} \PY{n}{res32\PYZus{}results\PYZus{}n1}\PY{p}{]}
\PY{n}{eigenvector} \PY{o}{=} \PY{n}{vectors}\PY{p}{[}\PY{n}{locmin}\PY{p}{]}
\PY{n}{listvecs} \PY{o}{=} \PY{n}{np}\PY{o}{.}\PY{n}{append}\PY{p}{(}\PY{n}{listvecs}\PY{p}{,}\PY{p}{[}\PY{n}{eigenvector}\PY{p}{]}\PY{p}{,}\PY{n}{axis}\PY{o}{=}\PY{l+m+mi}{0}\PY{p}{)}
\PY{n}{np}\PY{o}{.}\PY{n}{savetxt}\PY{p}{(}\PY{l+s+s1}{\PYZsq{}}\PY{l+s+s1}{HO\PYZus{}phimax5\PYZus{}nq5\PYZus{}2.dat}\PY{l+s+s1}{\PYZsq{}}\PY{p}{,} \PY{n}{listvecs}\PY{p}{)}
\end{Verbatim}
\end{tcolorbox}
%

\subsection{Feynman clock implementation}
\noindent
Although the workflow for implementing the Feynman clock algorithm is  similar to the one detailed in the previous section, 
the QUBO matrix is different and the corresponding functions are modified.
The main difference is in the handling of the real and imaginary parts of the Hamiltonian ({\tt hamRe} and {\tt hamIm}):
\begin{tcolorbox}[breakable, size=fbox, boxrule=1pt, pad at break*=1mm,colback=cellbackground, colframe=cellborder,fontupper=\footnotesize]
\prompt{In}{incolor}{24}{\boxspacing}
\begin{Verbatim}[commandchars=\\\{\}]
\PY{k}{def} \PY{n+nf}{hermitian\PYZus{}embedding}\PY{p}{(}\PY{n}{dimH}\PY{p}{,} \PY{n}{hamRe}\PY{p}{,} \PY{n}{hamIm}\PY{p}{,} \PY{n}{Kval}\PY{p}{,} \PY{n}{etaval}\PY{p}{,} \PY{n}{tokenval}\PY{p}{)}\PY{p}{:}

    \PY{n}{Hlambdare} \PY{o}{=} \PY{n}{defaultdict}\PY{p}{(}\PY{n+nb}{float}\PY{p}{)}
    \PY{n}{Hlambdaim} \PY{o}{=} \PY{n}{defaultdict}\PY{p}{(}\PY{n+nb}{float}\PY{p}{)}
    
    \PY{k}{for} \PY{n}{num} \PY{o+ow}{in} \PY{n+nb}{range}\PY{p}{(}\PY{l+m+mi}{1}\PY{p}{,} \PY{n}{dimH}\PY{o}{+}\PY{l+m+mi}{1}\PY{p}{)}\PY{p}{:}
        \PY{n}{Hlambdare}\PY{p}{[}\PY{n}{num}\PY{p}{,}\PY{n}{num}\PY{p}{]} \PY{o}{\PYZhy{}}\PY{o}{=} \PY{n}{etaval}
    
    \PY{k}{for} \PY{n}{num1} \PY{o+ow}{in} \PY{n+nb}{range}\PY{p}{(}\PY{l+m+mi}{1}\PY{p}{,} \PY{n}{dimH}\PY{o}{+}\PY{l+m+mi}{1}\PY{p}{)}\PY{p}{:}
        \PY{k}{for} \PY{n}{num2} \PY{o+ow}{in} \PY{n+nb}{range}\PY{p}{(}\PY{l+m+mi}{1}\PY{p}{,} \PY{n}{dimH}\PY{o}{+}\PY{l+m+mi}{1}\PY{p}{)}\PY{p}{:}
            \PY{n}{Hlambdare}\PY{p}{[}\PY{n}{num1}\PY{p}{,}\PY{n}{num2}\PY{p}{]} \PY{o}{+}\PY{o}{=} \PY{n}{hamRe}\PY{p}{[}\PY{n}{num1}\PY{o}{\PYZhy{}}\PY{l+m+mi}{1}\PY{p}{,}\PY{n}{num2}\PY{o}{\PYZhy{}}\PY{l+m+mi}{1}\PY{p}{]}
            \PY{n}{Hlambdaim}\PY{p}{[}\PY{n}{num1}\PY{p}{,}\PY{n}{num2}\PY{p}{]} \PY{o}{+}\PY{o}{=} \PY{n}{hamIm}\PY{p}{[}\PY{n}{num1}\PY{o}{\PYZhy{}}\PY{l+m+mi}{1}\PY{p}{,}\PY{n}{num2}\PY{o}{\PYZhy{}}\PY{l+m+mi}{1}\PY{p}{]}
        
    \PY{n}{delta} \PY{o}{=} \PY{n}{np}\PY{o}{.}\PY{n}{zeros}\PY{p}{(}\PY{p}{(}\PY{n}{Kval}\PY{o}{+}\PY{l+m+mi}{1}\PY{p}{,}\PY{n}{Kval}\PY{o}{+}\PY{l+m+mi}{1}\PY{p}{)}\PY{p}{,}\PY{n+nb}{int}\PY{p}{)}
    \PY{n}{i} \PY{o}{=} \PY{n}{np}\PY{o}{.}\PY{n}{arange}\PY{p}{(}\PY{n}{Kval}\PY{o}{+}\PY{l+m+mi}{1}\PY{p}{)}
    \PY{n}{delta}\PY{p}{[}\PY{n}{i}\PY{p}{,}\PY{n}{i}\PY{p}{]} \PY{o}{=} \PY{l+m+mi}{1}
    
    \PY{n}{acenter} \PY{o}{=} \PY{p}{[}\PY{l+m+mi}{0}\PY{p}{]}\PY{o}{*}\PY{l+m+mi}{2}\PY{o}{*}\PY{n}{dimH}
    
    \PY{n}{zoom} \PY{o}{=} \PY{l+m+mi}{0}
    \PY{n}{Q} \PY{o}{=} \PY{n}{defaultdict}\PY{p}{(}\PY{n+nb}{float}\PY{p}{)}
    \PY{k}{for} \PY{n}{alpha} \PY{o+ow}{in} \PY{n+nb}{range}\PY{p}{(}\PY{l+m+mi}{1}\PY{p}{,}\PY{n}{dimH}\PY{o}{+}\PY{l+m+mi}{1}\PY{p}{)}\PY{p}{:}
        \PY{k}{for} \PY{n}{beta} \PY{o+ow}{in} \PY{n+nb}{range}\PY{p}{(}\PY{l+m+mi}{1}\PY{p}{,}\PY{n}{dimH}\PY{o}{+}\PY{l+m+mi}{1}\PY{p}{)}\PY{p}{:}
            \PY{k}{for} \PY{n}{i}\PY{p}{,}\PY{n}{j} \PY{o+ow}{in} \PY{n}{itertools}\PY{o}{.}\PY{n}{product}\PY{p}{(}\PY{n+nb}{range}\PY{p}{(}\PY{l+m+mi}{1}\PY{p}{,}\PY{l+m+mi}{2}\PY{o}{*}\PY{n}{Kval}\PY{o}{+}\PY{l+m+mi}{1}\PY{p}{)}\PY{p}{,}\PY{n+nb}{range}\PY{p}{(}\PY{l+m+mi}{1}\PY{p}{,}\PY{l+m+mi}{2}\PY{o}{*}\PY{n}{Kval}\PY{o}{+}\PY{l+m+mi}{1}\PY{p}{)}\PY{p}{)}\PY{p}{:}
                \PY{n}{n} \PY{o}{=} \PY{l+m+mi}{2}\PY{o}{*}\PY{n}{Kval}\PY{o}{*}\PY{p}{(}\PY{n}{alpha}\PY{o}{\PYZhy{}}\PY{l+m+mi}{1}\PY{p}{)} \PY{o}{+} \PY{n}{i}
                \PY{n}{m} \PY{o}{=} \PY{l+m+mi}{2}\PY{o}{*}\PY{n}{Kval}\PY{o}{*}\PY{p}{(}\PY{n}{beta}\PY{o}{\PYZhy{}}\PY{l+m+mi}{1}\PY{p}{)} \PY{o}{+} \PY{n}{j}
                \PY{k}{if} \PY{n}{i}\PY{o}{\PYZlt{}}\PY{o}{=}\PY{n}{Kval} \PY{o+ow}{and} \PY{n}{j}\PY{o}{\PYZlt{}}\PY{o}{=}\PY{n}{Kval}\PY{p}{:}
                    \PY{n}{Q}\PY{p}{[}\PY{n}{n}\PY{p}{,}\PY{n}{m}\PY{p}{]} \PY{o}{+}\PY{o}{=} \PY{n}{Hlambdare}\PY{p}{[}\PY{n}{alpha}\PY{p}{,}\PY{n}{beta}\PY{p}{]}\PY{o}{*}\PY{l+m+mi}{2}\PY{o}{*}\PY{o}{*}\PY{p}{(}\PY{n}{i}\PY{o}{+}\PY{n}{j}\PY{o}{\PYZhy{}}\PY{l+m+mi}{2}\PY{o}{*}\PY{n}{Kval}\PY{o}{\PYZhy{}}\PY{l+m+mi}{2}\PY{o}{*}\PY{n}{zoom}\PY{p}{)}\PY{o}{*}\PY{p}{(}\PY{o}{\PYZhy{}}\PY{l+m+mi}{1}\PY{p}{)}\PY{o}{*}\PY{o}{*}\PY{p}{(}\PY{n}{delta}\PY{p}{[}\PY{n}{i}\PY{p}{,}\PY{n}{Kval}\PY{p}{]}\PY{o}{+}\PY{n}{delta}\PY{p}{[}\PY{n}{j}\PY{p}{,}\PY{n}{Kval}\PY{p}{]}\PY{p}{)}
                    \PY{k}{if} \PY{n}{alpha}\PY{o}{==}\PY{n}{beta} \PY{o+ow}{and} \PY{n}{n}\PY{o}{==}\PY{n}{m}\PY{p}{:}
                        \PY{k}{for} \PY{n}{loop} \PY{o+ow}{in} \PY{n+nb}{range}\PY{p}{(}\PY{l+m+mi}{1}\PY{p}{,}\PY{n}{dimH}\PY{o}{+}\PY{l+m+mi}{1}\PY{p}{)}\PY{p}{:}
                            \PY{n}{Q}\PY{p}{[}\PY{n}{n}\PY{p}{,}\PY{n}{m}\PY{p}{]}\PY{o}{+}\PY{o}{=}\PY{l+m+mi}{2}\PY{o}{*}\PY{n}{acenter}\PY{p}{[}\PY{p}{(}\PY{n}{loop}\PY{o}{\PYZhy{}}\PY{l+m+mi}{1}\PY{p}{)}\PY{o}{*}\PY{l+m+mi}{2}\PY{p}{]}\PY{o}{*}\PY{l+m+mi}{2}\PY{o}{*}\PY{o}{*}\PY{p}{(}\PY{n}{i}\PY{o}{\PYZhy{}}\PY{n}{Kval}\PY{o}{\PYZhy{}}\PY{n}{zoom}\PY{p}{)}\PY{o}{*}\PY{n}{Hlambdare}\PY{p}{[}\PY{n}{loop}\PY{p}{,}\PY{n}{beta}\PY{p}{]}\PY{o}{*}\PY{p}{(}\PY{o}{\PYZhy{}}\PY{l+m+mi}{1}\PY{p}{)}\PY{o}{*}\PY{o}{*}\PY{p}{(}\PY{n}{delta}\PY{p}{[}\PY{n}{i}\PY{p}{,}\PY{n}{Kval}\PY{p}{]}\PY{p}{)}
                            \PY{n}{Q}\PY{p}{[}\PY{n}{n}\PY{p}{,}\PY{n}{m}\PY{p}{]}\PY{o}{\PYZhy{}}\PY{o}{=}\PY{n}{acenter}\PY{p}{[}\PY{p}{(}\PY{n}{loop}\PY{o}{\PYZhy{}}\PY{l+m+mi}{1}\PY{p}{)}\PY{o}{*}\PY{l+m+mi}{2}\PY{o}{+}\PY{l+m+mi}{1}\PY{p}{]}\PY{o}{*}\PY{l+m+mi}{2}\PY{o}{*}\PY{o}{*}\PY{p}{(}\PY{n}{i}\PY{o}{\PYZhy{}}\PY{n}{Kval}\PY{o}{\PYZhy{}}\PY{n}{zoom}\PY{p}{)}\PY{o}{*}\PY{n}{Hlambdaim}\PY{p}{[}\PY{n}{alpha}\PY{p}{,}\PY{n}{loop}\PY{p}{]}\PY{o}{*}\PY{p}{(}\PY{o}{\PYZhy{}}\PY{l+m+mi}{1}\PY{p}{)}\PY{o}{*}\PY{o}{*}\PY{p}{(}\PY{n}{delta}\PY{p}{[}\PY{n}{i}\PY{p}{,}\PY{n}{Kval}\PY{p}{]}\PY{p}{)}
                            \PY{n}{Q}\PY{p}{[}\PY{n}{n}\PY{p}{,}\PY{n}{m}\PY{p}{]}\PY{o}{+}\PY{o}{=}\PY{n}{acenter}\PY{p}{[}\PY{p}{(}\PY{n}{loop}\PY{o}{\PYZhy{}}\PY{l+m+mi}{1}\PY{p}{)}\PY{o}{*}\PY{l+m+mi}{2}\PY{o}{+}\PY{l+m+mi}{1}\PY{p}{]}\PY{o}{*}\PY{l+m+mi}{2}\PY{o}{*}\PY{o}{*}\PY{p}{(}\PY{n}{i}\PY{o}{\PYZhy{}}\PY{n}{Kval}\PY{o}{\PYZhy{}}\PY{n}{zoom}\PY{p}{)}\PY{o}{*}\PY{n}{Hlambdaim}\PY{p}{[}\PY{n}{loop}\PY{p}{,}\PY{n}{beta}\PY{p}{]}\PY{o}{*}\PY{p}{(}\PY{o}{\PYZhy{}}\PY{l+m+mi}{1}\PY{p}{)}\PY{o}{*}\PY{o}{*}\PY{p}{(}\PY{n}{delta}\PY{p}{[}\PY{n}{i}\PY{p}{,}\PY{n}{Kval}\PY{p}{]}\PY{p}{)}
                \PY{k}{elif} \PY{n}{i}\PY{o}{\PYZlt{}}\PY{o}{=}\PY{n}{Kval} \PY{o+ow}{and} \PY{n}{j}\PY{o}{\PYZgt{}}\PY{n}{Kval}\PY{p}{:}
                    \PY{n}{jj} \PY{o}{=} \PY{n}{j} \PY{o}{\PYZhy{}} \PY{n}{Kval}
                    \PY{n}{Q}\PY{p}{[}\PY{n}{n}\PY{p}{,}\PY{n}{m}\PY{p}{]} \PY{o}{\PYZhy{}}\PY{o}{=} \PY{n}{Hlambdaim}\PY{p}{[}\PY{n}{alpha}\PY{p}{,}\PY{n}{beta}\PY{p}{]}\PY{o}{*}\PY{l+m+mi}{2}\PY{o}{*}\PY{o}{*}\PY{p}{(}\PY{n}{i}\PY{o}{+}\PY{n}{jj}\PY{o}{\PYZhy{}}\PY{l+m+mi}{2}\PY{o}{*}\PY{n}{Kval}\PY{o}{\PYZhy{}}\PY{l+m+mi}{2}\PY{o}{*}\PY{n}{zoom}\PY{p}{)}\PY{o}{*}\PY{p}{(}\PY{o}{\PYZhy{}}\PY{l+m+mi}{1}\PY{p}{)}\PY{o}{*}\PY{o}{*}\PY{p}{(}\PY{n}{delta}\PY{p}{[}\PY{n}{i}\PY{p}{,}\PY{n}{Kval}\PY{p}{]}\PY{o}{+}\PY{n}{delta}\PY{p}{[}\PY{n}{jj}\PY{p}{,}\PY{n}{Kval}\PY{p}{]}\PY{p}{)}
                \PY{k}{elif} \PY{n}{n}\PY{o}{\PYZgt{}}\PY{n}{Kval} \PY{o+ow}{and} \PY{n}{m}\PY{o}{\PYZlt{}}\PY{o}{=}\PY{n}{Kval}\PY{p}{:}
                    \PY{n}{ii} \PY{o}{=} \PY{n}{i} \PY{o}{\PYZhy{}} \PY{n}{Kval}
                    \PY{n}{Q}\PY{p}{[}\PY{n}{n}\PY{p}{,}\PY{n}{m}\PY{p}{]} \PY{o}{+}\PY{o}{=} \PY{n}{Hlambdaim}\PY{p}{[}\PY{n}{alpha}\PY{p}{,}\PY{n}{beta}\PY{p}{]}\PY{o}{*}\PY{l+m+mi}{2}\PY{o}{*}\PY{o}{*}\PY{p}{(}\PY{n}{ii}\PY{o}{+}\PY{n}{j}\PY{o}{\PYZhy{}}\PY{l+m+mi}{2}\PY{o}{*}\PY{n}{Kval}\PY{o}{\PYZhy{}}\PY{l+m+mi}{2}\PY{o}{*}\PY{n}{zoom}\PY{p}{)}\PY{o}{*}\PY{p}{(}\PY{o}{\PYZhy{}}\PY{l+m+mi}{1}\PY{p}{)}\PY{o}{*}\PY{o}{*}\PY{p}{(}\PY{n}{delta}\PY{p}{[}\PY{n}{ii}\PY{p}{,}\PY{n}{Kval}\PY{p}{]}\PY{o}{+}\PY{n}{delta}\PY{p}{[}\PY{n}{j}\PY{p}{,}\PY{n}{Kval}\PY{p}{]}\PY{p}{)}
                \PY{k}{else}\PY{p}{:}
                    \PY{n}{ii} \PY{o}{=} \PY{n}{i} \PY{o}{\PYZhy{}} \PY{n}{Kval}
                    \PY{n}{jj} \PY{o}{=} \PY{n}{j} \PY{o}{\PYZhy{}} \PY{n}{Kval}                    
                    \PY{n}{Q}\PY{p}{[}\PY{n}{n}\PY{p}{,}\PY{n}{m}\PY{p}{]} \PY{o}{+}\PY{o}{=} \PY{n}{Hlambdare}\PY{p}{[}\PY{n}{alpha}\PY{p}{,}\PY{n}{beta}\PY{p}{]}\PY{o}{*}\PY{l+m+mi}{2}\PY{o}{*}\PY{o}{*}\PY{p}{(}\PY{n}{ii}\PY{o}{+}\PY{n}{jj}\PY{o}{\PYZhy{}}\PY{l+m+mi}{2}\PY{o}{*}\PY{n}{Kval}\PY{o}{\PYZhy{}}\PY{l+m+mi}{2}\PY{o}{*}\PY{n}{zoom}\PY{p}{)}\PY{o}{*}\PY{p}{(}\PY{o}{\PYZhy{}}\PY{l+m+mi}{1}\PY{p}{)}\PY{o}{*}\PY{o}{*}\PY{p}{(}\PY{n}{delta}\PY{p}{[}\PY{n}{ii}\PY{p}{,}\PY{n}{Kval}\PY{p}{]}\PY{o}{+}\PY{n}{delta}\PY{p}{[}\PY{n}{jj}\PY{p}{,}\PY{n}{Kval}\PY{p}{]}\PY{p}{)}
                    \PY{k}{if} \PY{n}{alpha}\PY{o}{==}\PY{n}{beta} \PY{o+ow}{and} \PY{n}{ii}\PY{o}{==}\PY{n}{jj}\PY{p}{:}
                        \PY{k}{for} \PY{n}{loop} \PY{o+ow}{in} \PY{n+nb}{range}\PY{p}{(}\PY{l+m+mi}{1}\PY{p}{,}\PY{n}{dimH}\PY{o}{+}\PY{l+m+mi}{1}\PY{p}{)}\PY{p}{:}
                            \PY{n}{Q}\PY{p}{[}\PY{n}{n}\PY{p}{,}\PY{n}{m}\PY{p}{]}\PY{o}{+}\PY{o}{=}\PY{l+m+mi}{2}\PY{o}{*}\PY{n}{acenter}\PY{p}{[}\PY{p}{(}\PY{n}{loop}\PY{o}{\PYZhy{}}\PY{l+m+mi}{1}\PY{p}{)}\PY{o}{*}\PY{l+m+mi}{2}\PY{o}{+}\PY{l+m+mi}{1}\PY{p}{]}\PY{o}{*}\PY{l+m+mi}{2}\PY{o}{*}\PY{o}{*}\PY{p}{(}\PY{n}{ii}\PY{o}{\PYZhy{}}\PY{n}{Kval}\PY{o}{\PYZhy{}}\PY{n}{zoom}\PY{p}{)}\PY{o}{*}\PY{n}{Hlambdare}\PY{p}{[}\PY{n}{loop}\PY{p}{,}\PY{n}{beta}\PY{p}{]}\PY{o}{*}\PY{p}{(}\PY{o}{\PYZhy{}}\PY{l+m+mi}{1}\PY{p}{)}\PY{o}{*}\PY{o}{*}\PY{p}{(}\PY{n}{delta}\PY{p}{[}\PY{n}{ii}\PY{p}{,}\PY{n}{Kval}\PY{p}{]}\PY{p}{)}
                            \PY{n}{Q}\PY{p}{[}\PY{n}{n}\PY{p}{,}\PY{n}{m}\PY{p}{]}\PY{o}{\PYZhy{}}\PY{o}{=}\PY{n}{acenter}\PY{p}{[}\PY{p}{(}\PY{n}{loop}\PY{o}{\PYZhy{}}\PY{l+m+mi}{1}\PY{p}{)}\PY{o}{*}\PY{l+m+mi}{2}\PY{p}{]}\PY{o}{*}\PY{l+m+mi}{2}\PY{o}{*}\PY{o}{*}\PY{p}{(}\PY{n}{ii}\PY{o}{\PYZhy{}}\PY{n}{Kval}\PY{o}{\PYZhy{}}\PY{n}{zoom}\PY{p}{)}\PY{o}{*}\PY{n}{Hlambdaim}\PY{p}{[}\PY{n}{loop}\PY{p}{,}\PY{n}{beta}\PY{p}{]}\PY{o}{*}\PY{p}{(}\PY{o}{\PYZhy{}}\PY{l+m+mi}{1}\PY{p}{)}\PY{o}{*}\PY{o}{*}\PY{p}{(}\PY{n}{delta}\PY{p}{[}\PY{n}{ii}\PY{p}{,}\PY{n}{Kval}\PY{p}{]}\PY{p}{)}
                            \PY{n}{Q}\PY{p}{[}\PY{n}{n}\PY{p}{,}\PY{n}{m}\PY{p}{]}\PY{o}{+}\PY{o}{=}\PY{n}{acenter}\PY{p}{[}\PY{p}{(}\PY{n}{loop}\PY{o}{\PYZhy{}}\PY{l+m+mi}{1}\PY{p}{)}\PY{o}{*}\PY{l+m+mi}{2}\PY{p}{]}\PY{o}{*}\PY{l+m+mi}{2}\PY{o}{*}\PY{o}{*}\PY{p}{(}\PY{n}{ii}\PY{o}{\PYZhy{}}\PY{n}{Kval}\PY{o}{\PYZhy{}}\PY{n}{zoom}\PY{p}{)}\PY{o}{*}\PY{n}{Hlambdaim}\PY{p}{[}\PY{n}{alpha}\PY{p}{,}\PY{n}{loop}\PY{p}{]}\PY{o}{*}\PY{p}{(}\PY{o}{\PYZhy{}}\PY{l+m+mi}{1}\PY{p}{)}\PY{o}{*}\PY{o}{*}\PY{p}{(}\PY{n}{delta}\PY{p}{[}\PY{n}{ii}\PY{p}{,}\PY{n}{Kval}\PY{p}{]}\PY{p}{)}
    
    \PY{n}{start} \PY{o}{=} \PY{n}{time}\PY{o}{.}\PY{n}{time}\PY{p}{(}\PY{p}{)}
    \PY{n}{sampler} \PY{o}{=} \PY{n}{DWaveSampler}\PY{p}{(}\PY{n}{solver}\PY{o}{=}\PY{p}{\PYZob{}}\PY{l+s+s1}{\PYZsq{}}\PY{l+s+s1}{topology\PYZus{}\PYZus{}type\PYZus{}\PYZus{}eq}\PY{l+s+s1}{\PYZsq{}}\PY{p}{:}\PY{l+s+s1}{\PYZsq{}}\PY{l+s+s1}{pegasus}\PY{l+s+s1}{\PYZsq{}}\PY{p}{\PYZcb{}}\PY{p}{,}\PY{n}{token}\PY{o}{=}\PY{n}{tokenval}\PY{p}{)}
    \PY{n}{embedding} \PY{o}{=} \PY{n}{find\PYZus{}embedding}\PY{p}{(}\PY{n}{Q}\PY{p}{,} \PY{n}{sampler}\PY{o}{.}\PY{n}{edgelist}\PY{p}{)}
    \PY{n}{end} \PY{o}{=} \PY{n}{time}\PY{o}{.}\PY{n}{time}\PY{p}{(}\PY{p}{)}
    \PY{n+nb}{print}\PY{p}{(}\PY{l+s+sa}{f}\PY{l+s+s2}{\PYZdq{}}\PY{l+s+s2}{Number of logical qubits: }\PY{l+s+si}{\PYZob{}}\PY{n+nb}{len}\PY{p}{(}\PY{n}{embedding}\PY{o}{.}\PY{n}{keys}\PY{p}{(}\PY{p}{)}\PY{p}{)}\PY{l+s+si}{\PYZcb{}}\PY{l+s+s2}{\PYZdq{}}\PY{p}{)}
    \PY{n+nb}{print}\PY{p}{(}\PY{l+s+sa}{f}\PY{l+s+s2}{\PYZdq{}}\PY{l+s+s2}{Number of physical qubits used in embedding: }\PY{l+s+si}{\PYZob{}}\PY{n+nb}{sum}\PY{p}{(}\PY{n+nb}{len}\PY{p}{(}\PY{n}{chain}\PY{p}{)} \PY{k}{for} \PY{n}{chain} \PY{o+ow}{in} \PY{n}{embedding}\PY{o}{.}\PY{n}{values}\PY{p}{(}\PY{p}{)}\PY{p}{)}\PY{l+s+si}{\PYZcb{}}\PY{l+s+s2}{\PYZdq{}}\PY{p}{)}
    \PY{n+nb}{print}\PY{p}{(}\PY{l+s+sa}{f}\PY{l+s+s2}{\PYZdq{}}\PY{l+s+s2}{Runtime of computing the embedding is }\PY{l+s+si}{\PYZob{}}\PY{n}{end} \PY{o}{\PYZhy{}} \PY{n}{start}\PY{l+s+si}{\PYZcb{}}\PY{l+s+s2}{\PYZdq{}}\PY{p}{)}
    \PY{k}{return}\PY{p}{(}\PY{n}{embedding}\PY{p}{)}
\end{Verbatim}
\end{tcolorbox}

\begin{tcolorbox}[breakable, size=fbox, boxrule=1pt, pad at break*=1mm,colback=cellbackground, colframe=cellborder,fontupper=\footnotesize]
\prompt{In}{incolor}{25}{\boxspacing}
\begin{Verbatim}[commandchars=\\\{\}]
\PY{k}{def} \PY{n+nf}{hermitian\PYZus{}energy}\PY{p}{(}\PY{n}{dimH}\PY{p}{,} \PY{n}{hamRe}\PY{p}{,} \PY{n}{hamIm}\PY{p}{,} \PY{n}{Kval}\PY{p}{,} \PY{n}{etaval}\PY{p}{,} \PY{n}{shots}\PY{p}{,} \PY{n}{zoommax}\PY{p}{,} \PY{n}{quantum}\PY{p}{,} \PY{n}{embedding}\PY{p}{,} \PY{n}{tokenval}\PY{p}{,} \PY{n}{chainval}\PY{p}{,} \PY{n}{jobid}\PY{p}{,}
                     \PY{n}{acenterval}\PY{p}{,} \PY{n}{zoomint}\PY{p}{)}\PY{p}{:}
    
    \PY{n}{Hlambdare} \PY{o}{=} \PY{n}{defaultdict}\PY{p}{(}\PY{n+nb}{float}\PY{p}{)}
    \PY{n}{Hlambdaim} \PY{o}{=} \PY{n}{defaultdict}\PY{p}{(}\PY{n+nb}{float}\PY{p}{)}
    
    \PY{k}{for} \PY{n}{num} \PY{o+ow}{in} \PY{n+nb}{range}\PY{p}{(}\PY{l+m+mi}{1}\PY{p}{,} \PY{n}{dimH}\PY{o}{+}\PY{l+m+mi}{1}\PY{p}{)}\PY{p}{:}
        \PY{n}{Hlambdare}\PY{p}{[}\PY{n}{num}\PY{p}{,}\PY{n}{num}\PY{p}{]} \PY{o}{\PYZhy{}}\PY{o}{=} \PY{n}{etaval}
    
    \PY{k}{for} \PY{n}{num1} \PY{o+ow}{in} \PY{n+nb}{range}\PY{p}{(}\PY{l+m+mi}{1}\PY{p}{,} \PY{n}{dimH}\PY{o}{+}\PY{l+m+mi}{1}\PY{p}{)}\PY{p}{:}
        \PY{k}{for} \PY{n}{num2} \PY{o+ow}{in} \PY{n+nb}{range}\PY{p}{(}\PY{l+m+mi}{1}\PY{p}{,} \PY{n}{dimH}\PY{o}{+}\PY{l+m+mi}{1}\PY{p}{)}\PY{p}{:}
            \PY{n}{Hlambdare}\PY{p}{[}\PY{n}{num1}\PY{p}{,}\PY{n}{num2}\PY{p}{]} \PY{o}{+}\PY{o}{=} \PY{n}{hamRe}\PY{p}{[}\PY{n}{num1}\PY{o}{\PYZhy{}}\PY{l+m+mi}{1}\PY{p}{,}\PY{n}{num2}\PY{o}{\PYZhy{}}\PY{l+m+mi}{1}\PY{p}{]}
            \PY{n}{Hlambdaim}\PY{p}{[}\PY{n}{num1}\PY{p}{,}\PY{n}{num2}\PY{p}{]} \PY{o}{+}\PY{o}{=} \PY{n}{hamIm}\PY{p}{[}\PY{n}{num1}\PY{o}{\PYZhy{}}\PY{l+m+mi}{1}\PY{p}{,}\PY{n}{num2}\PY{o}{\PYZhy{}}\PY{l+m+mi}{1}\PY{p}{]}
        
    \PY{n}{delta} \PY{o}{=} \PY{n}{np}\PY{o}{.}\PY{n}{zeros}\PY{p}{(}\PY{p}{(}\PY{n}{Kval}\PY{o}{+}\PY{l+m+mi}{1}\PY{p}{,}\PY{n}{Kval}\PY{o}{+}\PY{l+m+mi}{1}\PY{p}{)}\PY{p}{,}\PY{n+nb}{int}\PY{p}{)}
    \PY{n}{i} \PY{o}{=} \PY{n}{np}\PY{o}{.}\PY{n}{arange}\PY{p}{(}\PY{n}{Kval}\PY{o}{+}\PY{l+m+mi}{1}\PY{p}{)}
    \PY{n}{delta}\PY{p}{[}\PY{n}{i}\PY{p}{,}\PY{n}{i}\PY{p}{]} \PY{o}{=} \PY{l+m+mi}{1}
    
    \PY{n}{Cmat} \PY{o}{=} \PY{n}{np}\PY{o}{.}\PY{n}{zeros}\PY{p}{(}\PY{p}{(}\PY{n}{dimH}\PY{p}{,}\PY{n}{dimH}\PY{p}{)}\PY{p}{,}\PY{n}{dtype}\PY{o}{=}\PY{n}{np}\PY{o}{.}\PY{n}{complex\PYZus{}}\PY{p}{)}
    \PY{n}{Cmat} \PY{o}{=} \PY{n}{hamRe} \PY{o}{+} \PY{n}{hamIm} \PY{o}{*} \PY{l+m+mi}{1}\PY{n}{j}
    \PY{n}{Cvec} \PY{o}{=} \PY{n}{np}\PY{o}{.}\PY{n}{zeros}\PY{p}{(}\PY{n}{dimH}\PY{p}{,}\PY{n}{dtype}\PY{o}{=}\PY{n}{np}\PY{o}{.}\PY{n}{complex\PYZus{}}\PY{p}{)}
    
    \PY{n}{acenter} \PY{o}{=} \PY{n}{acenterval}
    
    \PY{n}{listen} \PY{o}{=} \PY{p}{[}\PY{p}{]}
    \PY{n}{listvec} \PY{o}{=} \PY{p}{[}\PY{p}{]}
    \PY{k}{for} \PY{n}{zoom} \PY{o+ow}{in} \PY{n+nb}{range}\PY{p}{(}\PY{n}{zoomint}\PY{p}{,}\PY{n}{zoommax}\PY{p}{)}\PY{p}{:}
        \PY{n}{Q} \PY{o}{=} \PY{n}{defaultdict}\PY{p}{(}\PY{n+nb}{float}\PY{p}{)}
        \PY{k}{for} \PY{n}{alpha} \PY{o+ow}{in} \PY{n+nb}{range}\PY{p}{(}\PY{l+m+mi}{1}\PY{p}{,}\PY{n}{dimH}\PY{o}{+}\PY{l+m+mi}{1}\PY{p}{)}\PY{p}{:}
        \PY{k}{for} \PY{n}{beta} \PY{o+ow}{in} \PY{n+nb}{range}\PY{p}{(}\PY{l+m+mi}{1}\PY{p}{,}\PY{n}{dimH}\PY{o}{+}\PY{l+m+mi}{1}\PY{p}{)}\PY{p}{:}
            \PY{k}{for} \PY{n}{i}\PY{p}{,}\PY{n}{j} \PY{o+ow}{in} \PY{n}{itertools}\PY{o}{.}\PY{n}{product}\PY{p}{(}\PY{n+nb}{range}\PY{p}{(}\PY{l+m+mi}{1}\PY{p}{,}\PY{l+m+mi}{2}\PY{o}{*}\PY{n}{Kval}\PY{o}{+}\PY{l+m+mi}{1}\PY{p}{)}\PY{p}{,}\PY{n+nb}{range}\PY{p}{(}\PY{l+m+mi}{1}\PY{p}{,}\PY{l+m+mi}{2}\PY{o}{*}\PY{n}{Kval}\PY{o}{+}\PY{l+m+mi}{1}\PY{p}{)}\PY{p}{)}\PY{p}{:}
                \PY{n}{n} \PY{o}{=} \PY{l+m+mi}{2}\PY{o}{*}\PY{n}{Kval}\PY{o}{*}\PY{p}{(}\PY{n}{alpha}\PY{o}{\PYZhy{}}\PY{l+m+mi}{1}\PY{p}{)} \PY{o}{+} \PY{n}{i}
                \PY{n}{m} \PY{o}{=} \PY{l+m+mi}{2}\PY{o}{*}\PY{n}{Kval}\PY{o}{*}\PY{p}{(}\PY{n}{beta}\PY{o}{\PYZhy{}}\PY{l+m+mi}{1}\PY{p}{)} \PY{o}{+} \PY{n}{j}
                \PY{k}{if} \PY{n}{i}\PY{o}{\PYZlt{}}\PY{o}{=}\PY{n}{Kval} \PY{o+ow}{and} \PY{n}{j}\PY{o}{\PYZlt{}}\PY{o}{=}\PY{n}{Kval}\PY{p}{:}
                    \PY{n}{Q}\PY{p}{[}\PY{n}{n}\PY{p}{,}\PY{n}{m}\PY{p}{]} \PY{o}{+}\PY{o}{=} \PY{n}{Hlambdare}\PY{p}{[}\PY{n}{alpha}\PY{p}{,}\PY{n}{beta}\PY{p}{]}\PY{o}{*}\PY{l+m+mi}{2}\PY{o}{*}\PY{o}{*}\PY{p}{(}\PY{n}{i}\PY{o}{+}\PY{n}{j}\PY{o}{\PYZhy{}}\PY{l+m+mi}{2}\PY{o}{*}\PY{n}{Kval}\PY{o}{\PYZhy{}}\PY{l+m+mi}{2}\PY{o}{*}\PY{n}{zoom}\PY{p}{)}\PY{o}{*}\PY{p}{(}\PY{o}{\PYZhy{}}\PY{l+m+mi}{1}\PY{p}{)}\PY{o}{*}\PY{o}{*}\PY{p}{(}\PY{n}{delta}\PY{p}{[}\PY{n}{i}\PY{p}{,}\PY{n}{Kval}\PY{p}{]}\PY{o}{+}\PY{n}{delta}\PY{p}{[}\PY{n}{j}\PY{p}{,}\PY{n}{Kval}\PY{p}{]}\PY{p}{)}
                    \PY{k}{if} \PY{n}{alpha}\PY{o}{==}\PY{n}{beta} \PY{o+ow}{and} \PY{n}{n}\PY{o}{==}\PY{n}{m}\PY{p}{:}
                        \PY{k}{for} \PY{n}{loop} \PY{o+ow}{in} \PY{n+nb}{range}\PY{p}{(}\PY{l+m+mi}{1}\PY{p}{,}\PY{n}{dimH}\PY{o}{+}\PY{l+m+mi}{1}\PY{p}{)}\PY{p}{:}
                            \PY{n}{Q}\PY{p}{[}\PY{n}{n}\PY{p}{,}\PY{n}{m}\PY{p}{]}\PY{o}{+}\PY{o}{=}\PY{l+m+mi}{2}\PY{o}{*}\PY{n}{acenter}\PY{p}{[}\PY{p}{(}\PY{n}{loop}\PY{o}{\PYZhy{}}\PY{l+m+mi}{1}\PY{p}{)}\PY{o}{*}\PY{l+m+mi}{2}\PY{p}{]}\PY{o}{*}\PY{l+m+mi}{2}\PY{o}{*}\PY{o}{*}\PY{p}{(}\PY{n}{i}\PY{o}{\PYZhy{}}\PY{n}{Kval}\PY{o}{\PYZhy{}}\PY{n}{zoom}\PY{p}{)}\PY{o}{*}\PY{n}{Hlambdare}\PY{p}{[}\PY{n}{loop}\PY{p}{,}\PY{n}{beta}\PY{p}{]}\PY{o}{*}\PY{p}{(}\PY{o}{\PYZhy{}}\PY{l+m+mi}{1}\PY{p}{)}\PY{o}{*}\PY{o}{*}\PY{p}{(}\PY{n}{delta}\PY{p}{[}\PY{n}{i}\PY{p}{,}\PY{n}{Kval}\PY{p}{]}\PY{p}{)}
                            \PY{n}{Q}\PY{p}{[}\PY{n}{n}\PY{p}{,}\PY{n}{m}\PY{p}{]}\PY{o}{\PYZhy{}}\PY{o}{=}\PY{n}{acenter}\PY{p}{[}\PY{p}{(}\PY{n}{loop}\PY{o}{\PYZhy{}}\PY{l+m+mi}{1}\PY{p}{)}\PY{o}{*}\PY{l+m+mi}{2}\PY{o}{+}\PY{l+m+mi}{1}\PY{p}{]}\PY{o}{*}\PY{l+m+mi}{2}\PY{o}{*}\PY{o}{*}\PY{p}{(}\PY{n}{i}\PY{o}{\PYZhy{}}\PY{n}{Kval}\PY{o}{\PYZhy{}}\PY{n}{zoom}\PY{p}{)}\PY{o}{*}\PY{n}{Hlambdaim}\PY{p}{[}\PY{n}{alpha}\PY{p}{,}\PY{n}{loop}\PY{p}{]}\PY{o}{*}\PY{p}{(}\PY{o}{\PYZhy{}}\PY{l+m+mi}{1}\PY{p}{)}\PY{o}{*}\PY{o}{*}\PY{p}{(}\PY{n}{delta}\PY{p}{[}\PY{n}{i}\PY{p}{,}\PY{n}{Kval}\PY{p}{]}\PY{p}{)}
                            \PY{n}{Q}\PY{p}{[}\PY{n}{n}\PY{p}{,}\PY{n}{m}\PY{p}{]}\PY{o}{+}\PY{o}{=}\PY{n}{acenter}\PY{p}{[}\PY{p}{(}\PY{n}{loop}\PY{o}{\PYZhy{}}\PY{l+m+mi}{1}\PY{p}{)}\PY{o}{*}\PY{l+m+mi}{2}\PY{o}{+}\PY{l+m+mi}{1}\PY{p}{]}\PY{o}{*}\PY{l+m+mi}{2}\PY{o}{*}\PY{o}{*}\PY{p}{(}\PY{n}{i}\PY{o}{\PYZhy{}}\PY{n}{Kval}\PY{o}{\PYZhy{}}\PY{n}{zoom}\PY{p}{)}\PY{o}{*}\PY{n}{Hlambdaim}\PY{p}{[}\PY{n}{loop}\PY{p}{,}\PY{n}{beta}\PY{p}{]}\PY{o}{*}\PY{p}{(}\PY{o}{\PYZhy{}}\PY{l+m+mi}{1}\PY{p}{)}\PY{o}{*}\PY{o}{*}\PY{p}{(}\PY{n}{delta}\PY{p}{[}\PY{n}{i}\PY{p}{,}\PY{n}{Kval}\PY{p}{]}\PY{p}{)}
                \PY{k}{elif} \PY{n}{i}\PY{o}{\PYZlt{}}\PY{o}{=}\PY{n}{Kval} \PY{o+ow}{and} \PY{n}{j}\PY{o}{\PYZgt{}}\PY{n}{Kval}\PY{p}{:}
                    \PY{n}{jj} \PY{o}{=} \PY{n}{j} \PY{o}{\PYZhy{}} \PY{n}{Kval}
                    \PY{n}{Q}\PY{p}{[}\PY{n}{n}\PY{p}{,}\PY{n}{m}\PY{p}{]} \PY{o}{\PYZhy{}}\PY{o}{=} \PY{n}{Hlambdaim}\PY{p}{[}\PY{n}{alpha}\PY{p}{,}\PY{n}{beta}\PY{p}{]}\PY{o}{*}\PY{l+m+mi}{2}\PY{o}{*}\PY{o}{*}\PY{p}{(}\PY{n}{i}\PY{o}{+}\PY{n}{jj}\PY{o}{\PYZhy{}}\PY{l+m+mi}{2}\PY{o}{*}\PY{n}{Kval}\PY{o}{\PYZhy{}}\PY{l+m+mi}{2}\PY{o}{*}\PY{n}{zoom}\PY{p}{)}\PY{o}{*}\PY{p}{(}\PY{o}{\PYZhy{}}\PY{l+m+mi}{1}\PY{p}{)}\PY{o}{*}\PY{o}{*}\PY{p}{(}\PY{n}{delta}\PY{p}{[}\PY{n}{i}\PY{p}{,}\PY{n}{Kval}\PY{p}{]}\PY{o}{+}\PY{n}{delta}\PY{p}{[}\PY{n}{jj}\PY{p}{,}\PY{n}{Kval}\PY{p}{]}\PY{p}{)}
                \PY{k}{elif} \PY{n}{n}\PY{o}{\PYZgt{}}\PY{n}{Kval} \PY{o+ow}{and} \PY{n}{m}\PY{o}{\PYZlt{}}\PY{o}{=}\PY{n}{Kval}\PY{p}{:}
                    \PY{n}{ii} \PY{o}{=} \PY{n}{i} \PY{o}{\PYZhy{}} \PY{n}{Kval}
                    \PY{n}{Q}\PY{p}{[}\PY{n}{n}\PY{p}{,}\PY{n}{m}\PY{p}{]} \PY{o}{+}\PY{o}{=} \PY{n}{Hlambdaim}\PY{p}{[}\PY{n}{alpha}\PY{p}{,}\PY{n}{beta}\PY{p}{]}\PY{o}{*}\PY{l+m+mi}{2}\PY{o}{*}\PY{o}{*}\PY{p}{(}\PY{n}{ii}\PY{o}{+}\PY{n}{j}\PY{o}{\PYZhy{}}\PY{l+m+mi}{2}\PY{o}{*}\PY{n}{Kval}\PY{o}{\PYZhy{}}\PY{l+m+mi}{2}\PY{o}{*}\PY{n}{zoom}\PY{p}{)}\PY{o}{*}\PY{p}{(}\PY{o}{\PYZhy{}}\PY{l+m+mi}{1}\PY{p}{)}\PY{o}{*}\PY{o}{*}\PY{p}{(}\PY{n}{delta}\PY{p}{[}\PY{n}{ii}\PY{p}{,}\PY{n}{Kval}\PY{p}{]}\PY{o}{+}\PY{n}{delta}\PY{p}{[}\PY{n}{j}\PY{p}{,}\PY{n}{Kval}\PY{p}{]}\PY{p}{)}
                \PY{k}{else}\PY{p}{:}
                    \PY{n}{ii} \PY{o}{=} \PY{n}{i} \PY{o}{\PYZhy{}} \PY{n}{Kval}
                    \PY{n}{jj} \PY{o}{=} \PY{n}{j} \PY{o}{\PYZhy{}} \PY{n}{Kval}                    
                    \PY{n}{Q}\PY{p}{[}\PY{n}{n}\PY{p}{,}\PY{n}{m}\PY{p}{]} \PY{o}{+}\PY{o}{=} \PY{n}{Hlambdare}\PY{p}{[}\PY{n}{alpha}\PY{p}{,}\PY{n}{beta}\PY{p}{]}\PY{o}{*}\PY{l+m+mi}{2}\PY{o}{*}\PY{o}{*}\PY{p}{(}\PY{n}{ii}\PY{o}{+}\PY{n}{jj}\PY{o}{\PYZhy{}}\PY{l+m+mi}{2}\PY{o}{*}\PY{n}{Kval}\PY{o}{\PYZhy{}}\PY{l+m+mi}{2}\PY{o}{*}\PY{n}{zoom}\PY{p}{)}\PY{o}{*}\PY{p}{(}\PY{o}{\PYZhy{}}\PY{l+m+mi}{1}\PY{p}{)}\PY{o}{*}\PY{o}{*}\PY{p}{(}\PY{n}{delta}\PY{p}{[}\PY{n}{ii}\PY{p}{,}\PY{n}{Kval}\PY{p}{]}\PY{o}{+}\PY{n}{delta}\PY{p}{[}\PY{n}{jj}\PY{p}{,}\PY{n}{Kval}\PY{p}{]}\PY{p}{)}
                    \PY{k}{if} \PY{n}{alpha}\PY{o}{==}\PY{n}{beta} \PY{o+ow}{and} \PY{n}{ii}\PY{o}{==}\PY{n}{jj}\PY{p}{:}
                        \PY{k}{for} \PY{n}{loop} \PY{o+ow}{in} \PY{n+nb}{range}\PY{p}{(}\PY{l+m+mi}{1}\PY{p}{,}\PY{n}{dimH}\PY{o}{+}\PY{l+m+mi}{1}\PY{p}{)}\PY{p}{:}
                            \PY{n}{Q}\PY{p}{[}\PY{n}{n}\PY{p}{,}\PY{n}{m}\PY{p}{]}\PY{o}{+}\PY{o}{=}\PY{l+m+mi}{2}\PY{o}{*}\PY{n}{acenter}\PY{p}{[}\PY{p}{(}\PY{n}{loop}\PY{o}{\PYZhy{}}\PY{l+m+mi}{1}\PY{p}{)}\PY{o}{*}\PY{l+m+mi}{2}\PY{o}{+}\PY{l+m+mi}{1}\PY{p}{]}\PY{o}{*}\PY{l+m+mi}{2}\PY{o}{*}\PY{o}{*}\PY{p}{(}\PY{n}{ii}\PY{o}{\PYZhy{}}\PY{n}{Kval}\PY{o}{\PYZhy{}}\PY{n}{zoom}\PY{p}{)}\PY{o}{*}\PY{n}{Hlambdare}\PY{p}{[}\PY{n}{loop}\PY{p}{,}\PY{n}{beta}\PY{p}{]}\PY{o}{*}\PY{p}{(}\PY{o}{\PYZhy{}}\PY{l+m+mi}{1}\PY{p}{)}\PY{o}{*}\PY{o}{*}\PY{p}{(}\PY{n}{delta}\PY{p}{[}\PY{n}{ii}\PY{p}{,}\PY{n}{Kval}\PY{p}{]}\PY{p}{)}
                            \PY{n}{Q}\PY{p}{[}\PY{n}{n}\PY{p}{,}\PY{n}{m}\PY{p}{]}\PY{o}{\PYZhy{}}\PY{o}{=}\PY{n}{acenter}\PY{p}{[}\PY{p}{(}\PY{n}{loop}\PY{o}{\PYZhy{}}\PY{l+m+mi}{1}\PY{p}{)}\PY{o}{*}\PY{l+m+mi}{2}\PY{p}{]}\PY{o}{*}\PY{l+m+mi}{2}\PY{o}{*}\PY{o}{*}\PY{p}{(}\PY{n}{ii}\PY{o}{\PYZhy{}}\PY{n}{Kval}\PY{o}{\PYZhy{}}\PY{n}{zoom}\PY{p}{)}\PY{o}{*}\PY{n}{Hlambdaim}\PY{p}{[}\PY{n}{loop}\PY{p}{,}\PY{n}{beta}\PY{p}{]}\PY{o}{*}\PY{p}{(}\PY{o}{\PYZhy{}}\PY{l+m+mi}{1}\PY{p}{)}\PY{o}{*}\PY{o}{*}\PY{p}{(}\PY{n}{delta}\PY{p}{[}\PY{n}{ii}\PY{p}{,}\PY{n}{Kval}\PY{p}{]}\PY{p}{)}
                            \PY{n}{Q}\PY{p}{[}\PY{n}{n}\PY{p}{,}\PY{n}{m}\PY{p}{]}\PY{o}{+}\PY{o}{=}\PY{n}{acenter}\PY{p}{[}\PY{p}{(}\PY{n}{loop}\PY{o}{\PYZhy{}}\PY{l+m+mi}{1}\PY{p}{)}\PY{o}{*}\PY{l+m+mi}{2}\PY{p}{]}\PY{o}{*}\PY{l+m+mi}{2}\PY{o}{*}\PY{o}{*}\PY{p}{(}\PY{n}{ii}\PY{o}{\PYZhy{}}\PY{n}{Kval}\PY{o}{\PYZhy{}}\PY{n}{zoom}\PY{p}{)}\PY{o}{*}\PY{n}{Hlambdaim}\PY{p}{[}\PY{n}{alpha}\PY{p}{,}\PY{n}{loop}\PY{p}{]}\PY{o}{*}\PY{p}{(}\PY{o}{\PYZhy{}}\PY{l+m+mi}{1}\PY{p}{)}\PY{o}{*}\PY{o}{*}\PY{p}{(}\PY{n}{delta}\PY{p}{[}\PY{n}{ii}\PY{p}{,}\PY{n}{Kval}\PY{p}{]}\PY{p}{)}
        
        \PY{k}{if} \PY{n}{quantum}\PY{o}{==}\PY{l+m+mi}{1}\PY{p}{:}
            \PY{n}{Q\PYZus{}values} \PY{o}{=} \PY{n}{Q}\PY{o}{.}\PY{n}{values}\PY{p}{(}\PY{p}{)}
            \PY{n}{chainstrength} \PY{o}{=} \PY{n}{chainval}\PY{o}{*}\PY{n+nb}{max}\PY{p}{(}\PY{n+nb}{max}\PY{p}{(}\PY{n}{Q\PYZus{}values}\PY{p}{)}\PY{p}{,}\PY{n+nb}{abs}\PY{p}{(}\PY{n+nb}{min}\PY{p}{(}\PY{n}{Q\PYZus{}values}\PY{p}{)}\PY{p}{)}\PY{p}{)}
            \PY{n}{sampler} \PY{o}{=} \PY{n}{FixedEmbeddingComposite}\PY{p}{(}\PY{n}{DWaveSampler}\PY{p}{(}\PY{n}{solver}\PY{o}{=}\PY{p}{\PYZob{}}\PY{l+s+s1}{\PYZsq{}}\PY{l+s+s1}{topology\PYZus{}\PYZus{}type\PYZus{}\PYZus{}eq}\PY{l+s+s1}{\PYZsq{}}\PY{p}{:}\PY{l+s+s1}{\PYZsq{}}\PY{l+s+s1}{pegasus}\PY{l+s+s1}{\PYZsq{}}\PY{p}{\PYZcb{}}\PY{p}{,}\PY{n}{token}\PY{o}{=}\PY{n}{tokenval}\PY{p}{)}\PY{p}{,}
                                              \PY{n}{embedding}\PY{p}{)}
            \PY{n}{sampleset} \PY{o}{=} \PY{n}{sampler}\PY{o}{.}\PY{n}{sample\PYZus{}qubo}\PY{p}{(}\PY{n}{Q}\PY{p}{,}\PY{n}{num\PYZus{}reads}\PY{o}{=}\PY{n}{shots}\PY{p}{,}\PY{n}{chain\PYZus{}strength}\PY{o}{=}\PY{n}{chainstrength}\PY{p}{,}\PY{n}{annealing\PYZus{}time}\PY{o}{=}\PY{l+m+mi}{20}\PY{p}{,}
                                            \PY{n}{label}\PY{o}{=}\PY{n}{jobid}\PY{p}{)}
            \PY{n}{rawoutput} \PY{o}{=} \PY{n}{sampleset}\PY{o}{.}\PY{n}{aggregate}\PY{p}{(}\PY{p}{)}
        \PY{k}{else}\PY{p}{:}
            \PY{n}{sampler} \PY{o}{=} \PY{n}{SimulatedAnnealingSampler}\PY{p}{(}\PY{p}{)}
            \PY{n}{sampleset} \PY{o}{=} \PY{n}{sampler}\PY{o}{.}\PY{n}{sample\PYZus{}qubo}\PY{p}{(}\PY{n}{Q}\PY{p}{,}\PY{n}{num\PYZus{}reads}\PY{o}{=}\PY{n}{shots}\PY{p}{)}
            \PY{n}{rawoutput} \PY{o}{=} \PY{n}{sampleset}\PY{o}{.}\PY{n}{aggregate}\PY{p}{(}\PY{p}{)}

        \PY{n}{minimumevalue} \PY{o}{=} \PY{l+m+mf}{100.0}
        \PY{n}{minimuma} \PY{o}{=} \PY{p}{[}\PY{p}{]}
        \PY{n}{minimumunita} \PY{o}{=} \PY{p}{[}\PY{p}{]}
        \PY{n}{warning} \PY{o}{=} \PY{l+m+mi}{0}
        \PY{n}{chaincount} \PY{o}{=} \PY{l+m+mi}{0}

        \PY{k}{for} \PY{n}{irow} \PY{o+ow}{in} \PY{n+nb}{range}\PY{p}{(}\PY{n+nb}{len}\PY{p}{(}\PY{n}{rawoutput}\PY{o}{.}\PY{n}{record}\PY{p}{)}\PY{p}{)}\PY{p}{:}
            \PY{k}{if} \PY{n}{quantum}\PY{o}{==}\PY{l+m+mi}{1}\PY{p}{:}
                \PY{n}{chain} \PY{o}{=} \PY{n}{rawoutput}\PY{o}{.}\PY{n}{record}\PY{p}{[}\PY{n}{irow}\PY{p}{]}\PY{p}{[}\PY{l+m+mi}{3}\PY{p}{]}
            \PY{n}{numoc} \PY{o}{=} \PY{n}{rawoutput}\PY{o}{.}\PY{n}{record}\PY{p}{[}\PY{n}{irow}\PY{p}{]}\PY{p}{[}\PY{l+m+mi}{2}\PY{p}{]}
            \PY{n}{a} \PY{o}{=} \PY{p}{[}\PY{p}{]}
            \PY{k}{for} \PY{n}{alphaminus1} \PY{o+ow}{in} \PY{n+nb}{range}\PY{p}{(}\PY{l+m+mi}{2}\PY{o}{*}\PY{n}{dimH}\PY{p}{)}\PY{p}{:}
                \PY{n}{a}\PY{o}{.}\PY{n}{append}\PY{p}{(}\PY{l+m+mi}{0}\PY{p}{)}
                \PY{k}{for} \PY{n}{kminus1} \PY{o+ow}{in} \PY{n+nb}{range}\PY{p}{(}\PY{n}{Kval}\PY{o}{\PYZhy{}}\PY{l+m+mi}{1}\PY{p}{)}\PY{p}{:}
                    \PY{n}{i} \PY{o}{=} \PY{n}{Kval}\PY{o}{*}\PY{n}{alphaminus1} \PY{o}{+} \PY{n}{kminus1}
                    \PY{n}{a}\PY{p}{[}\PY{n}{alphaminus1}\PY{p}{]} \PY{o}{+}\PY{o}{=} \PY{l+m+mi}{2}\PY{o}{*}\PY{o}{*}\PY{p}{(}\PY{l+m+mi}{1}\PY{o}{+}\PY{n}{kminus1}\PY{o}{\PYZhy{}}\PY{n}{Kval}\PY{o}{\PYZhy{}}\PY{n}{zoom}\PY{p}{)}\PY{o}{*}\PY{n}{rawoutput}\PY{o}{.}\PY{n}{record}\PY{p}{[}\PY{n}{irow}\PY{p}{]}\PY{p}{[}\PY{l+m+mi}{0}\PY{p}{]}\PY{p}{[}\PY{n}{i}\PY{p}{]}
                \PY{n}{i} \PY{o}{=} \PY{n}{Kval}\PY{o}{*}\PY{n}{alphaminus1} \PY{o}{+} \PY{n}{Kval} \PY{o}{\PYZhy{}} \PY{l+m+mi}{1}
                \PY{n}{a}\PY{p}{[}\PY{n}{alphaminus1}\PY{p}{]} \PY{o}{+}\PY{o}{=} \PY{n}{acenter}\PY{p}{[}\PY{n}{alphaminus1}\PY{p}{]}\PY{o}{\PYZhy{}}\PY{l+m+mi}{2}\PY{o}{*}\PY{o}{*}\PY{p}{(}\PY{o}{\PYZhy{}}\PY{n}{zoom}\PY{p}{)}\PY{o}{*}\PY{n}{rawoutput}\PY{o}{.}\PY{n}{record}\PY{p}{[}\PY{n}{irow}\PY{p}{]}\PY{p}{[}\PY{l+m+mi}{0}\PY{p}{]}\PY{p}{[}\PY{n}{i}\PY{p}{]}
            \PY{n}{anorm} \PY{o}{=} \PY{n}{np}\PY{o}{.}\PY{n}{sqrt}\PY{p}{(}\PY{n+nb}{sum}\PY{p}{(}\PY{n}{a}\PY{p}{[}\PY{n}{i}\PY{p}{]}\PY{o}{*}\PY{o}{*}\PY{l+m+mi}{2} \PY{k}{for} \PY{n}{i} \PY{o+ow}{in} \PY{n+nb}{range}\PY{p}{(}\PY{l+m+mi}{2}\PY{o}{*}\PY{n}{dimH}\PY{p}{)}\PY{p}{)}\PY{p}{)}
            \PY{k}{if} \PY{n}{anorm}\PY{o}{\PYZlt{}}\PY{l+m+mf}{1.0e\PYZhy{}6}\PY{p}{:}
                \PY{n}{warning} \PY{o}{+}\PY{o}{=} \PY{n}{numoc}
            \PY{k}{else}\PY{p}{:}
                \PY{n}{unita} \PY{o}{=} \PY{p}{[}\PY{n}{a}\PY{p}{[}\PY{n}{i}\PY{p}{]}\PY{o}{/}\PY{n}{anorm} \PY{k}{for} \PY{n}{i} \PY{o+ow}{in} \PY{n+nb}{range}\PY{p}{(}\PY{l+m+mi}{2}\PY{o}{*}\PY{n}{dimH}\PY{p}{)}\PY{p}{]}
                \PY{n}{Cvec} \PY{o}{=} \PY{n}{np}\PY{o}{.}\PY{n}{array}\PY{p}{(}\PY{n}{unita}\PY{p}{[}\PY{p}{:}\PY{p}{:}\PY{l+m+mi}{2}\PY{p}{]}\PY{p}{)} \PY{o}{+} \PY{n}{np}\PY{o}{.}\PY{n}{array}\PY{p}{(}\PY{n}{unita}\PY{p}{[}\PY{l+m+mi}{1}\PY{p}{:}\PY{p}{:}\PY{l+m+mi}{2}\PY{p}{]}\PY{p}{)} \PY{o}{*} \PY{l+m+mi}{1}\PY{n}{j}
                \PY{n}{evalue} \PY{o}{=} \PY{n}{np}\PY{o}{.}\PY{n}{real}\PY{p}{(}\PY{n}{np}\PY{o}{.}\PY{n}{matmul}\PY{p}{(}\PY{n}{np}\PY{o}{.}\PY{n}{conj}\PY{p}{(}\PY{n}{Cvec}\PY{p}{)}\PY{p}{,}\PY{n}{np}\PY{o}{.}\PY{n}{matmul}\PY{p}{(}\PY{n}{Cmat}\PY{p}{,}\PY{n}{Cvec}\PY{p}{)}\PY{p}{)}\PY{p}{)}
                \PY{k}{if} \PY{n}{quantum}\PY{o}{==}\PY{l+m+mi}{1}\PY{p}{:}
                    \PY{k}{if} \PY{n}{chain}\PY{o}{\PYZgt{}}\PY{l+m+mf}{1.0e\PYZhy{}6}\PY{p}{:}
                        \PY{n}{chaincount} \PY{o}{+}\PY{o}{=} \PY{l+m+mi}{1}
                \PY{n}{minimumevalue} \PY{o}{=} \PY{n+nb}{min}\PY{p}{(}\PY{n}{evalue}\PY{p}{,}\PY{n}{minimumevalue}\PY{p}{)}
                \PY{k}{if} \PY{n}{evalue}\PY{o}{==}\PY{n}{minimumevalue}\PY{p}{:}
                    \PY{n}{minimuma} \PY{o}{=} \PY{n}{a}
                    \PY{n}{minimumunita} \PY{o}{=} \PY{n}{unita}
                    \PY{k}{if} \PY{n}{quantum}\PY{o}{==}\PY{l+m+mi}{1}\PY{p}{:}
                        \PY{n}{minimumchain} \PY{o}{=} \PY{n}{chain}
                
        \PY{n}{acenter} \PY{o}{=} \PY{n}{minimumunita}
        \PY{n}{Cvec} \PY{o}{=} \PY{n}{np}\PY{o}{.}\PY{n}{zeros}\PY{p}{(}\PY{n}{dimH}\PY{p}{,}\PY{n}{dtype}\PY{o}{=}\PY{n}{np}\PY{o}{.}\PY{n}{complex\PYZus{}}\PY{p}{)}
        \PY{n}{Cvec} \PY{o}{=} \PY{n}{np}\PY{o}{.}\PY{n}{array}\PY{p}{(}\PY{n}{minimumunita}\PY{p}{[}\PY{p}{:}\PY{p}{:}\PY{l+m+mi}{2}\PY{p}{]}\PY{p}{)} \PY{o}{+} \PY{n}{np}\PY{o}{.}\PY{n}{array}\PY{p}{(}\PY{n}{minimumunita}\PY{p}{[}\PY{l+m+mi}{1}\PY{p}{:}\PY{p}{:}\PY{l+m+mi}{2}\PY{p}{]}\PY{p}{)} \PY{o}{*} \PY{l+m+mi}{1}\PY{n}{j}
        \PY{n}{listvec}\PY{o}{.}\PY{n}{append}\PY{p}{(}\PY{n}{Cvec}\PY{p}{)}
        \PY{n}{listen}\PY{o}{.}\PY{n}{append}\PY{p}{(}\PY{n}{np}\PY{o}{.}\PY{n}{real}\PY{p}{(}\PY{n}{np}\PY{o}{.}\PY{n}{matmul}\PY{p}{(}\PY{n}{np}\PY{o}{.}\PY{n}{conj}\PY{p}{(}\PY{n}{Cvec}\PY{p}{)}\PY{p}{,}\PY{n}{np}\PY{o}{.}\PY{n}{matmul}\PY{p}{(}\PY{n}{Cmat}\PY{p}{,}\PY{n}{Cvec}\PY{p}{)}\PY{p}{)}\PY{p}{)}\PY{p}{)}
    \PY{k}{return}\PY{p}{(}\PY{n}{listen}\PY{p}{,}\PY{n}{listvec}\PY{p}{)}
\end{Verbatim}
\end{tcolorbox}

As an example, we show how to compute the time evolution of the four-neutrino systems.
Using $\delta t=1.1$ and two time slices $n_T=2$,
``raw results'' are first generated, and then one iterative step is applied to reduce the uncertainty. 
We start by importing the Hamiltonian constructed using {\tt Mathematica}, and build the Feynman clock Hamiltonian:
\begin{tcolorbox}[breakable, size=fbox, boxrule=1pt, pad at break*=1mm,colback=cellbackground, colframe=cellborder,fontupper=\footnotesize]
\prompt{In}{incolor}{26}{\boxspacing}
\begin{Verbatim}[commandchars=\\\{\}]
\PY{n}{ham} \PY{o}{=} \PY{n}{np}\PY{o}{.}\PY{n}{genfromtxt}\PY{p}{(}\PY{l+s+s1}{\PYZsq{}}\PY{l+s+s1}{hamiltonian\PYZus{}Neutrinos.dat}\PY{l+s+s1}{\PYZsq{}}\PY{p}{)}
\PY{n}{dimMat} \PY{o}{=} \PY{l+m+mi}{16}
\PY{n}{initvec} \PY{o}{=} \PY{p}{[}\PY{l+m+mi}{0}\PY{p}{,} \PY{l+m+mi}{0}\PY{p}{,} \PY{l+m+mi}{0}\PY{p}{,} \PY{l+m+mi}{1}\PY{p}{,} \PY{l+m+mi}{0}\PY{p}{,} \PY{l+m+mi}{0}\PY{p}{,} \PY{l+m+mi}{0}\PY{p}{,} \PY{l+m+mi}{0}\PY{p}{,} \PY{l+m+mi}{0}\PY{p}{,} \PY{l+m+mi}{0}\PY{p}{,} \PY{l+m+mi}{0}\PY{p}{,} \PY{l+m+mi}{0}\PY{p}{,} \PY{l+m+mi}{0}\PY{p}{,} \PY{l+m+mi}{0}\PY{p}{,} \PY{l+m+mi}{0}\PY{p}{,} \PY{l+m+mi}{0}\PY{p}{]}
\end{Verbatim}
\end{tcolorbox}

\begin{tcolorbox}[breakable, size=fbox, boxrule=1pt, pad at break*=1mm,colback=cellbackground, colframe=cellborder,fontupper=\footnotesize]
\prompt{In}{incolor}{27}{\boxspacing}
\begin{Verbatim}[commandchars=\\\{\}]
\PY{n}{evSteps} \PY{o}{=} \PY{l+m+mi}{2}
\PY{n}{timestep} \PY{o}{=} \PY{l+m+mf}{1.1}
\PY{n}{hamre} \PY{o}{=} \PY{n}{np}\PY{o}{.}\PY{n}{real}\PY{p}{(}\PY{n}{scipy}\PY{o}{.}\PY{n}{linalg}\PY{o}{.}\PY{n}{expm}\PY{p}{(}\PY{o}{\PYZhy{}}\PY{l+m+mi}{1}\PY{n}{j}\PY{o}{*}\PY{n}{timestep}\PY{o}{*}\PY{n}{ham}\PY{p}{)}\PY{p}{)}
\PY{n}{hamim} \PY{o}{=} \PY{n}{np}\PY{o}{.}\PY{n}{imag}\PY{p}{(}\PY{n}{scipy}\PY{o}{.}\PY{n}{linalg}\PY{o}{.}\PY{n}{expm}\PY{p}{(}\PY{o}{\PYZhy{}}\PY{l+m+mi}{1}\PY{n}{j}\PY{o}{*}\PY{n}{timestep}\PY{o}{*}\PY{n}{ham}\PY{p}{)}\PY{p}{)}
\PY{n}{c0} \PY{o}{=} \PY{n}{np}\PY{o}{.}\PY{n}{identity}\PY{p}{(}\PY{n}{dimMat}\PY{p}{)}\PY{o}{\PYZhy{}}\PY{n}{np}\PY{o}{.}\PY{n}{outer}\PY{p}{(}\PY{n}{initvec}\PY{p}{,}\PY{n}{initvec}\PY{p}{)}

\PY{n}{topRe}\PY{o}{=}\PY{p}{[}\PY{n}{c0}\PY{o}{+}\PY{l+m+mf}{0.5}\PY{o}{*}\PY{n}{np}\PY{o}{.}\PY{n}{identity}\PY{p}{(}\PY{n}{dimMat}\PY{p}{)}\PY{p}{,}\PY{o}{\PYZhy{}}\PY{l+m+mf}{0.5}\PY{o}{*}\PY{n}{np}\PY{o}{.}\PY{n}{transpose}\PY{p}{(}\PY{n}{hamre}\PY{p}{)}\PY{p}{]}
\PY{n}{bottomRe}\PY{o}{=}\PY{p}{[}\PY{o}{\PYZhy{}}\PY{l+m+mf}{0.5}\PY{o}{*}\PY{n}{hamre}\PY{p}{,}\PY{l+m+mf}{0.5}\PY{o}{*}\PY{n}{np}\PY{o}{.}\PY{n}{identity}\PY{p}{(}\PY{n}{dimMat}\PY{p}{)}\PY{p}{]}
\PY{n}{midRe}\PY{o}{=}\PY{p}{[}\PY{o}{\PYZhy{}}\PY{l+m+mf}{0.5}\PY{o}{*}\PY{n}{hamre}\PY{p}{,}\PY{n}{np}\PY{o}{.}\PY{n}{identity}\PY{p}{(}\PY{n}{dimMat}\PY{p}{)}\PY{p}{,}\PY{o}{\PYZhy{}}\PY{l+m+mf}{0.5}\PY{o}{*}\PY{n}{np}\PY{o}{.}\PY{n}{transpose}\PY{p}{(}\PY{n}{hamre}\PY{p}{)}\PY{p}{]}
\PY{n}{arrayRe} \PY{o}{=} \PY{p}{[}\PY{p}{]}
\PY{n}{topIm}\PY{o}{=}\PY{p}{[}\PY{n}{np}\PY{o}{.}\PY{n}{zeros}\PY{p}{(}\PY{p}{(}\PY{n}{dimMat}\PY{p}{,} \PY{n}{dimMat}\PY{p}{)}\PY{p}{)}\PY{p}{,}\PY{l+m+mf}{0.5}\PY{o}{*}\PY{n}{np}\PY{o}{.}\PY{n}{transpose}\PY{p}{(}\PY{n}{hamim}\PY{p}{)}\PY{p}{]}
\PY{n}{bottomIm}\PY{o}{=}\PY{p}{[}\PY{o}{\PYZhy{}}\PY{l+m+mf}{0.5}\PY{o}{*}\PY{n}{hamim}\PY{p}{,}\PY{n}{np}\PY{o}{.}\PY{n}{zeros}\PY{p}{(}\PY{p}{(}\PY{n}{dimMat}\PY{p}{,} \PY{n}{dimMat}\PY{p}{)}\PY{p}{)}\PY{p}{]}
\PY{n}{midIm}\PY{o}{=}\PY{p}{[}\PY{o}{\PYZhy{}}\PY{l+m+mf}{0.5}\PY{o}{*}\PY{n}{hamim}\PY{p}{,}\PY{n}{np}\PY{o}{.}\PY{n}{zeros}\PY{p}{(}\PY{p}{(}\PY{n}{dimMat}\PY{p}{,} \PY{n}{dimMat}\PY{p}{)}\PY{p}{)}\PY{p}{,}\PY{l+m+mf}{0.5}\PY{o}{*}\PY{n}{np}\PY{o}{.}\PY{n}{transpose}\PY{p}{(}\PY{n}{hamim}\PY{p}{)}\PY{p}{]}
\PY{n}{arrayIm} \PY{o}{=} \PY{p}{[}\PY{p}{]}

\PY{k}{if} \PY{n}{evSteps} \PY{o}{==} \PY{l+m+mi}{2}\PY{p}{:}
    \PY{n}{arrayRe}\PY{o}{=}\PY{p}{[}\PY{n}{topRe}\PY{p}{,}\PY{n}{bottomRe}\PY{p}{]}
    \PY{n}{arrayIm}\PY{o}{=}\PY{p}{[}\PY{n}{topIm}\PY{p}{,}\PY{n}{bottomIm}\PY{p}{]}
\PY{k}{else}\PY{p}{:}
    \PY{k}{for} \PY{n}{i} \PY{o+ow}{in} \PY{n+nb}{range}\PY{p}{(}\PY{n}{evSteps}\PY{o}{\PYZhy{}}\PY{l+m+mi}{2}\PY{p}{)}\PY{p}{:}
        \PY{n}{topRe}\PY{o}{.}\PY{n}{append}\PY{p}{(}\PY{n}{np}\PY{o}{.}\PY{n}{zeros}\PY{p}{(}\PY{p}{(}\PY{n}{dimMat}\PY{p}{,} \PY{n}{dimMat}\PY{p}{)}\PY{p}{)}\PY{p}{)}
        \PY{n}{bottomRe}\PY{o}{.}\PY{n}{insert}\PY{p}{(}\PY{l+m+mi}{0}\PY{p}{,}\PY{n}{np}\PY{o}{.}\PY{n}{zeros}\PY{p}{(}\PY{p}{(}\PY{n}{dimMat}\PY{p}{,} \PY{n}{dimMat}\PY{p}{)}\PY{p}{)}\PY{p}{)}
        \PY{n}{topIm}\PY{o}{.}\PY{n}{append}\PY{p}{(}\PY{n}{np}\PY{o}{.}\PY{n}{zeros}\PY{p}{(}\PY{p}{(}\PY{n}{dimMat}\PY{p}{,} \PY{n}{dimMat}\PY{p}{)}\PY{p}{)}\PY{p}{)}
        \PY{n}{bottomIm}\PY{o}{.}\PY{n}{insert}\PY{p}{(}\PY{l+m+mi}{0}\PY{p}{,}\PY{n}{np}\PY{o}{.}\PY{n}{zeros}\PY{p}{(}\PY{p}{(}\PY{n}{dimMat}\PY{p}{,} \PY{n}{dimMat}\PY{p}{)}\PY{p}{)}\PY{p}{)}
    \PY{k}{for} \PY{n}{i} \PY{o+ow}{in} \PY{n+nb}{range}\PY{p}{(}\PY{n}{evSteps}\PY{o}{\PYZhy{}}\PY{l+m+mi}{3}\PY{p}{)}\PY{p}{:}
        \PY{n}{midRe}\PY{o}{.}\PY{n}{append}\PY{p}{(}\PY{n}{np}\PY{o}{.}\PY{n}{zeros}\PY{p}{(}\PY{p}{(}\PY{n}{dimMat}\PY{p}{,} \PY{n}{dimMat}\PY{p}{)}\PY{p}{)}\PY{p}{)}
        \PY{n}{midIm}\PY{o}{.}\PY{n}{append}\PY{p}{(}\PY{n}{np}\PY{o}{.}\PY{n}{zeros}\PY{p}{(}\PY{p}{(}\PY{n}{dimMat}\PY{p}{,} \PY{n}{dimMat}\PY{p}{)}\PY{p}{)}\PY{p}{)}
        
    \PY{n}{arrayRe}\PY{o}{.}\PY{n}{append}\PY{p}{(}\PY{n}{topRe}\PY{p}{)}
    \PY{n}{arrayIm}\PY{o}{.}\PY{n}{append}\PY{p}{(}\PY{n}{topIm}\PY{p}{)}
    \PY{k}{for} \PY{n}{n} \PY{o+ow}{in} \PY{n+nb}{range}\PY{p}{(}\PY{n}{evSteps}\PY{o}{\PYZhy{}}\PY{l+m+mi}{2}\PY{p}{)}\PY{p}{:}
        \PY{n}{arrayRe}\PY{o}{.}\PY{n}{append}\PY{p}{(}\PY{n}{midRe}\PY{p}{[}\PY{o}{\PYZhy{}}\PY{n}{n}\PY{p}{:}\PY{p}{]}\PY{o}{+}\PY{n}{midRe}\PY{p}{[}\PY{p}{:}\PY{o}{\PYZhy{}}\PY{n}{n}\PY{p}{]}\PY{p}{)}
        \PY{n}{arrayIm}\PY{o}{.}\PY{n}{append}\PY{p}{(}\PY{n}{midIm}\PY{p}{[}\PY{o}{\PYZhy{}}\PY{n}{n}\PY{p}{:}\PY{p}{]}\PY{o}{+}\PY{n}{midIm}\PY{p}{[}\PY{p}{:}\PY{o}{\PYZhy{}}\PY{n}{n}\PY{p}{]}\PY{p}{)}
    \PY{n}{arrayRe}\PY{o}{.}\PY{n}{append}\PY{p}{(}\PY{n}{bottomRe}\PY{p}{)}
    \PY{n}{arrayIm}\PY{o}{.}\PY{n}{append}\PY{p}{(}\PY{n}{bottomIm}\PY{p}{)}
    
\PY{n}{Feynhamre}\PY{o}{=}\PY{n}{np}\PY{o}{.}\PY{n}{block}\PY{p}{(}\PY{n}{arrayRe}\PY{p}{)}
\PY{n}{Feynhamim}\PY{o}{=}\PY{n}{np}\PY{o}{.}\PY{n}{block}\PY{p}{(}\PY{n}{arrayIm}\PY{p}{)}
\end{Verbatim}
\end{tcolorbox}
Then, the embedding of the clock Hamiltonian is determined, and subsequently used to compute the ground-state energy and wavefunction (again, using $N_{\rm run}=20$ samples for error estimation).
\begin{tcolorbox}[breakable, size=fbox, boxrule=1pt, pad at break*=1mm,colback=cellbackground, colframe=cellborder,fontupper=\footnotesize]
\prompt{In}{incolor}{28}{\boxspacing}
\begin{Verbatim}[commandchars=\\\{\}]
\PY{n}{emNeut} \PY{o}{=} \PY{n}{hermitian\PYZus{}embedding}\PY{p}{(}\PY{n}{dimMat}\PY{o}{*}\PY{n}{evSteps}\PY{p}{,} \PY{n}{Feynhamre}\PY{p}{,} \PY{n}{Feynhamim}\PY{p}{,} \PY{l+m+mi}{2}\PY{p}{,} \PY{l+m+mi}{0}\PY{p}{,} \PY{n}{tokenval}\PY{p}{)}
\end{Verbatim}
\end{tcolorbox}

\begin{tcolorbox}[breakable, size=fbox, boxrule=1pt, pad at break*=1mm,colback=cellbackground, colframe=cellborder,fontupper=\footnotesize]
\prompt{In}{incolor}{29}{\boxspacing}
\begin{Verbatim}[commandchars=\\\{\}]
\PY{n}{res11\PYZus{}results} \PY{o}{=} \PY{p}{[}\PY{p}{]}
\PY{k}{for} \PY{n}{val} \PY{o+ow}{in} \PY{n}{np}\PY{o}{.}\PY{n}{arange}\PY{p}{(}\PY{l+m+mi}{20}\PY{p}{)}\PY{p}{:}
    \PY{n}{res11} \PY{o}{=} \PY{n}{hermitian\PYZus{}energy}\PY{p}{(}\PY{n}{dimMat}\PY{o}{*}\PY{n}{evSteps}\PY{p}{,} \PY{n}{Feynhamre}\PY{p}{,} \PY{n}{Feynhamim}\PY{p}{,} \PY{l+m+mi}{2}\PY{p}{,} \PY{l+m+mi}{0}\PY{p}{,} \PY{l+m+mi}{1000}\PY{p}{,} \PY{l+m+mi}{15}\PY{p}{,} \PY{l+m+mi}{1}\PY{p}{,} \PY{n}{emNeut}\PY{p}{,} \PY{n}{tokenval}\PY{p}{,} \PY{l+m+mf}{0.2}\PY{p}{,} \PY{l+s+s1}{\PYZsq{}}\PY{l+s+s1}{neut}\PY{l+s+s1}{\PYZsq{}}\PY{p}{,}
                             \PY{p}{[}\PY{l+m+mi}{0}\PY{p}{]}\PY{o}{*}\PY{l+m+mi}{2}\PY{o}{*}\PY{n}{dimMat}\PY{o}{*}\PY{n}{evSteps}\PY{p}{,} \PY{l+m+mi}{0}\PY{p}{)}
    \PY{n}{res11\PYZus{}results}\PY{o}{.}\PY{n}{append}\PY{p}{(}\PY{n}{res11}\PY{p}{)}
\end{Verbatim}
\end{tcolorbox}
The resulting energies and wavefunctions are saved, and the wavefunction with the lowest energy is selected to be used as the starting point of a new anneal, with $z^{\rm init}=4$.
\begin{tcolorbox}[breakable, size=fbox, boxrule=1pt, pad at break*=1mm,colback=cellbackground, colframe=cellborder,fontupper=\footnotesize]
\prompt{In}{incolor}{30}{\boxspacing}
\begin{Verbatim}[commandchars=\\\{\}]
\PY{k}{for} \PY{n}{zoom} \PY{o+ow}{in} \PY{n+nb}{range}\PY{p}{(}\PY{l+m+mi}{15}\PY{p}{)}\PY{p}{:}
    \PY{n}{energy} \PY{o}{=} \PY{p}{[}\PY{n}{i}\PY{p}{[}\PY{l+m+mi}{0}\PY{p}{]}\PY{p}{[}\PY{n}{zoom}\PY{p}{]} \PY{k}{for} \PY{n}{i} \PY{o+ow}{in} \PY{n}{res11\PYZus{}results}\PY{p}{]}
    \PY{n}{np}\PY{o}{.}\PY{n}{savetxt}\PY{p}{(}\PY{l+s+s1}{\PYZsq{}}\PY{l+s+s1}{Neutrinos\PYZus{}eta0\PYZus{}z}\PY{l+s+s1}{\PYZsq{}}\PY{o}{+}\PY{n+nb}{str}\PY{p}{(}\PY{n}{zoom}\PY{p}{)}\PY{o}{+}\PY{l+s+s1}{\PYZsq{}}\PY{l+s+s1}{\PYZus{}K2\PYZus{}shots1000\PYZus{}dt1.1\PYZus{}T2\PYZus{}state0\PYZus{}en.dat}\PY{l+s+s1}{\PYZsq{}}\PY{p}{,}\PY{n}{energy}\PY{p}{)}
    \PY{n}{vectors} \PY{o}{=} \PY{p}{[}\PY{n}{i}\PY{p}{[}\PY{l+m+mi}{1}\PY{p}{]}\PY{p}{[}\PY{n}{zoom}\PY{p}{]} \PY{k}{for} \PY{n}{i} \PY{o+ow}{in} \PY{n}{res11\PYZus{}results}\PY{p}{]}
    \PY{n}{np}\PY{o}{.}\PY{n}{savetxt}\PY{p}{(}\PY{l+s+s1}{\PYZsq{}}\PY{l+s+s1}{Neutrinos\PYZus{}eta0\PYZus{}z}\PY{l+s+s1}{\PYZsq{}}\PY{o}{+}\PY{n+nb}{str}\PY{p}{(}\PY{n}{zoom}\PY{p}{)}\PY{o}{+}\PY{l+s+s1}{\PYZsq{}}\PY{l+s+s1}{\PYZus{}K2\PYZus{}shots1000\PYZus{}dt1.1\PYZus{}T2\PYZus{}state0\PYZus{}vec\PYZus{}re.dat}\PY{l+s+s1}{\PYZsq{}}\PY{p}{,}\PY{n}{np}\PY{o}{.}\PY{n}{real}\PY{p}{(}\PY{n}{vectors}\PY{p}{)}\PY{p}{)}    
    \PY{n}{np}\PY{o}{.}\PY{n}{savetxt}\PY{p}{(}\PY{l+s+s1}{\PYZsq{}}\PY{l+s+s1}{Neutrinos\PYZus{}eta0\PYZus{}z}\PY{l+s+s1}{\PYZsq{}}\PY{o}{+}\PY{n+nb}{str}\PY{p}{(}\PY{n}{zoom}\PY{p}{)}\PY{o}{+}\PY{l+s+s1}{\PYZsq{}}\PY{l+s+s1}{\PYZus{}K2\PYZus{}shots1000\PYZus{}dt1.1\PYZus{}T2\PYZus{}state0\PYZus{}vec\PYZus{}im.dat}\PY{l+s+s1}{\PYZsq{}}\PY{p}{,}\PY{n}{np}\PY{o}{.}\PY{n}{imag}\PY{p}{(}\PY{n}{vectors}\PY{p}{)}\PY{p}{)}
\end{Verbatim}
\end{tcolorbox}

\begin{tcolorbox}[breakable, size=fbox, boxrule=1pt, pad at break*=1mm,colback=cellbackground, colframe=cellborder,fontupper=\footnotesize]
\prompt{In}{incolor}{31}{\boxspacing}
\begin{Verbatim}[commandchars=\\\{\}]
\PY{n}{energy} \PY{o}{=} \PY{n+nb}{list}\PY{p}{(}\PY{n}{np}\PY{o}{.}\PY{n}{loadtxt}\PY{p}{(}\PY{l+s+s1}{\PYZsq{}}\PY{l+s+s1}{Neutrinos\PYZus{}eta0\PYZus{}z14\PYZus{}K2\PYZus{}shots1000\PYZus{}dt1.1\PYZus{}T2\PYZus{}state0\PYZus{}en.dat}\PY{l+s+s1}{\PYZsq{}}\PY{p}{)}\PY{p}{)}
\PY{n}{locmin} \PY{o}{=} \PY{n}{energy}\PY{o}{.}\PY{n}{index}\PY{p}{(}\PY{n+nb}{min}\PY{p}{(}\PY{n}{energy}\PY{p}{)}\PY{p}{)}
\PY{n}{wvfR} \PY{o}{=} \PY{n+nb}{list}\PY{p}{(}\PY{n}{np}\PY{o}{.}\PY{n}{loadtxt}\PY{p}{(}\PY{l+s+s1}{\PYZsq{}}\PY{l+s+s1}{Neutrinos\PYZus{}eta0\PYZus{}z14\PYZus{}K2\PYZus{}shots1000\PYZus{}dt1.1\PYZus{}T2\PYZus{}state0\PYZus{}vec\PYZus{}re.dat}\PY{l+s+s1}{\PYZsq{}}\PY{p}{)}\PY{p}{)}\PY{p}{[}\PY{n}{locmin}\PY{p}{]}
\PY{n}{wvfI} \PY{o}{=} \PY{n+nb}{list}\PY{p}{(}\PY{n}{np}\PY{o}{.}\PY{n}{loadtxt}\PY{p}{(}\PY{l+s+s1}{\PYZsq{}}\PY{l+s+s1}{Neutrinos\PYZus{}eta0\PYZus{}z14\PYZus{}K2\PYZus{}shots1000\PYZus{}dt1.1\PYZus{}T2\PYZus{}state0\PYZus{}vec\PYZus{}im.dat}\PY{l+s+s1}{\PYZsq{}}\PY{p}{)}\PY{p}{)}\PY{p}{[}\PY{n}{locmin}\PY{p}{]}
\PY{n}{acenter} \PY{o}{=} \PY{n}{np}\PY{o}{.}\PY{n}{transpose}\PY{p}{(}\PY{n}{np}\PY{o}{.}\PY{n}{concatenate}\PY{p}{(}\PY{p}{(}\PY{p}{[}\PY{n}{wvfR}\PY{p}{]}\PY{p}{,} \PY{p}{[}\PY{n}{wvfI}\PY{p}{]}\PY{p}{)}\PY{p}{)}\PY{p}{)}\PY{o}{.}\PY{n}{flatten}\PY{p}{(}\PY{p}{)}
\end{Verbatim}
\end{tcolorbox}

\begin{tcolorbox}[breakable, size=fbox, boxrule=1pt, pad at break*=1mm,colback=cellbackground, colframe=cellborder,fontupper=\footnotesize]
\prompt{In}{incolor}{32}{\boxspacing}
\begin{Verbatim}[commandchars=\\\{\}]
\PY{n}{res11b\PYZus{}results} \PY{o}{=} \PY{p}{[}\PY{p}{]}
\PY{k}{for} \PY{n}{val} \PY{o+ow}{in} \PY{n}{np}\PY{o}{.}\PY{n}{arange}\PY{p}{(}\PY{l+m+mi}{20}\PY{p}{)}\PY{p}{:}
    \PY{n}{res11} \PY{o}{=} \PY{n}{hermitian\PYZus{}energy}\PY{p}{(}\PY{n}{dimMat}\PY{o}{*}\PY{n}{evSteps}\PY{p}{,} \PY{n}{Feynhamre}\PY{p}{,} \PY{n}{Feynhamim}\PY{p}{,} \PY{l+m+mi}{2}\PY{p}{,} \PY{l+m+mi}{0}\PY{p}{,} \PY{l+m+mi}{1000}\PY{p}{,} \PY{l+m+mi}{15}\PY{o}{+}\PY{l+m+mi}{4}\PY{p}{,} \PY{l+m+mi}{1}\PY{p}{,} \PY{n}{emNeut}\PY{p}{,} \PY{n}{tokenval}\PY{p}{,} \PY{l+m+mf}{0.2}\PY{p}{,} \PY{l+s+s1}{\PYZsq{}}\PY{l+s+s1}{neut}\PY{l+s+s1}{\PYZsq{}}\PY{p}{,}
                             \PY{n}{acenter}\PY{p}{,} \PY{l+m+mi}{4}\PY{p}{)}
    \PY{n}{res11b\PYZus{}results}\PY{o}{.}\PY{n}{append}\PY{p}{(}\PY{n}{res11}\PY{p}{)}
\end{Verbatim}
\end{tcolorbox}

\begin{tcolorbox}[breakable, size=fbox, boxrule=1pt, pad at break*=1mm,colback=cellbackground, colframe=cellborder,fontupper=\footnotesize]
\prompt{In}{incolor}{33}{\boxspacing}
\begin{Verbatim}[commandchars=\\\{\}]
\PY{k}{for} \PY{n}{zoom} \PY{o+ow}{in} \PY{n+nb}{range}\PY{p}{(}\PY{l+m+mi}{15}\PY{p}{)}\PY{p}{:}
    \PY{n}{energy} \PY{o}{=} \PY{p}{[}\PY{n}{i}\PY{p}{[}\PY{l+m+mi}{0}\PY{p}{]}\PY{p}{[}\PY{n}{zoom}\PY{p}{]} \PY{k}{for} \PY{n}{i} \PY{o+ow}{in} \PY{n}{res11b\PYZus{}results}\PY{p}{]}
    \PY{n}{np}\PY{o}{.}\PY{n}{savetxt}\PY{p}{(}\PY{l+s+s1}{\PYZsq{}}\PY{l+s+s1}{Neutrinos\PYZus{}eta0\PYZus{}z}\PY{l+s+s1}{\PYZsq{}}\PY{o}{+}\PY{n+nb}{str}\PY{p}{(}\PY{n}{zoom}\PY{p}{)}\PY{o}{+}\PY{l+s+s1}{\PYZsq{}}\PY{l+s+s1}{\PYZus{}K2\PYZus{}shots1000\PYZus{}dt1.1\PYZus{}T2\PYZus{}zinit4\PYZus{}state0\PYZus{}en.dat}\PY{l+s+s1}{\PYZsq{}}\PY{p}{,}\PY{n}{energy}\PY{p}{)}
    \PY{n}{vectors} \PY{o}{=} \PY{p}{[}\PY{n}{i}\PY{p}{[}\PY{l+m+mi}{1}\PY{p}{]}\PY{p}{[}\PY{n}{zoom}\PY{p}{]} \PY{k}{for} \PY{n}{i} \PY{o+ow}{in} \PY{n}{res11b\PYZus{}results}\PY{p}{]}
    \PY{n}{np}\PY{o}{.}\PY{n}{savetxt}\PY{p}{(}\PY{l+s+s1}{\PYZsq{}}\PY{l+s+s1}{Neutrinos\PYZus{}eta0\PYZus{}z}\PY{l+s+s1}{\PYZsq{}}\PY{o}{+}\PY{n+nb}{str}\PY{p}{(}\PY{n}{zoom}\PY{p}{)}\PY{o}{+}\PY{l+s+s1}{\PYZsq{}}\PY{l+s+s1}{\PYZus{}K2\PYZus{}shots1000\PYZus{}dt1.1\PYZus{}T2\PYZus{}zinit4\PYZus{}state0\PYZus{}vec\PYZus{}re.dat}\PY{l+s+s1}{\PYZsq{}}\PY{p}{,}\PY{n}{np}\PY{o}{.}\PY{n}{real}\PY{p}{(}\PY{n}{vectors}\PY{p}{)}\PY{p}{)}    
    \PY{n}{np}\PY{o}{.}\PY{n}{savetxt}\PY{p}{(}\PY{l+s+s1}{\PYZsq{}}\PY{l+s+s1}{Neutrinos\PYZus{}eta0\PYZus{}z}\PY{l+s+s1}{\PYZsq{}}\PY{o}{+}\PY{n+nb}{str}\PY{p}{(}\PY{n}{zoom}\PY{p}{)}\PY{o}{+}\PY{l+s+s1}{\PYZsq{}}\PY{l+s+s1}{\PYZus{}K2\PYZus{}shots1000\PYZus{}dt1.1\PYZus{}T2\PYZus{}zinit4\PYZus{}state0\PYZus{}vec\PYZus{}im.dat}\PY{l+s+s1}{\PYZsq{}}\PY{p}{,}\PY{n}{np}\PY{o}{.}\PY{n}{imag}\PY{p}{(}\PY{n}{vectors}\PY{p}{)}\PY{p}{)}
\end{Verbatim}
\end{tcolorbox}



\end{document}